\documentstyle[eqsecnum,aps,prb,multicol,epsf]{revtex}

\def\top#1{\vskip #1\begin{picture}(290,80)(80,500)\thinlines \put(
65,500){\line( 1, 0){255}}\put(320,500){\line( 0, 1){
5}}\end{picture}}
\def\bottom#1{\vskip #1\begin{picture}(290,80)(80,500)\thinlines \put(
330,500){\line( 1, 0){255}}\put(330,500){\line( 0, -1){
5}}\end{picture}}

\begin{document}
\tighten

\title{Smectic Liquid Crystals in Random Environments}
\author{Leo Radzihovsky}
\address{Department of Physics, University of Colorado,
Boulder, CO 80309}
\author{John Toner} \address{Department of Physics,
Materials Science Institute, and Institute of Theoretical Science,
University of Oregon, Eugene, OR 97403-1274}

\date{\today}
\maketitle
\begin{abstract}

We study smectic liquid crystals in random environments, e.g.,
aerogel. A low temperature analysis reveals that even arbitrarily weak
{\em quenched} disorder (i.e., arbitrarily low aerogel density)
destroys translational (smectic) order, in agreement with recent
experimental results.  A harmonic approximation to the {\em elastic}
energy suggests that there may be no ``smectic Bragg glass'' phase in
this system: even at zero temperature, it is riddled with dislocation
loops induced by the quenched disorder.  This result would imply the
destruction of orientational (nematic) order as well, and that the
thermodynamically sharp NA transition is destroyed by disorder.  We
show, however, that the anharmonic elastic terms neglected in the
above approximate treatment {\em are} important (i.e., are
``relevant'' in the renormalization group sense), and may, indeed,
stabilize the smectic Bragg glass and the sharp phase transition into
it. However, they do {\em not} alter our conclusion that translational
(smectic) order is always destroyed.  In contrast, we expect that {\em
weak annealed} disorder should have no {\em qualitative} effects on
the smectic order.
\end{abstract}

\begin{multicols}{2}
\section{Introduction}
\label{introduction}

\subsection{Motivation and background}
\label{motivation_background}
The effect of quenched disorder on condensed matter systems is an
important and challenging problem that continues to be actively
investigated, because of its relevance to real systems, which always
contain some amount of random inhomogeneity. Recently, much attention
has focused on the random field XY model as a minimal model of a broad
class of systems such as disordered Josephson junction
arrays\cite{JJ}, roughening of crystal surfaces growing on a
disordered substrate\cite{TD}, the pinning of Wigner crystals, vortex
lattices in superconductors\cite{VG} or charge density
waves\cite{Gruner}. Another system that is considerably less well
understood is the superfluid transition, e.g., He$^4$ in the random,
``fractal--like'' environment of aerogel\cite{HeAerogel}. In apparent
contradiction to the Harris criterion\cite{Harris}, the disorder
modifies the critical properties of the transition (including the
critical exponents).

The central question addressed in all of the above studies is: does a
distinct low temperature ``glass'' phase exist in any of these
disordered systems?  And if so, what (if any) {\em static} property
distinguishes it from the high temperature, conventional thermally
disordered phase?  The answer to the second question is clearly
subtle: arguments dating back to Larkin\cite{Larkin} show that it is
impossible to have long-ranged translational order in three dimensions
in a randomly pinned elastic medium.  Hence, the glass cannot be
distinguished from the high temperature ``liquid'' phase by
long-ranged translational order.

In the context of pinned vortex lattices in dirty type II
superconductors, M. P. A. Fisher's\cite{VG} original argument for the
existence a vortex glass phase is based on a two (1+1)-dimensional
model in which the random pinning is relevant at low temperatures but
becomes irrelevant at higher temperatures.\cite{CardyOstlund} The low
temperature, pinning relevant phase is then identified as the glass
phase, and the temperature at which the pinning becomes irrelevant is
the thermodynamically sharp glass transition temperature.

Unfortunately, this simple scenario cannot be directly carried over to
three dimensions for vortex lattices (though, as we shall show later,
it almost can for smectics in aerogel).  In three-dimensional vortex
lattices (and other pinned isotropic elastic media listed in the
opening paragraph), the aforementioned Larkin argument shows that
disorder is {\em always} relevant in $d=3$.  Hence, its relevance
cannot be used as a criterion for distinguishing the glass and liquid
phases.

So what can?  One appealing proposal is the ``Bragg
glass''\cite{GL,GH,DSFisher} picture, in which the glass, while being
elastically disordered is topologically ordered, and is therefore
distinguished from the liquid by being free of unbound dislocation
loops, which proliferate in the liquid.  The glass transition is then
identified as an ``unbinding'' of the dislocation loops.

This transition is then, qualitatively, very similar to the melting of
the flux lattice in the absence of disorder, which can also be thought
of as an unbinding of dislocation loops.  The only difference is that
in the glass problem, the flux lattice is translationally disordered
{\em both} above and {\em below} the transition.  The absence of
defects below the transition, however, means that the low temperature
solid phase still has a finite shear modulus, leading to glassy
behavior.

Another related system that has been experimentally investigated is
nematic liquid crystals in aerogel near and below the pure system's
Nematic--Smectic-A (NA) transition temperature.\cite{review} The
mean-field theory of a bulk, pure smectic liquid crystal looks like
that of an XY model, with a complex scalar order parameter $\psi$
characterizing the smectic density wave. As was first noted by de
Gennes\cite{deGennes,dGP} a more precise model must include the
coupling of $\psi$ to the nematic director fluctuations $\delta{\bf
n}$, which takes the model out of the XY universality
class.\cite{dGP,HLM,Toner} Although the agreement between experiments
and theory is far from perfect, the {\em theoretical} consensus is
that the scaling near the NA transition {\em should} cross over from
that of a three dimensional XY transition line, (as in a neutral
superfluid) to inverted XY-like behavior with anisotropic X-ray
correlation length exponents.\cite{Toner}

The goal of this paper is to investigate the effects of disorder on
the liquid crystal phase diagram in the vicinity of and below the NA
transition. Does the NA transition survive in the presence of even
weak disorder?  If so, what is the nature of the low-temperature
phase? A condensed report of some of our results, the details and
extensions of which are presented in this paper, has appeared in
recent publications.\cite{aerogelPRL,anomPRL}.

Our interest in this problem was stimulated by recent
experiments\cite{review,Bellini,Clark,Wu,Kutnjak} that aim to answer
these very questions.  In these experiments, a liquid crystal that
exhibits a bulk Nematic--Smectic-A transition is introduced into an
aerogel.  X-ray scattering measurements show that this system {\em
never} develops true smectic long-ranged (or even quasi-long-ranged)
order: the translational correlation length $\xi^X$ remains finite at
all temperatures.  This length smoothly increases as temperature is
lowered across the bulk NA transition temperature, monotonically and
slowly rising to some finite asymptotic low temperature limit, which
does {\em not} appear to be associated with any natural length scale
of the aerogel itself. Our theory predicts this X-ray correlation
length as the scale at which the pinning disorder energy begins to
dominate over the smectic elastic energy, consistent with these
experimental observations; the smectic correlation length is {\em not}
simply ``pore'' size of the aerogel.\cite{review} This behavior is
also in sharp contrast to the fast rise of the smectic correlation
length to the nominal pore size at the bulk transition temperature
that one would expect in more regularly porous materials, which can be
understood as a bulk-like sharp transition cut-off by the finite pore
size. The specific heat in aerogel was also measured and was found to
exhibit a broadened but very well defined peak, near but slightly
shifted down from the bulk transition temperature, T$_{NA}$. These
experimental observations suggest the destruction of the NA transition
by the disorder imposed by the fractal aerogel environment.\cite{sr}
However, in contrast, recent experiments\cite{GermanoAerosil}, which
study the NA transition in even lower density aerosils, appear to show
a true resolution limited heat capacity singularity, while displaying
a {\em finite} X-ray smectic correlation length indicative of
short-range translational order. These latter experimental findings
appear to support the idea that the Nematic--Smectic-A transition in
pure systems, when confined in a low density quenched random
structure, is replaced by a new phase transition into a novel
thermodynamically stable phase with a finite smectic correlation
length in both the high and low temperature phases.

\subsection{Summary, interpretation and consequences of the results}
\label{summary}
The main conclusion of our work is that, consistent with these and
many other experiments,\cite{review,Bellini,Clark,Wu,Kutnjak} the
three dimensional smectic phase, as defined by the existence of
quasi-long-ranged translational order, is unstable to arbitrarily weak
{\em quenched} disorder (i.e., arbitrarily low aerogel or aerosil
density).  Furthermore, we find that a new {\it orientationally}
ordered low temperature ``smectic Bragg glass'' (SBG) phase replaces
the smectic phase, and a thermodynamically sharp Nematic-to-SBG
(N-SBG) transition, can survive in the presence of arbitrarily weak
disorder, if and only if two {\it universal} positive definite
anomalous exponents $\eta_B$ and $\eta_K$ satisfy the bounds
\begin{mathletters}
\begin{eqnarray}
\eta_K  + \eta_B&<& 2\;,
\label{bound_orientations}\\
\eta_K&<&1 \;,
\label{bound_dislocations}\\
\eta_B  + 5\eta_K&>& 4\;.
\label{bound_etadelta}
\end{eqnarray}
\label{bounds12}
\end{mathletters}
where the bounds in Eqs.\ref{bound_orientations},b come from the
requirement of long-ranged orientational order and the condition for
dislocations to remain confined, respectively. The region in the
$\eta_B, \eta_K$ plane that satisfies these three bounds is
illustrated in Fig.1.
\begin{figure}[bth]
{\centering
\setlength{\unitlength}{1mm}
\begin{picture}(150,60)(0,0)
\put(-20,-57){\begin{picture}(150,0)(0,0)
\includegraphics{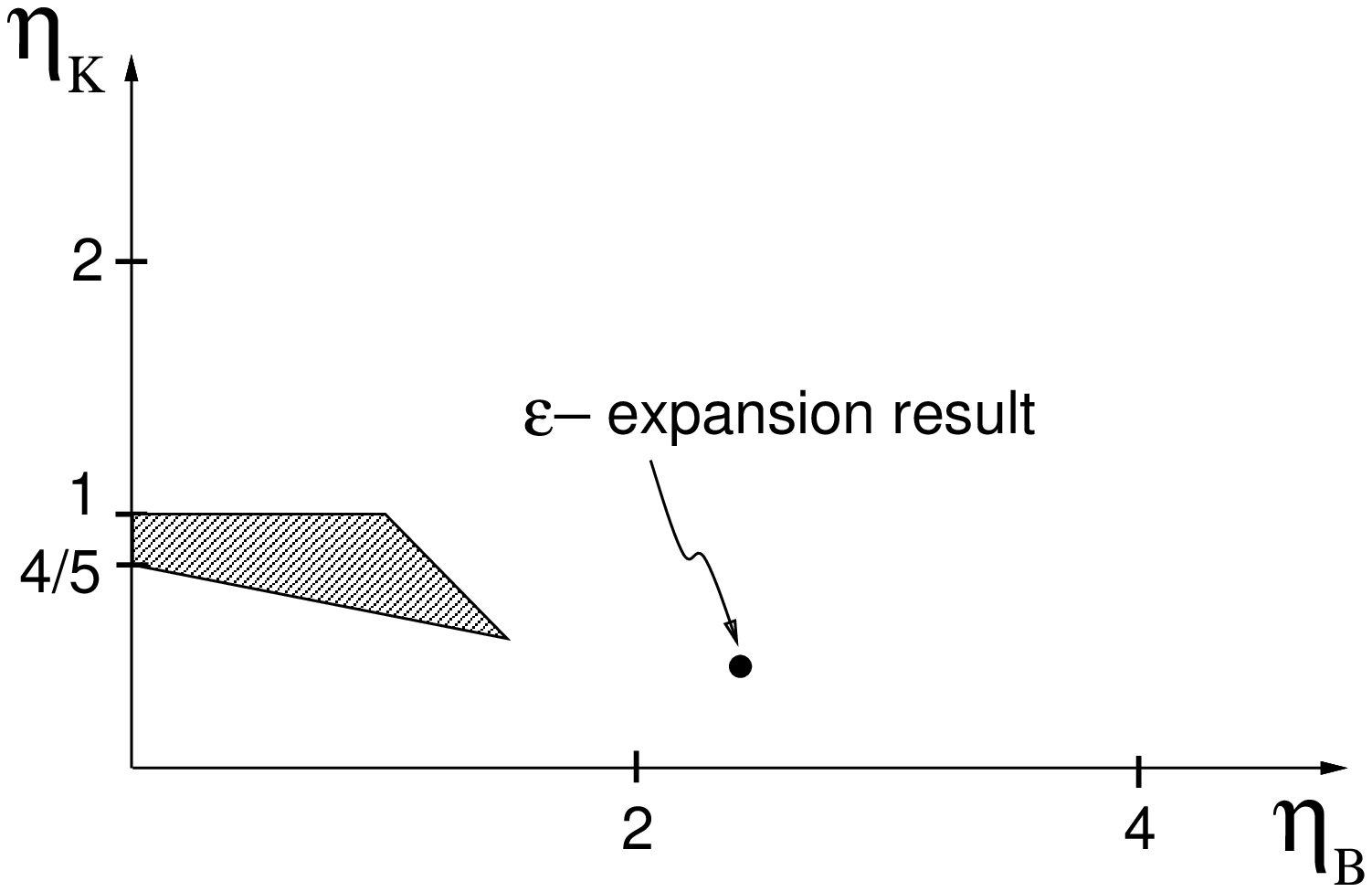}
\end{picture}}
\end{picture}}
Fig.1. {The region indicates those values of $\eta_B$ and $\eta_K$ for
which, in three dimensions, the long-range orientationally ordered
``smectic Bragg glass'' phase is stable for sufficiently small
disorder (sufficiently low density aerogel).}
\label{bounds}
\end{figure}
These exponents are {\it universal}; i.e., the same for {\it all}
smectics in low density {\em quenched} disordered media.
Unfortunately, we have only been able to calculate them in a rather
poor approximation: a $d = 5-\epsilon$ expansion.  Since we are
interested in $d = 3$, the ostensibly small parameter $\epsilon$ in
this expansion is $= 2$, and so the expansion is expected to be rather
poor.  However, taking this expansion as our best estimate (since it
is our {\it only} estimate) of the values of $\eta_B$ and $\eta_K$, we
obtain, in $d = 3$, $\eta_B = {12/5}$ and $\eta_K = {2/5}$.  These
values, which are illustrated by the dot labeled
``$\epsilon$-expansion'' in Fig.1, violate the bounds in
Eq.\ref{bounds12}, and, hence, imply that the smectic Bragg glass
phase does {\it not} exist, and that the thermodynamically sharp NA
transition is destroyed by the presence of any disorder, no matter how
weak.  

However, since the expansion parameter in this calculation
$\epsilon=5-d=2$ in the physical case of $d=3$, this argument that the
transition and the SBG are destroyed is utterly
uncompelling. Fortunately, experiments (which we will describe
shortly) well {\em below} the pure transition temperature $T_{NA}$ can
measure the exponents $\eta_B$ and $\eta_K$. With these numbers in
hand, the bounds given in Eqs.\ref{bounds12} would then provide an
unambiguous prediction as to whether or not the transition to the SBG
is {\em always} destroyed by disorder.

The physical significance of these exponents is quite intriguing: they
reflect ``anomalous elasticity'', which, in this context, refers to
the fact that the smectic bend modulus $K$ and the layer compression
modulus $B$ are {\it not} constants in the randomly pinned smectic,
but, rather, singular functions of the wavevector $\bf k$ under
consideration. $K({\bf k})$ and $B({\bf k})$ diverge and vanish
respectively, as ${\bf k} \rightarrow 0$, according to the scaling
laws:
\begin{mathletters}
\begin{eqnarray}
K({\bf k})&=&K\left(k_\perp\xi^{NL}_\perp\right)^{-\eta_K}
f_K(k_z\xi^{NL}_{z}/(k_\perp\xi^{NL}_\perp)^\zeta) \;,
\label{K}\\
B({\bf k})&=&B\left(k_\perp\xi^{NL}_\perp\right)^{\eta_B}
f_B(k_z\xi^{NL}_{z}/(k_\perp\xi^{NL}_\perp)^\zeta)\;.
\label{B}
\end{eqnarray}
\label{BK}
\end{mathletters}
Here $k_z$ and $k_{\perp}$ denote the projections of $\bf k$ along and
perpendicular to the mean normal to the smectic layers, respectively,
and $K$ and $B$ without arguments $\bf k$ will here and hereafter
denote the ``bare'' values of the smectic elastic moduli, i.e., their
values in the pure (bulk) smectic. The universal exponent $\zeta$ is
determined by $\eta_B$ and $\eta_K$ through Eq.\ref{zeta1}.

The ``non-linear'' length scales $\xi_\perp^{NL}$ and $\xi_z^{NL}$ are
the distances in the $\perp$ and $z$ directions respectively, beyond
which the disorder-driven anomalous elasticity is manifest. That is,
Eqs.\ref{BK} only hold for $k_\perp\xi_\perp^{NL}<<1$ and
$k_z\xi_z^{NL}<<1$. If either of these limits is violated, both $K$
and $B$ are $\bf k$ independent (up to logarithms), as in pure
smectics.\cite{GP} In three dimensions these nonlinear length scales
are given by
\begin{mathletters}
\begin{eqnarray}
\xi_\perp^{NL}&=&{\left(64\pi\over3\right)^{1/2}}
{K^{5/4}\over B^{1/4}\Delta_h^{1/2}}\;,\label{xi_perp_NL}\\
\xi_z^{NL}&=&{64\pi\over3}{K^2\over\Delta_h}\;.\label{xi_z_NL}
\end{eqnarray}
\label{xi_zp_NL}
\end{mathletters}
Expressions for these lengths for general dimensionality $d\geq3$ are
given in Eqs.\ref{xiNLperp} and \ref{z-perp_relation} of
Sec.\ref{anom_elasticity}. In the above $\Delta_h$ is a measure of the
strength of one of two types of disorder, (the other random ``field''
disorder $\Delta_V$ being less important in three dimensions) whose
relation to various parameters of the aerogel is given in
Sec.\ref{model}. For now, it suffices to say that $\Delta_h$ is a
monotonically increasing function of the aerogel density, and we
expect it to be a smooth, analytic, non-singular, finite and
non-vanishing function of temperature through the bulk NA transition
temperature $T_{NA}$. Since $K$ is likewise well-behaved through
$T_{NA}$, $\xi_z^{NL}$ is as well, a fact that we will make use of
shortly.

Although we have been unable to calculate the scaling functions
$f_K(x)$ and $f_B(x)$ exactly, we do know that they have the simple
property of making $K({\bf k})$ and $B({\bf k})$ independent of $k_z$
when the scaling argument
$k_z\xi^{NL}_{z}/(k_\perp\xi^{NL}_\perp)^\zeta$ is $<<1$, and
independent of $k_\perp$ in the opposite limit. This implies
\begin{mathletters}
\begin{eqnarray}
K({\bf k})&\propto&
\left\{\begin{array}{lr}
k_\perp^{-\eta_K}, & \mbox{for}\;k_z\xi^{NL}_{z}
<<(k_\perp\xi^{NL}_\perp)^\zeta\\
k_z^{-\eta_K/\zeta}, & \mbox{for}\;k_z\xi^{NL}_{z}
>>(k_\perp\xi^{NL}_\perp)^\zeta
\end{array} \right.
\label{Klimits}\\
B({\bf k})&\propto&
\left\{\begin{array}{lr}
k_\perp^{\eta_B}, & \mbox{for}\;k_z\xi^{NL}_{z}
<<(k_\perp\xi^{NL}_\perp)^\zeta\\
k_z^{\eta_B/\zeta}, & \mbox{for}\;k_z\xi^{NL}_{z}
>>(k_\perp\xi^{NL}_\perp)^\zeta
\end{array} \right.
\label{Blimits}
\end{eqnarray}
\end{mathletters}

This strong wavevector dependence of $K$ and $B$ (driven by disorder)
is caused by the same mechanism that leads to the much weaker
(logarithmic) but still singular divergence and vanishing of $K$ and
$B$ (driven by thermal fluctuations) that occurs in disorder-free
smectics\cite{GP}: the anharmonic elasticity of large fluctuations in
the smectic layers.

As we will demonstrate in Sec.\ref{anom_elasticity}, the effects are
larger in the pinned smectic because the disorder induces layer
roughness that is much larger than that due to thermal fluctuations,
thereby leading to the stronger diverging and vanishing of $K$ and $B$
found here.

The stability of the smectic Bragg glass phase depends on $\eta_K$ and
$\eta_B$ because the elastic moduli $K$ and $B$ determine {\it both}
the size of the orientational fluctuations and the stability of the
phase against the unbinding of dislocations.  Requiring that real
space orientational fluctuations remain finite leads to
Eq.\ref{bound_orientations}, while dislocations remain bound only if
Eq.\ref{bound_dislocations} is satisfied. If the bounds
Eq.\ref{bounds12} are satisfied for a real smectic liquid crystal,
confined inside a low density aerogel, then the resulting $\rho_A-T$
phase diagram will be topologically identical to that displayed in
Fig.2.

It is important to note that all of the results quoted here are {\em
equilibrium} results. In contrast, in other pinned elastic media (e.g.,
Abrikosov flux lattices in dirty superconductors and charge density
waves in anisotropic metals), it is often difficult to observe the
true equilibrium behavior due to the extremely slow dynamics of those
systems, which cause them to drop out of equilibrium, as evidenced by
strong hysteretic effects.\cite{VG,Gruner}

In this sense, smectics, and liquid crystals in general, appear to be
far better systems for investigating the {\em equilibrium} effects of
quenched disorder, since they typically do {\em not} exhibit
hysteresis. This suggests that liquid crystals have intrinsically
faster dynamics than, e.g., vortex lattices in superconductors, or
charge density wave systems in metals.

That this should be so is hardly surprising: liquid crystals are,
after all, ``liquids'' in the sense that, even in the translationally
ordered phases like smectics, molecules are quite free to move
around. In contrast, the atoms in the conventional crystalline solids
are essentially locked into lattice sites, from which they can only
escape by thermally activated hopping over substantial energy
barriers. Even at room temperature, this hopping is extremely slow;
the fact that even ``high-T$_c$'' superconductors are much colder than
room temperature exacerbates the problem further.

So, because of both their liquid-like {\em micro}structure and {\em
micro}dynamics, {\em and} the fact that they are at higher temperature
than, e.g., superconductors, one might have anticipated that it would
be far easier to observe the equilibrium effects of quenched pinning
in liquid crystals than in other pinned elastic media studied to date.
This belief is supported by experiments,\cite{DynamicsI-N_Wu,Bellini}
and so we believe that the {\em equilibrium} results we obtain here
should be directly testable in experiments.

\begin{figure}[bth]
{\centering
\setlength{\unitlength}{1mm}
\begin{picture}(150,60)(0,0)
\put(-20,-57){\begin{picture}(150,0)(0,0)
\includegraphics{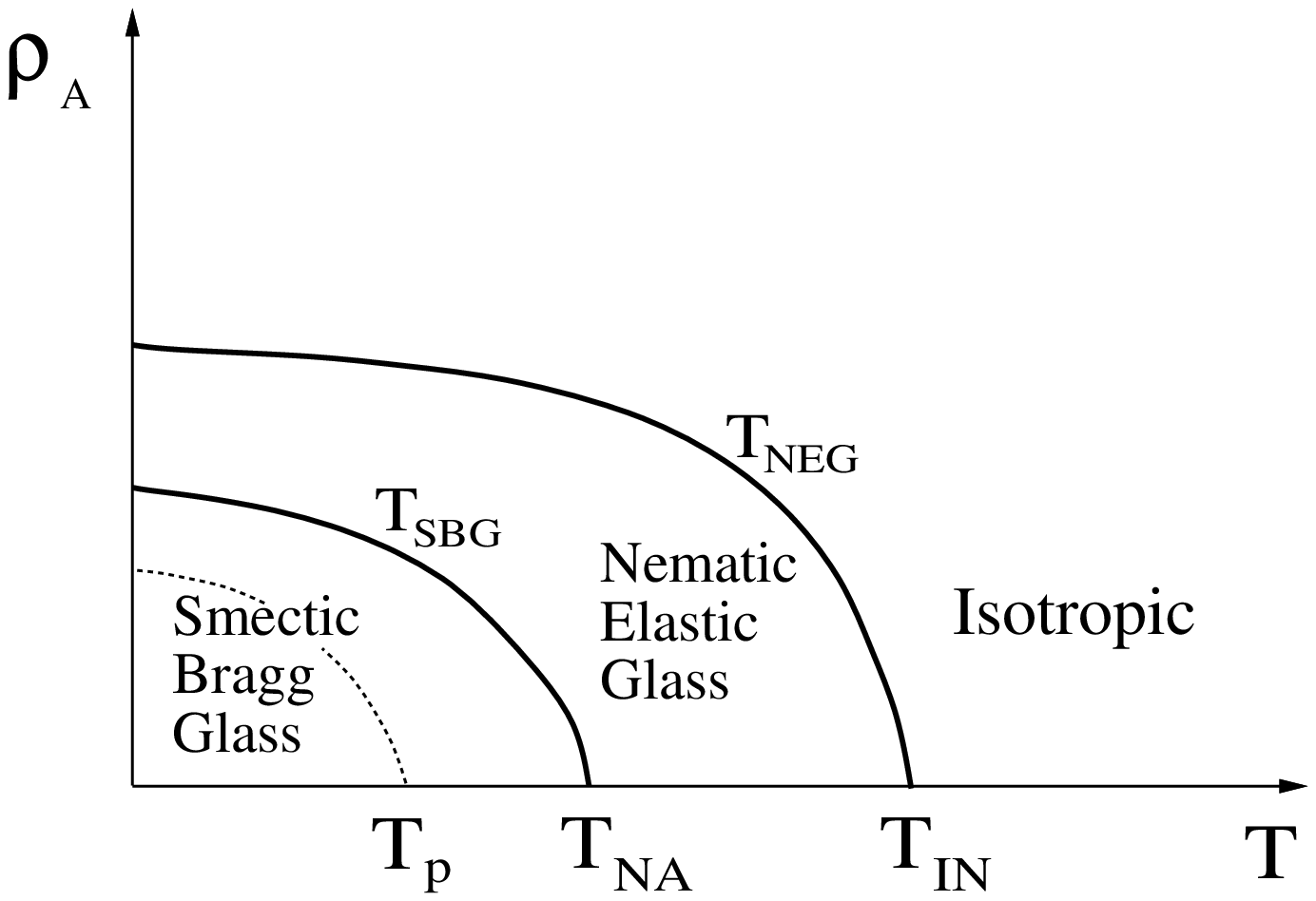}
\end{picture}}
\end{picture}}
Fig.2. {A schematic (aerogel density $\rho_A$\ -\ temperature $T$)
phase diagram for a smectic liquid crystal, confined inside a low
density aerogel, valid {\em if and only if} all of the bounds,
Eq.\ref{bounds12} are satisfied. As discussed in the text the
``smectic Bragg glass'' (SBG) and ``nematic elastic glass'' (NEG)
phases are, respectively, translationally and orientationally
disordered, but are topologically ordered, and are therefore distinct
from each other and from the fully disordered ``isotropic'' phase. The
dotted line within the SBG phase represents a remnant of the disorder
driven pinning transition of Fig.6, rounded by the anharmonic
elasticity studied in
Secs.\ref{anom_elasticity}-\ref{random_field_3-5}.}
\label{phase_diagram}
\end{figure}

The theoretical analysis that leads to these conclusions is quite
interesting and novel. Our first-principles analysis of smectics in
aerogel demonstrates that the random pinning induced by the aerogel
leads to only two potentially relevant types of random perturbations
to the smectic: a random {\it positional} field disorder (hereafter
referred to simply as the ``random field'' and designated by
$\Delta_V$), which represents the aerogel's tendency to force the
smectic layers to sit at particular {\it positions}, and a random {\it
orientational} field disorder (hereafter called the ``tilt'' field,
with designation $\Delta_h$), reflecting the aerogel's proclivity
for particular {\it orientations} of the nematogens and smectic
layers.  If we ignore the anharmonic effects that lead to anomalous
elasticity, we find that the response of the smectic-A phase to the
random field disorder in {\em three} dimensions is in very close
mathematical analogy to that of the XY-model in {\em two}
dimensions.\cite{CardyOstlund} While this suggests that a nontrivial
glassy phase might replace the smectic-A phase in analogy with the
experience with the 2d XY-model,\cite{CardyOstlund} the presence of
{\em tilt} disorder (which is always generated by random field)
actually destroys this phase.  Even in the regime where the random
field by itself has no effect at long scales, the tilt disorder leads
to short-range smectic order parameter $\psi$ correlations, which fall
off exponentially in the direction of the layer normal and as a
Gaussian within the smectic layers (within a purely {\em harmonic}
elastic model).  This absence of long-ranged order in the elastic
model is a strong indication of its limitation and that dislocation
defects are likely to proliferate. Focusing on the most important part
of randomness, the tilt disorder, we find, in the approximation of
ignoring elastic anharmonicities, that disorder {\em always} creates
dislocations.  The formalism we use to demonstrate this is new,
powerful, and potentially applicable to a wide variety of candidate
``Bragg glass'' systems\cite{RTgauge_glass}.

Once dislocations are present, the phase is best characterized as a
nematic in a random tilt field.  However, subsequent examination of
orientational fluctuations in this nematic lead to the conclusion that
tilt disorder destroys the orientational order of the smectic layers
as well. Whether an orientationally and translationally disordered low
temperature phase distinct from the conventionally disordered high
temperature isotropic phase is a ``nematic elastic glass'' (NEG),
separated from it by a {\em disclination} unbinding transition, is the
subject of an active current investigation.\cite{RTunpublished}

Including anharmonic elastic effects considerably modifies this
picture.  Although spatial dimensions $d$ greater than three are
clearly irrelevant to experiment, it is conceptually extremely useful
(and fun!)  to generalize our model to arbitrary spatial dimensions.
We find that for $d > 7$, both the random field and the tilt disorder
are irrelevant, and a stable, translationally ordered, non-glassy
smectic phase exists for sufficiently weak disorder.  For $5 < d < 7$,
tilt disorder remains irrelevant, but the random field and the
anharmonic terms both become relevant, destroying smectic
translational order and leading instead to a stable smectic Bragg
glass phase.  As in previously studied ``vortex glass'' models of
pinned elastic media\cite{VillainFernandez,GL}, here too we find that
in all spatial dimensions $5 < d < 7$, real-space positional
fluctuations diverge logarithmically with system size:
\begin{equation}
\langle u^2({\bf r})\rangle= 
c_d\ln\left[L_\perp f_u(\lambda L_z/L_\perp^2)\right]\;,
\label{u2superuniversal}
\end{equation}
where $L_{\perp(z)}$ is the linear spatial extent of the system in the
$\perp$ ($z$) direction, 
\begin{equation}
\lambda\equiv\sqrt{K\over B}\;,
\label{lambda_definition}
\end{equation}
$f_u(x)$ is a universal scaling function and $c_d$ is universal,
dimension-dependent constant proportional to $7-d$ for $d$ near
$7$. As a result, translational correlations throughout the range of
spatial dimensions $5 < d < 7$ decay algebraically,
\begin{equation}
\langle\rho^{\ast}_G \left({\bf r} \right) \rho_G \left({\bf 0}
\right)\rangle \propto r^{- \eta (d)}_{\perp} f_d(\lambda z/r^2_{\perp})
\label{rho_scale}
\end{equation}
with the exponent $\eta(d)$ a {\it universal} function of $d$, and
$f_d(x)$ a {\it universal} ($d$-dependent) scaling
function.\cite{comment1}

{\it Unlike} the vortex glass case, however, here we find anomalous
elasticity in this dimension range as well, obtaining
\begin{mathletters}
\begin{eqnarray}
K({\bf k})&\propto&\left|\ln\left[k_{\perp} f_K(k_z/\lambda
k^2_{\perp}) \right]\right|^{\gamma_K(d)}
\label{Kfrg}\;,\\
B({\bf k})&\propto&\left|\ln\left[k_{\perp}
f_B(k_z/\lambda k^2_{\perp})\right]\right|^{-\gamma_B(d)}
\label{Bfrg}\;,
\end{eqnarray}
\label{BKfrg}
\end{mathletters}
where $f_K(x)$ and $f_B(x)$ are universal scaling functions, and we
have, rather remarkably, calculated the universal exponents
$\gamma_K(d)$ and $\gamma_B(d)$ {\it exactly} for all $d$ in this
range, $5 < d < 7$, finding
\begin{mathletters}
\begin{eqnarray}
\gamma_K(d)&=&{1\over2}\left({-d^2 + 12d - 23 \over d^2 + 6d - 13}\right)
\label{gammaK}\;,\\
\gamma_B(d)&=&{3\over2}\left({d^2 - 1 \over d^2 + 6d - 13} \right)
\label{gammaB}\;.
\end{eqnarray}
\label{gammaBK}
\end{mathletters}

Note that both the fluctuations of $\langle u^2({\bf r})\rangle
\propto\ln(L)$ and the anomalous behavior of $K({\bf k})$ and $B({\bf
k})$ are very weakly divergent functions of system size $L$ and $\bf
k$ respectively, depending only logarithmically on these quantities.
For $d < 5$, these weak logarithmic divergences are overwhelmed by
power-law divergences caused by the {\it tilt} disorder, with $K({\bf
k})$ and $B({\bf k})$ diverging and vanishing, respectively, according
to Eq.\ref{BK}, with,
\begin{mathletters}
\begin{eqnarray}
\eta_K &=& {\epsilon \over 5} + O\left(\epsilon^2 \right)
\label{etaK}\;,\\
\eta_B &=& {6 \epsilon \over 5} + 0 \left(\epsilon^2 \right)
\label{etaB}\;,
\end{eqnarray}
\label{etaBK}
\end{mathletters}
where $\epsilon\equiv 5 - d$, and
\begin{mathletters}
\begin{eqnarray}
\langle u^2({\bf r})\rangle&=& L^{2 \chi}_{\perp}
\tilde{f}_u\left({\left(L_z/\xi_z^{NL}\right) \over
\left(L_{\perp}/\xi_{\perp}^{NL}\right)^{\zeta}} \right)
\label{u_fluc}\;,\\
&\propto&\left\{\begin{array}{lr}
L_z^{2\chi/\zeta}, & \mbox{for}\;L_z/\xi^{NL}_{z}
<<(L_\perp/\xi^{NL}_\perp)^\zeta\\
L_\perp^{2\chi}, & \mbox{for}\;L_z/\xi^{NL}_{z}
>>(L_\perp/\xi^{NL}_\perp)^\zeta\\
\end{array} \right.
\label{uulimits}
\end{eqnarray}
\label{uu_fluc_limits}
\end{mathletters}
with the universal roughness $\chi$ and anisotropy $\zeta$ exponents
given by
\begin{mathletters}
\begin{eqnarray}
\chi &=& {\eta_B + \eta_K \over 2}\;,
\label{chi}\\
\zeta &=& 2 - {\eta_B + \eta_K \over 2}
\label{zeta1}\;.
\end{eqnarray}
\label{chi_zeta}
\end{mathletters}
Here $\tilde{f}_u(x)$ is another universal scaling function.

This result Eq.\ref{uu_fluc_limits} for $\langle u^2({\bf r})\rangle$
leads to a quantitative, experimentally testable prediction for the
X-ray correlation length $\xi^X$, obtained by equating $\langle
u^2({\bf r})\rangle = a^2$, where $a$ is the smectic layer spacing,
and solving for $L_z$, with $L_\perp\rightarrow\infty$.  This solution
for $L_z$ is the X-ray correlation length that will be obtained by
scattering off a powder sample (which probably means all samples,
since, as discussed earlier, there is probably no long-ranged
orientational order in $d = 3$, given our best estimates of $\eta_B$
and $\eta_K$).  The value of $\xi^X$ so obtained is
\begin{equation}
\xi^{X}=\left\{\begin{array}{lr}
\xi_z^{NL}\left({a\over\lambda}\right)^{\zeta/\chi}, 
&\;\lambda<<a\\
\xi_z^{NL}\left({a\over\lambda}\right)^{2}, 
&\;\lambda>>a\\
\end{array} \right.
\label{xiX}
\end{equation}

This relation gives a way of experimentally measuring the exponents
$\zeta$ and $\chi$: since the bare compression modulus $B(T)$ of the
pure smectic is a strong function of temperature $T$ near the bulk
smectic NA transition temperature $T_{NA}$, vanishing according to
$B(T)\propto|T-T_{NA}|^{\tilde{\phi}}$, and since, furthermore,
$\xi_z^X$, $a$, and $K$ vary smoothly through $T_{NA}$, a plot of
$\ln{\xi^X}$ versus $\ln|T-T_{NA}|$ should yield a straight line for
$T$ near $T_{NA}$, with a slope ${\tilde{\phi}}\zeta/2\chi$, {\em
provided} that we stay far enough below $T_{NA}$ that $\lambda<<a$, so
that the first expression for $\xi^X$ in Eq.\ref{xiX} applies. In
deriving this result, we have used the fact that
$\lambda=\sqrt{K/B}\propto|T-T_{NA}|^{-{\tilde{\phi}}/2}$. This
prediction for the X-ray correlation length, Eq.\ref{xiX}, is
schematically displayed in Fig.3.

A detailed renormalization group analysis\cite{RTunpublished} of the
critical behavior near the pure NA transition (details of which will
be presented elsewhere) shows that Eq.\ref{xiX} breaks down altogether
at the temperature $T_*$, which lies below (and for weak disorder very
close to) the pure $T_{NA}$. The corresponding reduced temperature
$t_*\equiv(T_{NA}-T_*)/T_{NA}$ obeys
\begin{equation}
\xi_{crit.}(t_*)=\xi^X\;,\label{xi=xi}
\end{equation}
where $\xi^X$ is given by the second line of Eq.\ref{xiX}, and we have
defined a length scale $\xi_{crit.}(t)$ derived entirely from
properties of the {\em pure} system:
\begin{equation}
\xi_{crit.}(t)=\xi_\perp^0\left({\xi_\perp^{pure}(t)
\over\xi^0_\perp}\right)^{2+\tilde{\phi}/2\nu_\perp}\;,
\label{xi_crit}
\end{equation}
where $\xi_\perp^{pure}(t)$ is the X-ray correlation length
within the layers of the {\em pure} system at a reduced temperature
$t\equiv(T-T_{NA})/T_{NA}$ {\em above} the pure transition temperature
$T_{NA}$. Here $\nu_\perp$ is the X-ray correlation length exponent
for the pure system in the $\perp$ direction, defined via
$\xi_\perp^{pure}(t)\propto t^{-\nu_\perp}$.

\begin{figure}[bth]
{\centering \setlength{\unitlength}{1mm}
\begin{picture}(150,60)(0,0)
\put(-20,-57){\begin{picture}(150,0)(0,0)
\includegraphics{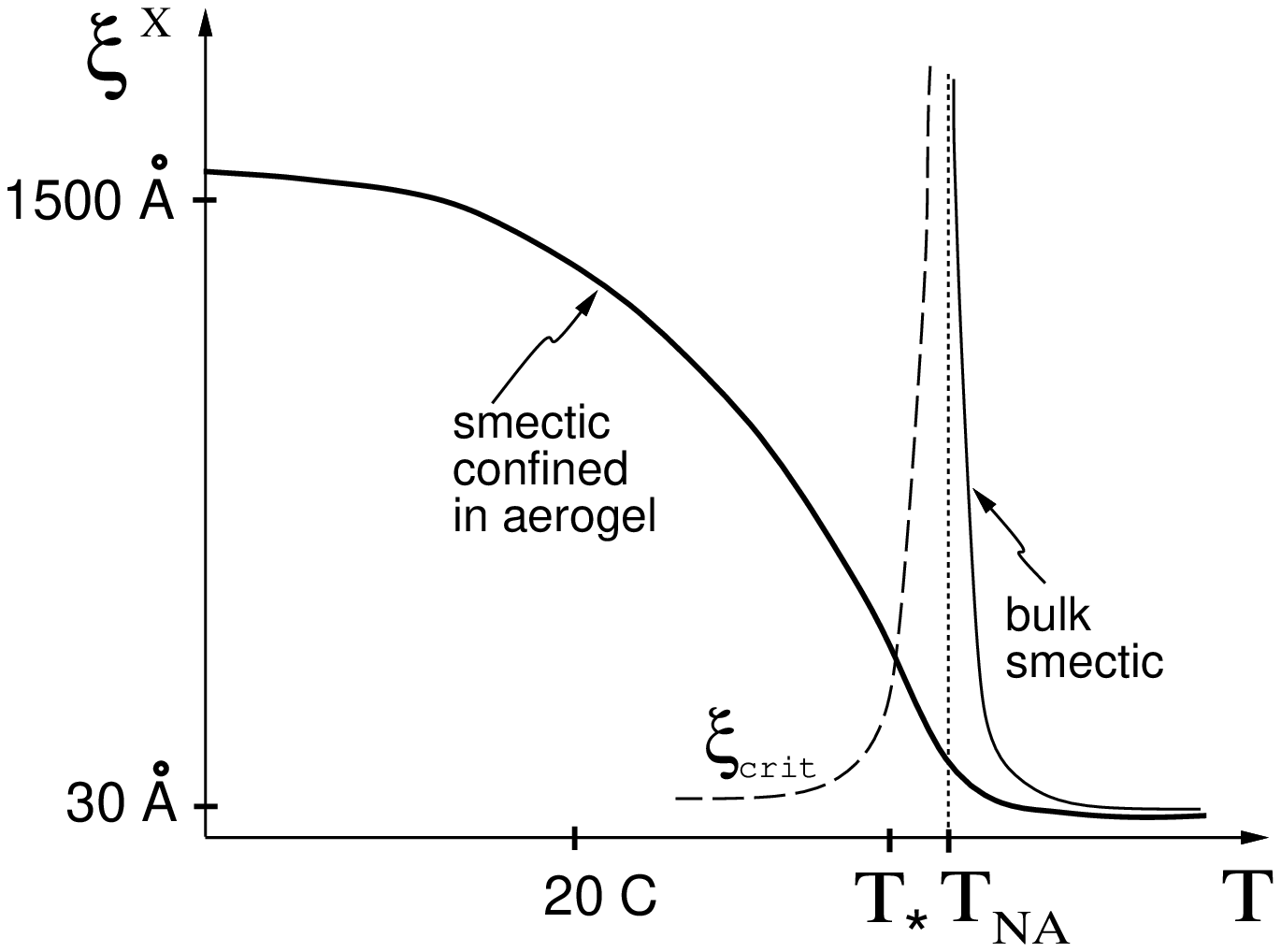}
\end{picture}}
\end{picture}}
Fig.3. {Finite, temperature-dependent X-ray correlation length
$\xi^X(T)$ for a smectic liquid crystal, confined inside a low density
aerogel. The essential features of the X-ray correlation length that
we predict are: (i) saturation as $T\rightarrow0$ of $\xi^X(T)$ at a
{\em finite} $\rho_A$-dependent value for {\em arbitrarily low}
aerogel density (diverging as $\rho_A\rightarrow0$), (ii) power-law
scaling with the {\em disorder-free} smectic bulk modulus $B(T)$ (see
Eq.\ref{xiX}), (iii) crossover from the exponent $\zeta/\chi$ to $2$
in Eq.\ref{xiX} as $T_{NA}$ is approached from below and anomalous
elasticity becomes unimportant for $\xi^X$, (iv) crossover to genuine
critical behavior at the temperature $T_*$ at which $\xi_{crit.}(T_*)$
(indicated by the dotted curve below $T_{NA}$) $=\xi^X$. In the low
density aerogels (e.g., $\rho_A=0.08$g/cm$^3$) we find
$T_{NA}-T_*\approx 3^0 K$.\cite{Clark}}
\label{xiXfig}
\end{figure}

Theoretically\cite{Toner}, ${\tilde{\phi}}$ is expected to be given by
${\tilde{\phi}}=\nu_{XY}\approx0.67$.
Experimentally,\cite{NAnonuniversal} the situation is more
complicated, with ${\tilde{\phi}}$ showing no universality, for
reasons that are still unclear (at least to us). Hence, in extracting
$\zeta/2\chi$ by the above analysis, the {\em experimentally}
determined value of ${\tilde{\phi}}$ of the {\em particular} bulk
smectic that is confined in the aerogel should be used. One should
also be careful, in such a fit, to treat $T_{NA}$ as a fitting
parameter, since the presence of even low density aerogel can
presumable shift this temperature slightly.  Alternatively, one could
avoid these complications by studying the smectic outside the pure
system's critical regime, but still reasonably close to the pure
$T_{NA}$, since in this regime mean-field theory applies and
${\tilde{\phi}}=1$.\cite{noMFregime}

Once we know the ratio $\zeta/\chi$, we know $\zeta$ and $\chi$
themselves, since the relation $\zeta=2-\chi$, implicit in
Eqs.\ref{chi_zeta}, provides us with a second equation for the two
unknowns $\zeta$ and $\chi$. If the roughness exponent $\chi$ so
obtained is $>1$, we predict that orientational fluctuations diverge,
and hence, long-range orientational order is destroyed. As a result,
there will be no ``smectic Bragg glass'' phase, and hence, no
thermodynamically sharp transition associated with the pure NA
transition that occurs in the absence of disorder. In any event, the
prediction Eq.\ref{xiX} for $\lambda>>a$ will inevitably apply
sufficiently close to the pure $T_{NA}$, since $B(T\rightarrow
T_{NA})\rightarrow0$, while $K(T\rightarrow T_{NA})$ remains non-zero,
and, hence $\lambda(T\rightarrow T_{NA})\rightarrow\infty>>a$.

Should it prove that orientational order {\it is} possible in $d = 3$
(i.e., that the $d = 3$ values of the universal exponents $\eta_B$ and
$\eta_K$ {\it do} satisfy the inequalities in Eq.\ref{bounds12}, which
would imply that the SBG phase is stable), then single orientational
domain scattering {\it is}, in principle, possible.  In such
scattering, $\xi_z^X$, given by $\xi^X$, as given by Eq.\ref{xiX},
will be the correlation length along the mean normal to the smectic
layers.  The correlation length $\xi^X_{\perp}$ along the smectic
layers is then given by
\begin{equation}
\xi^{X}_\perp=\left\{\begin{array}{lr}
\xi_\perp^{NL}\left({a\over\lambda}\right)^{1/\chi}, 
&\;\lambda<<a\\
\xi_\perp^{NL}\left({a\over\lambda}\right), 
&\;\lambda>>a\\
\end{array} \right.
\label{xiXperp}
\end{equation}

These X-ray correlation lengths should {\em not}, however, be
interpreted as being the length scales beyond which smectic
correlations cease. In fact, smectic behavior, by which we mean, e.g.,
the anomalous smectic elasticity, Eq.\ref{BK}, persists out to much
longer lengths. Specifically, in the $\perp$ direction, it persists
out to the shorter of the two lengths $\xi_O$ and $\xi_\perp^D$.
$\xi_O$ is the distance over which orientational order {\em would} be
well-correlated in the absence of dislocations, while $\xi_\perp^D$ is
the distance (in the $\perp$ direction) below which dislocations {\em
would} remain bound in the absence of large orientational
fluctuations. In actuality, only the prediction of the {\em smaller}
of the two of these lengths is valid. We expect that when
$\xi_O<<\xi_\perp^D$, the large orientational fluctuations also induce
dislocation unbinding at about the same length $\xi_O$. On the other
hand, in the opposite ($\xi_O>>\xi_\perp^D$), limit the dislocation
unbinding occurs first at $\xi_\perp^D$, and on longer scales our
system is indistinguishable from a nematic in a random orientational
field. Whether the orientational order is stable or not in this case
must be re-evaluated within an effective random-field nematic model
using a Larkin\cite{Larkin} type of analysis.\cite{RTunpublished}

The ``orientational correlation length'' $\xi_O$ is given by
\begin{equation}
\xi_O=\xi_\perp^{NL}\left({\xi_z^{NL}
\over\lambda}\right)^{1\over2(\chi-1)}\;,
\label{xi_O}
\end{equation}
while the ``dislocation length'' scale $\xi_\perp^D$ in the $\perp$
direction, is given by
\begin{equation}
\xi_\perp^D=\xi_\perp^{NL}\left({\lambda T^2\xi_z^{NL}
\over c K^2 a^2 d^2}\right)^{1\over2(\eta_\kappa-1)}\;,
\label{xi_perpD}
\end{equation}
where $d$ is a microscopic length of order the layer spacing $a$, and
$c(T)$ is a dimensionless constant. $c(T)$ vanishes like $e^{-E_c/T}$
as $T\rightarrow0$ and diverges like $T/E_c(T)\propto |T_{NA}
-T|^{-\gamma_\tau}$, where $E_c$ is the ``core energy'' of a
dislocation line segment of length $d$, and $\gamma_\tau$ is the
``line tension'' critical exponent describing how $E_c(T)$ vanishes as
$T\rightarrow T_{NA}$ from below.  We expect the resulting divergence
of $c(T)$ to overwhelm the corresponding divergence of $\lambda(T)$,
leading to a $\xi_\perp^D$ that gets smaller as $T\rightarrow T_{NA}$
from below. Clearly then, sufficiently close to $T_{NA}$,
$\xi_{\perp}^D$ gets to be less than $\xi_{\perp}^{NL}$, the system
enters the strong-disorder regime, and the weak-disorder theory we
have presented here no longer applies.

In the $z$-direction, the orientational correlation length is also
given by $\xi_O$, while the dislocation length is given by
\begin{equation}
\xi_z^D=\xi_z^{NL}\left({\lambda T^2\xi_z^{NL}
\over c K^2 a^2 d^2}\right)^{\zeta\over2(\eta_\kappa-1)}\;.
\label{xi_zD}
\end{equation}

Note that the orientational correlation length given by Eq.\ref{xi_O}
is always much greater than the non-linear length $\xi_\perp^{NL}$ in
the weak disorder limit, in which
$\xi_\perp^{NL}\rightarrow\infty$. However, if we hold the disorder
strength fixed (i.e., hold $\Delta_h$ fixed), and approach the pure NA
transition from below, we will always, eventually at some temperature
$T_*< T_{NA}$ ($T_*\rightarrow T_{NA}$ in weak disorder limit) leave
this weak disorder regime (no matter how weak), since $\xi_z^{NL}$, as
given by Eq.\ref{xi_z_NL}, does not change much as $T\rightarrow
T_{NA}^-$, while $\lambda(T\rightarrow T_{NA}^-)\rightarrow\infty$. So
the ratio $\xi_O/\xi_\perp^{NL}$ decreases without bound as
$T\rightarrow T_{NA}^-$, and, hence eventually drops below $1$,
signaling our entry into the strong disorder regime for $T>T_*$ (see
Fig.3). This is not surprising: the smectic's resistance to the
perturbing effects of the disorder is provided, in part, by B; so when
$B\rightarrow0$, as it does as the pure NA transition is approached,
any disorder will eventually look ``strong'', and our theory will no
longer apply. As long as we are at temperatures $T$ far enough below
$T_{NA}$, however, our weak disorder results will apply. As we weaken
the disorder, by e.g., reducing aerogel density, we can apply our weak
disorder theory closer to $T_{NA}$.

The expression (\ref{xi_O}) for the orientational correlation length
$\xi_O$ only holds if the condition for the {\em stability} of the
long-ranged orientational order, Eq.\ref{bound_orientations}, which is
equivalent to $\chi<1$, is {\em violated} (i.e. Eq.\ref{xi_O} only
holds for $\chi>1$). If $\chi<1$, $\xi_O$ is infinite. Likewise,
Eqs.\ref{xi_perpD} and \ref{xi_zD} for $\xi_{\perp,z}^D$ only hold if
$\eta_K>1$ (i.e., if dislocations {\em are}, in fact, {\em unbound});
otherwise, $\xi_{\perp,z}^D$ are infinite. Clearly, if both bounds
$\chi<1$ and $\eta_K<1$ are satisfied, {\em all three} lengths are
infinite, smectic behavior holds out to arbitrarily large length
scales, and the smectic Bragg glass is a stable phase of weakly
disordered smectic liquid crystals.

A summary of the many length scales in our theory for three dimensions
and their hierarchy, in the weak disorder limit, is illustrated for
the case of $\lambda<<a$ in Fig.4
\begin{figure}[bth]
{\centering
\setlength{\unitlength}{1mm}
\begin{picture}(50,30)(0,0)
\put(-20,-57){\begin{picture}(50,30)(0,0)
\includegraphics{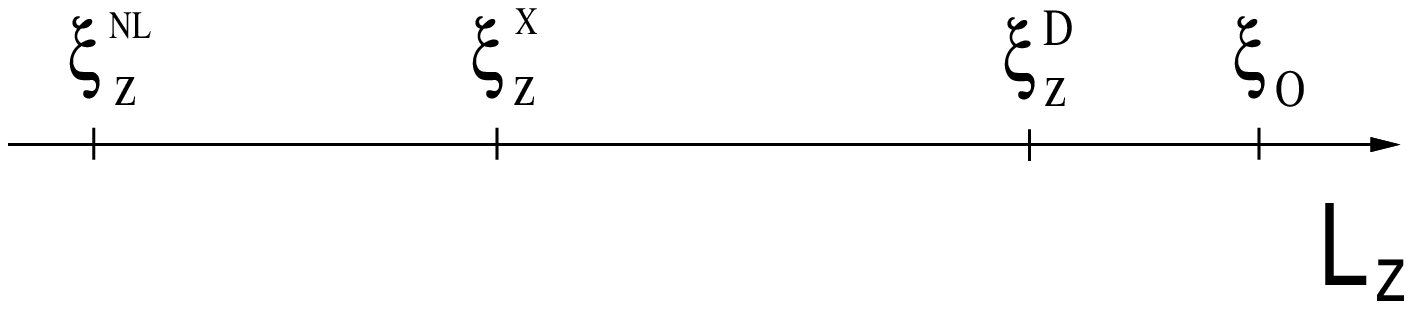}
\end{picture}}
\end{picture}}\\
Fig.4. {One possible hierarchy of important length scales in the
problem of a three-dimensional weakly disordered smectic, valid for
$\lambda<<a$, $\chi>1$, and $\eta_K>1$. If instead $\chi<1$ and
$\eta_K<1$, then the putative SBG phase is stable and $\xi_O$ and
$\xi_{\perp,z}^D$ are infinite.}
\label{lengthsPaper1}
\end{figure}
\noindent
and for the case of $\lambda>>a$ in Fig.5
\begin{figure}[bth]
{\centering
\setlength{\unitlength}{1mm}
\begin{picture}(50,30)(0,0)
\put(-20,-57){\begin{picture}(50,30)(0,0)
\includegraphics{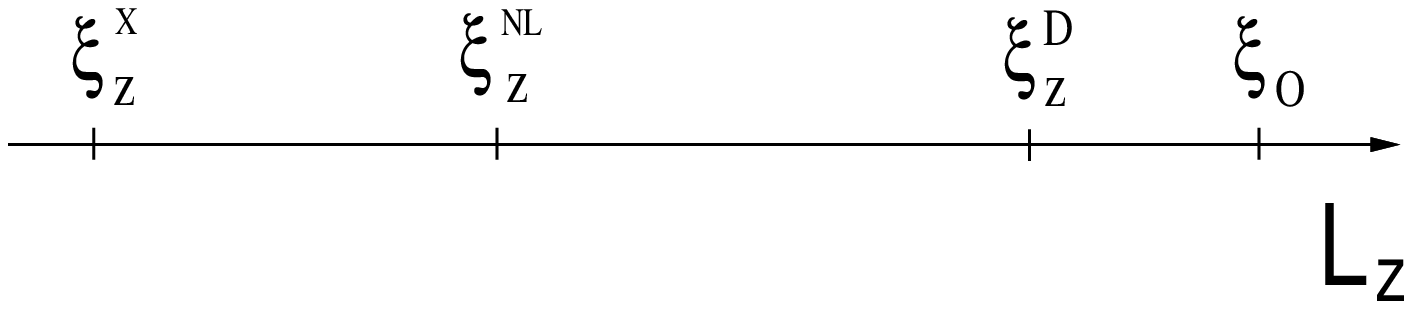}
\end{picture}}
\end{picture}}\\
Fig.5. {The hierarchy of important length scales in the problem of a
three-dimensional weakly disordered smectic, valid for $\lambda>>a$,
$\chi>1$, and $\eta_K>1$. If instead $\chi<1$ and $\eta_K<1$, then the
putative SBG phase is stable and $\xi_O$ and $\xi_{\perp,z}^D$ are
infinite.}
\label{lengthsPaper2}
\end{figure}

Just as the proposed ``Bragg glass'' phase of randomly pinned elastic
media is distinct from the liquid phase,\cite{GL,GH,DSFisher} here,
too, the absence of long-ranged orientational order does not preclude
a ``nematic elastic glass'' phase that would be separated from the
high temperature ``isotropic fluid'' phase by a thermodynamically
sharp equilibrium phase transition. That such a nematic elastic glass
phase may indeed exist is suggested by dynamic light scattering
experiments\cite{DynamicsI-N_Wu,Bellini} that show a dramatic slowing
down of director fluctuation relaxations in liquid crystals in aerogel
below a temperature $T_{NEG}$ near the bulk NI transition.

We are currently investigating theoretically the possibility of such a
nematic elastic glass phase, and initial results indicate that such
possibility is, in fact, allowed theoretically. Our results on this
subject will appear in a future publication.\cite{RTunpublished}

The low temperature analysis presented here is further supported by a
complementary approach that investigates the stability of the NA
transition to disorder from the high-temperature nematic
phase.\cite{RadzihovskyWard} The results of this latter work
substantiate our findings that in three dimensions the NA transition
and the smectic-A phase are destroyed. The rounded remnant of this
transition {\it can}, however, be studied utilizing self-consistent
methods previously successfully applied to pure systems below the
lower critical dimension.\cite{Bray} This method predicts a
Lorentzian-squared structure function with a correlation length that
monotonically, slowly increases through the bulk NA transition, in
good agreement with the X-ray measurements.\cite{Clark} The calculated
specific heat is also in qualitative agreement with the
experiments,\cite{Clark,Wu} exhibiting a broad, well-defined peak at a
temperature slightly lower than the bulk transition temperature
T$_{NA}$.

As we hope the introduction made clear, the phenomenology of this
system is extremely rich. Our organizational approach to presenting
the derivation of these results is the following: starting from the
full model, we first throw out all but the simplest effects; only
after developing a full understanding of this simplified model do we,
one at a time, reintroduce the complicating effects we had initially
thrown out, building gradually toward the complexity of the full
physical model.

Since many of these complicating effects prove to be important, this
approach has the drawback that some of the results derived for the
simplified models do {\em not}, in fact describe our physical system.
We have attempted to alert the reader everywhere results of a
simplified models differ from our actual predictions for the full
theory. Only the latter, of course, should be compared directly with
experiments.

The remainder of this paper is organized as follows: in Section
\ref{model}, we introduce and motivate our model for smectics in
disordered media, and discuss those aspects of the model specific to
aerogel.

In Section \ref{stability_harmonic_elastic}, we derive an elastic low
temperature model of a randomly pinned smectic, at first ignoring
anharmonic elasticity and dislocations. Using renormalization group
methods, we study the stability of the smectic phase within this model
and these approximations.

In Section \ref{tilt-only_model}, we use the results of this RG
treatment to calculate various smectic correlation functions, still
treating the elastic theory to harmonic order and ignoring
dislocations.

Section \ref{dislocations} incorporates the dislocations in the {\em
harmonic} theory, and shows that, in that theory, in three dimensions,
dislocations always proliferate, even in the presence of arbitrarily
weak disorder, thereby destroying, within the {\em harmonic}
approximation, both the smectic Bragg glass phase and the
thermodynamically sharp transition to it.

The effects of the previously ignored anharmonic elastic terms are
shown to be very important in Section \ref{anom_elasticity}, which
develops a renormalization group treatment of them in a model that
neglects both the random field and dislocations, but includes tilt
disorder.  The anomalous elasticity Eq \ref{BK} and the $u-u$
correlation functions that follow from it are derived in this section.
Given the importance of these elastic nonlinearities, we study their
effects on dislocation unbinding and orientational order in Section
\ref{anom_elasticity_dislocations_correlations} by incorporating
anomalous elasticity into the duality theory developed in Section
\ref{dislocations}. We thereby derive the bounds
Eq.(\ref{bound_orientations}) and Eq.(\ref{bound_dislocations}) on the
anomalous exponents $\eta_B$ and $\eta_K$ for the stability of the
smectic Bragg glass phase.  The irrelevance of the random field in a
full, anharmonic theory below $d = 5$ is demonstrated by a functional
renormalization group treatment in Section \ref{random_field_3-5}, in
which we also calculate exactly the exponents $\gamma_B(d)$ and
$\gamma_K(d)$ that govern the anomalous elasticity for $5 < d < 7$.
In Section \ref{failure}, we demonstrate that in $d=3$, dislocation
unbinding is {\it not} induced by the random field alone.

We conclude the main part of the manuscript with
Sec.\ref{experiments}, where we discuss interpretation of our results
in terms of past experiments and their implications for the future
experiments, as well as many remaining interesting and important
theoretical problems, some under current investigation.

The details of the analysis of the random field disorder in three
dimensions, the functional renormalization group analysis in higher
dimensions, an alternative derivation of the dislocation loop theory,
and a fluctuation-corrected mean-field treatment of the dual model of
randomly pinned smectic are presented in Appendices A, B, C and D,
respectively.

\section{Model}
\label{model}

Our theory of the disordered NA transition is based on the de Gennes
model. Near the mean-field transition from the nematic to the
smectic-A phase, the center-of-mass nematogen molecular density
$\rho({\bf r})$ (which is liquid-like in the nematic phase) begins to
develop strong fluctuations dominated by Fourier components that are
integer multiples of the smectic ordering wavevector ${\bf
q_0}=\hat{{\bf n}} 2\pi/a $ ($a$ is the layer spacing) parallel to the
nematic director $\hat{\bf n}$. We take the dominant lowest Fourier
component $\psi({\bf r})$ as the local (complex scalar) order
parameter which distinguishes the smectic-A from the nematic
phase\cite{deGennes}. It is related to the density $\rho({\bf r})$ by
\begin{equation}
\rho({\bf r})= \mbox{Re}[\rho_0+e^{i {\bf q_0}\cdot{\bf r}}\psi({\bf
r})]\;,
\label{rho}
\end{equation}
where $\rho_0$ is the mean density of the smectic.

As first discussed by de Gennes\cite{deGennes}, the effective Hamiltonian
functional $H_{dG}[\psi,\hat{\bf n}]$ that describes the NA transition at
long length scales, in bulk, pure liquid crystals, is:
\begin{eqnarray}
H_{dG}[\psi,\hat{\bf n}]&=&{1\over2}\int \! d^d r
\bigg[c_\perp|(\bbox{\nabla_\perp}-
i q_0 {\bf \delta n})\psi|^2
+c_\parallel|\bbox{\nabla_\parallel}\psi|^2\nonumber\\
&+& t_0|\psi|^2 +{1\over2} g_0 |\psi|^4\bigg]+H_F[\bf {n}]\;,
\label{HdG}
\end{eqnarray}
where $t_0\propto (T-T^{\rm pure}_{NA})/T^{\rm pure}_{NA}$, $T^{\rm
pure}_{NA}$ is the NA transition temperature in the pure (bulk)
system, $\delta\hat{\bf n}({\bf r}) \equiv \hat{\bf n}({\bf r}) -
\hat{\bf n}_0$ is the fluctuation of the local nematic director
$\hat{\bf n}({\bf r})$ away from its average value $\hat{\bf n}_0$,
which we take to be $\hat{\bf z}$, subscripts $\parallel$ and $\perp$
denote the directions parallel and transverse to $\hat{\bf n}_0$, and
$H_F[\bf n]$ is the Frank effective Hamiltonian that describes the
elasticity of the nematic order director:
\begin{eqnarray}
H_{F}[\hat{\bf n}] = \int \! d^d r\;
{1\over 2} \bigg[K_s(\bbox{\nabla}\cdot\hat{\bf n})^2+
K_t(\hat{\bf n} \cdot \bbox{\nabla} \times\hat{\bf n})^2 \nonumber\\
 + K_b(\hat{\bf n}\times\bbox{\nabla}\times\hat{\bf n})^2\bigg]\;,
\label{Frank}
\end{eqnarray}
where $K_s$, $K_t$, and $K_b$ are the bare elastic moduli for splay,
twist and bend of the nematic director field, respectively.

The minimal coupling between $\bf n$ and $\psi$ is enforced by the
requirement of global rotational invariance\cite{deGennes}.  It is
important to emphasize, however, that although the de Gennes
Hamiltonian $H_{dG}$ is closely analogous to that of a superconductor,
there are essential differences. The physical reality of the nematic
${\bf \delta n}$ and the smectic $\psi$ order parameters, in contrast
to the gauge ambiguity in the definition of the vector potential and
the superconducting order parameter, selects the liquid crystal
gauge\cite{DunnLubensky,Lubensky} ${\bf \delta n}\cdot\hat{\bf n_0}=0$
(since $|{\bf n_0}|^2=1$) as the preferred physical gauge. The strict
gauge invariance of $H_{dG}$ is already explicitly broken by the splay
term $K_s(\bbox{\nabla}\cdot\hat{\bf n})^2$ of the Frank Hamiltonian.
This is one source of distinction between a smectic-A liquid crystal
and a superconductor.

In fact, since the only true symmetry of the smectic-A liquid crystal
is invariance under {\it global} simultaneous rotation of the smectic
layers (equivalent to $\psi\rightarrow\psi e^{i
{\bbox{\theta}}\cdot{\bf r_\perp}}$) and the nematic director
${\bf\delta n}\rightarrow {\bf\delta n}+{\bbox{\theta}}$ accomplished
by a constant (in space) ${\bbox{\theta}}={\mbox const.}$, and {\em
not} arbitrary {\em local} gauge transformations, the requirement of
gauge invariance severely overrestricts the allowed effective
Hamiltonian. That is, the de Gennes model is more symmetric than is
required by the physics of the NA transition, and a more general
effective Hamiltonian will allow for terms that break nonlinear gauge
invariance. All of the extension terms that can be added to de Gennes
model are, unfortunately, irrelevant (which is probably why de Gennes
did not include them in his original formulation). However, in light
of the above remarks, the bending stiffness $\tilde{K}_s$ in the
smectic phase, (i.e. the coefficient of $(\bbox{\nabla}_\perp u)^2$)
can differ from the corresponding coefficient $K_s$ in the nematic
phase (the coefficient of $(\bbox{\nabla}\cdot\hat{\bf n})^2$), due to the
presence of terms like, e.g., $|\nabla^2_\perp \psi|^2$, in the smectic
Hamiltonian, which violates local gauge invariance, but {\em not}
global rotation invariance. This is in contrast to what is usually
assumed in all the standard treatments of this problem, which take
$\tilde{K}_s$ and ${K}_s$ to be the same.\cite{dGP} The importance of
this observation is that in mean-field theory it leads to a slope
discontinuity in $\tilde{K}_s(T)$ at the $T_{NA}^{mf}$, which is
likely to persist in the theory that includes fluctuations\cite{LR}.

Guided by these observations, we now proceed to construct the disorder
part of the effective Hamiltonian.  Because $\psi({\bf r})$ is related
to the smectic density $\rho({\bf r})$ via Eq.\ref{rho}, the
randomness can couple to it directly.  In particular, symmetry allows
a coupling of the form
\begin{eqnarray}
H_{d \rho}=\int d^d r \left [\frac{1}{2} \delta t({\bf r})
(\rho-\rho_0)^2 + U({\bf r}) \rho \right ]\;,
\label{Hrho}
\end{eqnarray}
where both the random $T_c$ ($\delta t({\bf r})$) and the random
potential $U({\bf r})$ are proportional to the local aerogel density
$\rho_A({\bf r})$.  We will take this aerogel density to be a quenched
random variable, and denote averages over it by a horizontal over-bar.

This means in particular that the autocorrelations of $\delta t({\bf
r})$ and $U({\bf r})$ will be proportional to those of the aerogel,
and, hence, long-ranged for fractal aerogel.\cite{commentGaussian}
That is, we expect
\begin{eqnarray}
\overline{U({\bf r})U({\bf r'})}&\equiv& C_U({\bf r}-{\bf
r'})\;,\nonumber\\
&=&\Gamma_U\overline{\rho_A({\bf r})\rho_A({\bf r'})}\;,\label{UU}\\
\overline{\delta t({\bf r})\delta t({\bf r'})}&\equiv&
C_t({\bf r}-{\bf r'})\;,\nonumber\\
&=&\Gamma_t\overline{\rho_A({\bf r})\rho_A({\bf r'})}\;,
\label{tt}
\end{eqnarray}
where $\Gamma_U$ and $\Gamma_t$ are coupling strengths that depend on
the microscopic physics of the aerogel-smectic interaction, but {\em
not}, at low aerogel density, on the aerogel density itself, nor on
the smectic density.

The fractal structure of aerogel over a range of length scales can be
determined from the aerogel density-density correlation function
$\overline{\rho_A({\bf r})\rho_A(0)}$ measured in an X-ray scattering
experiment. Writing the aerogel density $\rho_A$ in dimensionless
(volume fraction) units, the correlation function
$\overline{\rho_A({\bf r})\rho_A({\bf r'})}$ measures the probability
that, given that a point ${\bf r}$ is occupied, the point ${\bf r'}$
is also occupied. The probability that an arbitrary point ${\bf r}$ is
occupied is clearly proportional to the aerogel density
$\overline{\rho_A(L_f)}$, measured at (averaged over) a scale beyond
which the aerogel ceases to be fractal. We call this scale $L_f$. The
aerogel is also {\em not} a fractal for lengths smaller than the
microscopic length $a_f$ of the order the aerogel strand diameter. The
density of a structure which is a fractal on scales $a_f<r<L_f$, with
fractal dimension $d_F$, is, by definition, given by the ratio of the
number of occupied sites $N\sim (L_f/a_f)^{d_F}$ in a volume $V=L_f^d$
to this volume. Hence, the average aerogel density
$\rho_A\equiv\overline{\rho_A(L_f)}$ is given by
\begin{equation}
\overline{\rho_A(L_f)}\approx \left(\frac{a_f}{L_f} \right )^{d-d_F}\;,
\label{rhoA}
\end{equation}
Given that a site ${\bf r}$ is occupied, the conditional probability
$P({\bf r},{\bf r}')$ that a site ${\bf r}'$ is also occupied is,
roughly, the typical ``mass'' (actually total volume) of material
$M(|{\bf r}-{\bf r}'|)$ contained within a sphere of radius $|{\bf
r}-{\bf r}'|$ centered on a point ${\bf r}$ {\em that is also on the
fractal}, divided by the total volume $|{\bf r}-{\bf r}'|^d$ of that
sphere.  By definition, $M(|{\bf r}-{\bf r}'|)\approx a_f^d(|{\bf
r}-{\bf r}'|/a_f)^{d_F}$ (this is what we mean by the fractal dimension
$d_F$).  Hence, the conditional probability is $P({\bf r},{\bf
r}')=M(|{\bf r}-{\bf r}'|)/|{\bf r}-{\bf r}'|^d\sim(a_f/|{\bf r}-{\bf
r}'|)^{d-d_F}$.  Combining this with Eq.\ref{rhoA}, we obtain:
\begin{mathletters}
\begin{eqnarray}
\overline{\rho_A({\bf r})\rho_A({\bf r'})}
&\approx&\left(\frac{a_f}{L_f}
\right)^{d-d_F} \left(\frac{a_f}{|{\bf r}-{\bf r'}|}\right)^{d-d_F}\;,
\label{rhorho_a}\\
&\approx&\rho_A\left(\frac{a_f}{|{\bf r}-{\bf r'}|}\right)^{d-d_F}\;.
\label{rhorho_b}
\end{eqnarray}
\label{rhorho}
\end{mathletters}
We will see in a moment that these long-ranged correlations have no
effect on the long distance behavior of the smectic.

Combining the random field energy Eq.\ref{Hrho} with the relation
Eq.\ref{rho} between the smectic order parameter $\psi$ and the
density $\rho$, we obtain:
\begin{equation}
H_{d\rho}[\psi]=\int \! d^d r
{1\over 2}\bigg[\delta t({\bf r})|\psi|^2 + V({\bf r})\psi +
V^*({\bf r})\psi^*\bigg]\;,
\label{Hrhoii}
\end{equation}
where we have defined the {\em complex} random potential $V({\bf r})$,
which acts on the NA order parameter $\psi$ like a random magnetic
field acts on a spin, and is related to the potential disorder $U({\bf r})$
by
\begin{eqnarray}
V({\bf r}) \equiv U({\bf r})e^{iq_0z} \quad .
\label{Vdef}
\end{eqnarray}

Note that, despite the long-ranged correlations of $U({\bf r})$,
$V({\bf r})$ has only short-ranged correlations.  To see this,
consider a double Fourier transform of the $\overline{V({\bf
r})V^*({\bf r})}$ correlation function
\begin{eqnarray}
\overline{V({\bf k})V^*({\bf k'})}&\equiv&\int d^d r d^d r'  e^{-i{\bf
k}\cdot{\bf r}} e^{-i{\bf k'}\cdot{\bf r'}}
\overline{V({\bf r})V^*({\bf r'})} \;, \nonumber\\
&=&(2\pi)^d\delta^d({\bf k}+{\bf k'})C_V({\bf k})\;,
\end{eqnarray}
where
\begin{eqnarray}
C_V({\bf k})&=&\int d^d (r-r')
e^{i({\bf k}+q_0{\bf\hat{z}})\cdot({\bf r}-{\bf r'})}
\overline{U({\bf r})U({\bf r'})}\;,\nonumber\\
&=&C_U({\bf k}+q_0{\bf\hat{z}})\;,
\label{CV}
\end{eqnarray}
and $C_U({\bf k})$ is the Fourier transform of the $U-U$ correlation,
Eq.\ref{UU}.  Even though we expect this Fourier transformed
correlation function $C_U({\bf k})$ to diverge as ${\bf k} \rightarrow
0$, due to the power-law spatial correlations in $U({\bf r})$, there
is no reason for it to diverge as ${\bf k}\rightarrow
q_0{\bf\hat{z}}$, since the aerogel itself has no particular spatial
structure at the wavevector of the smectic ordering.  Hence, we see
from Eq.\ref{CV} that the $V-V$ autocorrelation function remains
finite as ${\bf k} \rightarrow 0$.  Thus, the correlations of $V({\bf
r})$ are short--ranged, and hence we can accurately capture the long
distance physics of the problem by taking those correlations to be
zero-ranged, and writing
\begin{equation}
\overline{V({\bf r})V^*({\bf r'})} =
W \delta^d({\bf r}-{\bf r'})\;,
\label{CV2}
\end{equation}
where
\begin{mathletters}
\begin{eqnarray}
W&=&C_U(q_0 \hat{{\bf z}})\;,\\
&=&\Gamma_U\left(\frac{a_f}{L_f}\right )^{d-d_F}
\left(\frac{1}{a_f q_0}\right)^{d_F}\;,
\end{eqnarray}
\label{CV2a}
\end{mathletters}
as can be readily seen by Fourier transforming Eq.\ref{rhorho}.

In addition to the density coupling of Eq.\ref{Hrho} and
Eq.\ref{Hrhoii}, we expect that the long nematogenic molecules that
make up the smectic will tend to line up with the randomly oriented
aerogel strands,\cite{commentLineup} whose orientational correlations
we expect to be short-ranged.  This interaction leads to an additional
{\em orientational} random coupling of the form:
\begin{eqnarray}
H_{d n}= {1\over 2}\int d^d r \big({\bf g}({\bf r})\cdot\hat{\bf n}\big)^2
\label{Hn}
\end{eqnarray}
where the quenched field ${\bf g}({\bf r})$ is random and short-range
correlated in direction, 
\begin{equation}
\overline{g_i({\bf r})g_j({\bf r'})}
=\Delta_g\delta^d({\bf r}-{\bf r'})\;\delta_{ij}\;,
\label{gi_corr}
\end{equation}
with strength $\Delta_g$ proportional to the local aerogel density.

As discussed in the Introduction, motivated by the experimental
observations, we assume that near the NA transition, in low-density
aerogel samples, the nematic order is well-developed. In this regime,
we can consider small disorder-driven deviations from perfect nematic
order $\hat{\bf n}_0={\bf\hat{z}}$ as small, writing $\hat{\bf n}({\bf
r})$ as
\begin{equation}
\hat{\bf n}({\bf r})={\hat{\bf z}}\sqrt{1-|\delta{\bf n}|^2}
+ \delta{\bf n}({\bf r})\;.
\end{equation}
Substituting this representation inside $H_{d n}$, Eq.\ref{Hn}, and,
given that $|\delta{\bf n}| \ll 1$, keeping only up to quadratic terms
in $\delta{\bf n}$, and dropping $\delta{\bf n}$-independent terms, we
obtain
\begin{eqnarray}
H_{d n}&\approx&{1\over2}\int d^d r\,\bigg[-g_z({\bf r})^2|\delta{\bf n}|^2 
+\left({\bf g}({\bf r})\cdot\delta{\bf n}\right)^2\nonumber\\
&+&2{\bf h}({\bf r})\cdot\delta{\bf n}\bigg]\;,
\label{Hn2}
\end{eqnarray}
where we have defined a quenched random tilt field
\begin{equation}
{\bf h}({\bf r}) \equiv g_z({\bf r}){\bf g}({\bf r})\;.
\label{h_def}
\end{equation}
It is easy to see that for {\em isotropic} disorder ${\bf g}({\bf
r})$, the terms quadratic in $\delta{\bf n}({\bf r})$, above, cancel
each other on average and therefore only make unimportant
contributions that are beyond quadratic order in $\delta{\bf n}({\bf
r})$. As a consequence, for such isotropic disorder, the effective
orientational disorder is given by the last term in Eq.\ref{Hn2}
\begin{equation}
H_{d n}\approx\int d^d r\; {\bf h}({\bf r})\cdot\delta{\bf n}\;,
\label{Hn1}
\end{equation}
%
%
%
We expect that random field ${\bf h}({\bf r})$ is related to the
aerogel density structure by the following correlation function
\begin{eqnarray}
\overline{{h}_i({\bf r}){h}_j({\bf r})}&=&
\overline{g_z({\bf r})g_i({\bf r})g_z({\bf r'})g_j({\bf r'})}\;,\nonumber\\
&=&\Gamma_h\overline{\rho^2_A({\bf r})\rho^2_A({\bf r'})
t_z({\bf r})t_i({\bf r})t_z({\bf r'})t_j({\bf r'})}\;,
\label{g_corr}
\end{eqnarray}
where ${\bf t}({\bf r})$ is the local tangent to the aerogel fibers,
and $\Gamma_h$ is a $\rho_A$-independent coupling constant that
depends on the microscopic aerogel-smectic interaction.

Since we expect this tangent ${\bf t}({\bf r})$ to have only
short--ranged correlations (with range of order the orientational
persistence length of the silica fibers), the above correlation
function of the tilt disorder should also be short--ranged.
Furthermore, it must be isotropic.  These considerations, taken
together with the fractal nature of the aerogel, and Eq.\ref{g_corr},
lead to the following form for the correlation function of the random
tilt-disorder ${\bf h}({\bf r})$
\begin{eqnarray}
\overline{h_i({\bf r})h_j({\bf r'})}&=&
\Gamma_h\left (\frac{a}{L} \right )^{d-d_F}\delta^d({\bf r}-{\bf r'})\;
\delta_{ij}\;,\nonumber \\
&\equiv&\Delta_h\delta^d({\bf r}-{\bf r'})\;\delta_{ij}\;,
\label{h_corr}
\end{eqnarray}
which is {\em short--ranged}. This defines the tilt field disorder
variance $\Delta_h$.

Thus our model of the NA transition in aerogel is characterized by the
effective Hamiltonian functional $H=H_{dG}+H_{d\rho}+H_{dn}$, with $H_{dG}$,
$H_{d\rho}$, and $H_{dn}$ given by Eqs.(\ref{HdG}), (\ref{Hrhoii}),
and (\ref{Hn1}), respectively. This total effective Hamiltonian must be
supplemented with correlation functions for the random $T_c$ ($\delta
t({\bf r})$), random field ($V({\bf r})$), and random tilt (${\bf
h}({\bf r})$) disorders, which are given by Eqs.\ref{tt}-\ref{rhorho},
\ref{CV2}-\ref{CV2a}, and \ref{h_corr}, respectively.

Finally, it is essential to keep in mind that, as discussed above,
even for fractal aerogel, although the correlations of the random
$T_c$ field $\delta t({\bf r})$ are long-ranged (power-law
correlated), those of the random field $V({\bf r})$ and the tilt ${\bf
h}({\bf r})$ are short-ranged, with $W << \Delta_h$.

\section{Smectic Phase and its Stability within the Harmonic Elastic Model}
\label{stability_harmonic_elastic}
In this section we study the disordered NA model, defined in the
preceding section, within the low-temperature phase. While it is
tempting to directly analyze the model $H=H_{dG}+H_{d\rho}+H_{dn}$
written in terms of the smectic order parameter $\psi$, we will not do
so here. Our motivation for this is two-fold: (i) As discussed in the
Introduction, even for the bulk (disorder-free) case the results
obtain through such a direct (de Gennes model) approach, have, so far,
failed to give predictions that agree with experiments even on the
{\em bulk} NA transition. Having not fully understood the difficulties
with the bulk NA transition, we hesitate to use such a high
temperature (``soft-spin'') analysis to study the significantly more
complicated NA transition in the presence of disorder (i.e., confined
inside the aerogel). (ii) While such a ``soft-spin'' approach is often
superior in understanding the critical properties of the {\em
transition} and the high temperature phase, it is significantly less
successful in the analysis of the low temperature phase. For example
it is {\em known} to {\em incorrectly} predict the lower critical
dimension in, e.g., the random field Ising model, as well as
completely missing the existence of the Kosterlitz-Thouless transition
in the XY-model. Since one of the main goals of the current initial
investigation of the disordered smectics problem is to ascertain the
stability and nature of the low temperature phase, we leave the direct
de Gennes model approach to this problem to subsequent
publications\cite{RadzihovskyWard}.

Instead, here, as a first analysis of the problem, we choose the
elastic, low-temperature approach in terms of the Goldstone phonon
mode $u$. The predictions of such an approach for the ordered pure
smectic phase are free of controversies, and are in agreement with
experiments on bulk smectics.

We will not, in either this or the next section (Section
\ref{tilt-only_model}), treat the full elastic model in all its
complexity, but rather, begin by simplifying the model by ignoring
{\em elastic anharmonicities} (i.e., terms higher than quadratic in
gradients of $u$).  We will also, in both this and the next section,
neglect the effects of dislocations, and focus on three spatial
dimensions.  All of these restrictions will be removed in later
sections of the paper, as we build up to the full complexity of the
disordered smectic system.

We begin by {\em assuming} the existence of smectic order and
investigate if/when this assumption is violated because of the
interaction of the smectic with the random environment of the aerogel.

Within the ordered smectic phase, the fluctuations are conveniently
described in terms of the fluctuations of the magnitude and phase of
$\psi$. It is easy to show that the fluctuations of the {\em
magnitude} of $\psi$ around the average value
$|\psi_0|=\sqrt{t_0/g_0}=\mbox{const.}$ are ``massive'', and can
therefore be safely integrated out of the partition function, leading
to only {\em finite}, unimportant shifts in the effective elastic
moduli. In contrast, the phase of $\psi$ is a U(1) massless Goldstone
mode, corresponding to spontaneously broken translational symmetry. It
is the essential low energy phonon degree of freedom of the smectic
phase, describing the local displacement of the smectic layers from
perfect periodic order. In accord with this discussion, deep within
the smectic phase, we can represent the smectic order parameter as
\begin{eqnarray}
\psi({\bf r})=|\psi_0|e^{i q_0 u({\bf r})}\;,
\end{eqnarray}
safely ignoring (actually integrating out the ``massive'')
fluctuations in the magnitude $|\psi_0|$ of $\psi$. It is important to
note that this can be done at any temperature {\em below} the
transition, without any {\em qualitative} consequences for phenomena
occuring on sufficiently long length scales, larger than a
well-defined crossover length $\xi_*(T)$. The elastic model is then
rigorously valid on length scales larger than $\xi_*(T)$ and breaks
down on shorter scales, and therefore, of course can only make
predictions about phenomena (e.g., length scales like $\xi^X$) larger
than $\xi_*(T)$. As $T\rightarrow T_{NA}^-$, $\xi_*(T)$ diverges and
the range of {\em length scales} about which the elastic model is able
to make predictions shrinks, being pushed out to infinite
scales. Consequently, for example, sufficiently close to $T_{NA}^-$,
i.e. for $T>T_*$ (see Fig.3), such that $\xi^X\rightarrow\xi_*$ from
above, our prediction for $\xi^X$, based on the weak disorder elastic
theory is no longer strictly valid.

Using this low-temperature ansatz inside the
$\psi$-dependent part of the effective Hamiltonian, given by
Eqs.\ref{HdG} and \ref{Hrhoii} and dropping constant terms, we find
\begin{eqnarray}
H[u,\delta{\bf n}]&=&\int\! d^d r \bigg[{B_\perp\over 2}
|{\bbox\nabla}_\perp u-\delta{\bf n}|^2+{B\over 2}(\partial_z
{u})^2\;\nonumber\\
+{K_s\over2}(\bbox{\nabla}&\cdot&\delta{\bf n})^2
+{K_t\over2}(\hat{\bf z} \cdot \bbox{\nabla} \times\delta{\bf n})^2
+ {K_b\over2}(\hat{\bf z}\times\bbox{\nabla}\times\delta{\bf n})^2\nonumber\\
+ {\bf h}({\bf r})&\cdot&\delta{\bf n}
-|\psi_0|U({\bf r})\cos[q_0\big(u({\bf r})+z\big)] \bigg]\;,\nonumber\\
&&\label{Hun}
\end{eqnarray}
where $B_\perp=c_{\perp}|\psi_0|^2 q_0^2$ and $B=c_{||}|\psi_0|^2
q_0^2$. We observe that the fluctuation mode $(\bbox{\nabla}_\perp
u-\delta{\bf n})$ is ``massive'' and leads to the Anderson-Higg's
mechanism, a hallmark of gauge theories.  As a consequence, after a
simple Gaussian integration over $\delta {\bf n}$, we find that at
long length scales, $\delta{\bf n}$ fluctuations are constrained to
follow $\bbox{\nabla}_\perp u$. The resulting effective elastic
Hamiltonian is then obtained by the replacement
\begin{equation}
\delta{\bf n}\rightarrow \bbox{\nabla}_\perp u\;,
\end{equation}
everywhere in the Eq.\ref{Hun}. This is valid in the long wavelength
limit, to quadratic order in gradients of $u$, and provided
dislocations are confined. We obtain,
\begin{eqnarray}
H[{u}]&=&\int \! d^d r \bigg[ {B\over 2}(\partial_z {u})^2 +
{K\over 2}(\nabla^2_\perp {u})^2 + {\bf h}({\bf
r})\cdot\bbox{\nabla_\perp} u\nonumber\\ &-& |\psi_0|U({\bf
r})\cos[q_0\big(u({\bf r})+z\big)] \bigg]\;,
\label{Hu}
\end{eqnarray}
or, equivalently, in terms of $V({\bf r})$
\begin{eqnarray}
H[{u}]=\int \! d^d r
\bigg[  {B\over 2}(\partial_z
{u})^2 + {K\over 2}(\nabla^2_\perp {u})^2
+ {\bf h}({\bf r})\cdot\bbox{\nabla_\perp} u\nonumber\\
-  |\psi_0|\left(V({\bf r})e^{i q_0 u({\bf r})}+
V^*({\bf r})e^{-i q_0 u({\bf r})}\right)
\bigg]\;.
\label{Huii}
\end{eqnarray}
In the above $K=K_s$, although, as discussed in the beginning of
Sec.\ref{model}, in a model of the NA transition that is more general
than the de Gennes model, $K$ and $K_s$ can differ by a {\em singular}
function of the reduced temperature $|T-T_{NA}|$.

To compute self-averaging quantities, (e.g., the disorder averaged
free energy) we employ the replica ``trick''\cite{Anderson}, which
allows us to work with a translationally invariant field theory at the
expense of introducing $n$ replica fields (with the $n\rightarrow 0$
limit to be taken at the end of the calculation). For the free energy
this procedure relies on the identity for the $\ln(x)$ function
\begin{equation}
{\overline F}=-T\overline{\ln Z}=-T\lim_{n\rightarrow
0}{\overline{Z^n}-1\over n}\;.
\label{replica}
\end{equation}
After replicating and integrating over the disorder using
Eqs.\ref{CV2}-\ref{CV2a}, \ref{h_corr}, we obtain
\begin{equation}
\overline{Z^n}=\int[d u_\alpha]e^{-H[u_\alpha]/T}\;.
\end{equation}
The effective translationally invariant replicated Hamiltonian
$H[u_\alpha]$ is given by
\begin{eqnarray}
H[u_\alpha]&=&\int d^d
r\bigg[\sum_{\alpha=1}^n\left({K\over2}(\nabla_\perp^2
u_\alpha)^2+{B\over2}(\partial_z u_\alpha)^2\right)\nonumber\\
&\hspace{-1.9cm}+&\hspace{-1cm}{1\over T}\hspace{-0.1cm}
\sum_{\alpha,\beta=1}^n\hspace{-0.1cm}
\left({\Delta_h\over4}|\bbox{\nabla_\perp}(u_\alpha-u_\beta)|^2 -
{\Delta}_V\cos[q_0(u_\alpha-u_\beta)]\right)\bigg]\;,\nonumber\\
\label{Hr}
\end{eqnarray}
where $\Delta_V=|\psi_0|^2W$. The statistical symmetry under global
rotation forces the disorder generated replica off-diagonal terms to
be invariant under $u_\alpha({\bf r})\rightarrow u_\alpha({\bf r})+
\bbox{\theta}\cdot {\bf r}_\perp$. In the replicated effective
Hamiltonian Eq.\ref{Hr} the nonlinearities only depend on the
difference between different replica fields and therefore do not
depend on the ``center of mass'' field $\sum_{\alpha=1}^n u_\alpha$,
which is therefore a noninteracting field. This implies an exact
result that $K$ and $B$ are not renormalized by the disorder in this
harmonic approximation.\cite{TD} $\Delta_h$, of course, will be
renormalized by the random-field nonlinearity.

It is easy to see that the random tilt disorder term with coefficient
$\Delta_h$ can be rewritten in Fourier space as
\begin{equation}
{{\Delta_h}\over2}q_\perp^2\sum_{\alpha,\beta}|u_\alpha-u_\beta|^2=
\sum_{\alpha,\beta}\Delta_h q_\perp^2 \left[-1+n\delta_{\alpha\beta}\right]
u_\alpha u_\beta\;,
\end{equation}
which, in the $n\rightarrow0$ limit leads to the quadratic part of the
Hamiltonian
\begin{equation}
H_0[u_\alpha]={1\over2}\int d^d
q\sum_{\alpha,\beta}^n\bigg[\left(K q_\perp^4+
B q_z^2\right)\delta_{\alpha\beta}-{\Delta_h\over T}
q_\perp^2\bigg]u_\alpha
u_\beta\;,
\label{Hr0}
\end{equation}
from which the propagator $G_{\alpha\beta}({\bf q})$ defined through
\begin{eqnarray}
\langle u_{\alpha}({\bf q})u_\beta({\bf q}\prime) \rangle = G_{\alpha
\beta}({\bf q}) \delta^d ({\bf q}+{\bf q}\prime) \,
\label{Propdef}
\end{eqnarray}
can be easily obtained,
\begin{equation}
G_{\alpha\beta}({\bf q})={T\delta_{\alpha\beta}\over K q_\perp^4+B
q_z^2}+{\Delta_h q_\perp^2\over(K q_\perp^4+B q_z^2)^2} ,
\label{Gab}
\end{equation}
using an identity for inverting matrices of the type
\begin{equation}
A_{\alpha\beta}=a\delta_{\alpha\beta}+b\;,
\label{Aab}
\end{equation}
namely:
\begin{eqnarray}
A^{-1}_{\alpha\beta}&=&{1\over a}\delta_{\alpha\beta}-{b\over a(a+b
n)}\;,\nonumber\\
&{\stackrel{=}{_{n\rightarrow 0}}}&\;
{1\over a}\delta_{\alpha\beta}-{b\over a^2}\;.
\label{Ainv}
\end{eqnarray}

In three dimensions, for $\Delta_h=\Delta_V=0$ (disorder-free liquid
crystal) at low temperatures $(T<T_{NA})$, the smectic-A phase is
described by a fixed plane, defined by the bare values of $K $ and
$B$.  Our initial goal is to establish how this fixed plane is
destabilized by the disorder, i.e. by the random field and random tilt
terms. The calculation is a generalization of that for the 2d
random-field XY model\cite{CardyOstlund,TD} to the anisotropic
elasticity of the smectic-A in three dimensions.  We employ the
standard momentum shell renormalization group
transformation\cite{Wilson}, by writing the displacement field as
\begin{equation}
u_\alpha({\bf r})=u^<_\alpha({\bf r})+u^>_\alpha({\bf r})\;,
\label{u_shell}
\end{equation} 
integrating perturbatively in $\Delta_V$ the high wavevector part
$u^>_\alpha({\bf r})$, nonvanishing inside a thin cylindrical momentum
shell
\begin{mathletters}
\begin{eqnarray}
\Lambda e^{-\ell}<&|{\bf q}_\perp|&<\Lambda\;,\\
-\infty<&q_z&<\infty\;,
\end{eqnarray}
\label{shell}
\end{mathletters}
and rescaling the lengths and long wavelength part of the fields with
\begin{mathletters}
\begin{eqnarray}
r_\perp&=&e^{\ell}r_\perp'\;,\label{r}\\
z&=&e^{\omega\ell}z'\;,\label{z}\\ 
u^<_\alpha({\bf r})&=&e^{\phi\ell}u_\alpha({\bf r'})\;,\label{u}
\end{eqnarray}
\label{rescale}
\end{mathletters}
so as to restore the ultraviolet cutoff back to $\Lambda$.  Because
the random-field nonlinearity is a periodic function, it is convenient
(but not necessary) to take the arbitrary field dimension $\phi=0$,
thereby preserving the period $2\pi/q_0$ under the renormalization
group transformation.\cite{q0flow} Under this transformation the
resulting effective Hamiltonian functional can be restored into its
original form Eq.\ref{Hr} with effective $\ell$-dependent
couplings. We relegate the details of these calculations to Appendix
A, and focus here on the results.  These can be succinctly summarized
in the renormalization group flow equations
\begin{mathletters}
\begin{eqnarray}
\frac{d\Delta_V(\ell)}{d\ell}&=&(2+\omega-\eta)\Delta_V - A_1\Delta_V^2\;,
\label{gamma_ell}\\
\frac{d K (\ell)}{d\ell}&=&(\omega-2)K \;,
\label{K_ell}\\
\frac{d B(\ell)}{d\ell}&=&(2-\omega)B\;,
\label{B_ell}\\
\frac{d\Delta_h(\ell)}{d\ell}&=&\omega\Delta_h +
A_2\Delta_V^2\;,\label{Deltah_ell}
\end{eqnarray}
\label{flows_random_field}
\end{mathletters}
where we have defined
\begin{mathletters}
\begin{eqnarray}
\eta &=& {q_0^2\,T\over 4\pi\sqrt{K  B}}\;,\label{eta}\\ 
A_1 &=& {q_0^4\over8\pi\Lambda^4\sqrt{K^3B}}\;,\label{A1}\\ 
A_2 &=& {c\,q_0^6\over4\pi\Lambda^6\sqrt{K^3B}}\;,\label{A2}
\end{eqnarray}
\end{mathletters}
and $c$ is a dimensionless number of order $1$. Note that
${d\eta(\ell)}/{d\ell}=0$, exactly.

As discussed above, the symmetry of the effective Hamiltonian $H$ in
Eq.\ref{Hr} guarantees that the flow equations for $K (\ell)$ and
$B(\ell)$ are {\em exact} (arising from simple length rescaling with
{\em no diagrammatic} corrections), ignoring (for now) the effects of
both anharmonic elastic terms and topological defect loops in $u$; the
latter become important at high temperatures, where they induce the NA
transition, by driving the bulk modulus $B$ to zero.\cite{Toner,Kflow}
It is convenient to choose the anisotropy exponent $\omega=2$, because
such a choice keeps $K (\ell)$ and $B(\ell)$,
Eqs.\ref{K_ell},\ref{B_ell}, fixed under the RG.  Although it {\em
appears} from Eq.\ref{gamma_ell} that the glass transition temperature
(relevance and irrelevance of $\Delta_V$) depends on the arbitrary
choice of $\omega$, it does not. This choice is actually completely
arbitrary and does not affect any {\em physical} quantities. This can
be seen by looking at the proper {\em dimensionless} coupling constant
\begin{equation}
\tilde{\Delta}_V\equiv A_1\Delta_V\;,
\end{equation}
whose recursion flow equation can be easily obtained by combining
Eqs.\ref{gamma_ell},\ref{K_ell}, and Eq.\ref{B_ell}
\begin{eqnarray}
\frac{d\tilde{\Delta}_V(\ell)}{d\ell}&=&(4-\eta)\tilde{\Delta}_V -
\tilde{\Delta}_V^2\;,\nonumber\\
&=&\big[4-\eta-\tilde{\Delta}_V\big]\tilde{\Delta}_V\;.
\label{gamma_tilde}
\end{eqnarray}
Obviously this flow equation is independent of the arbitrary rescaling
exponent $\omega$ and has the same form as that for $\Delta_V$,
Eq.\ref{gamma_ell}, with $\omega=2$.

From Eq.\ref{gamma_tilde}, we then find that for $\eta<4$ (large
elastic moduli and low temperature), or, equivalently, below the
pinning transition temperature $T_p$, given by
\begin{equation}
T_p={16\pi\sqrt{K B}\over q_0^2}\;,
\end{equation}
the smectic ($\Delta_V=0$) fixed plane is unstable to
disorder. However the initial runaway of disorder is halted by the
nonlinear terms in $\tilde{\Delta}_V$, which terminate the flow at a
new finite disorder fixed line,
\begin{equation}
\tilde{\Delta}_V^*(T)=4-\eta\;.
\label{g_fixed}
\end{equation}
This new fixed line then describes a randomly pinned, glassy smectic-A
phase, analogous to the 2d super-rough phase of crystal surface on a
random substrate\cite{TD} and the 1+1 vortex glass phase of flux-line
vortices (confined to a plane) in type II superconductors\cite{VG}.
This RG flow structure is summarized in Fig.6.
\begin{figure}[bth]
{\centering
\setlength{\unitlength}{1mm}
\begin{picture}(150,60)(0,0)
\put(-20,-57){\begin{picture}(150,0)(0,0)
\includegraphics{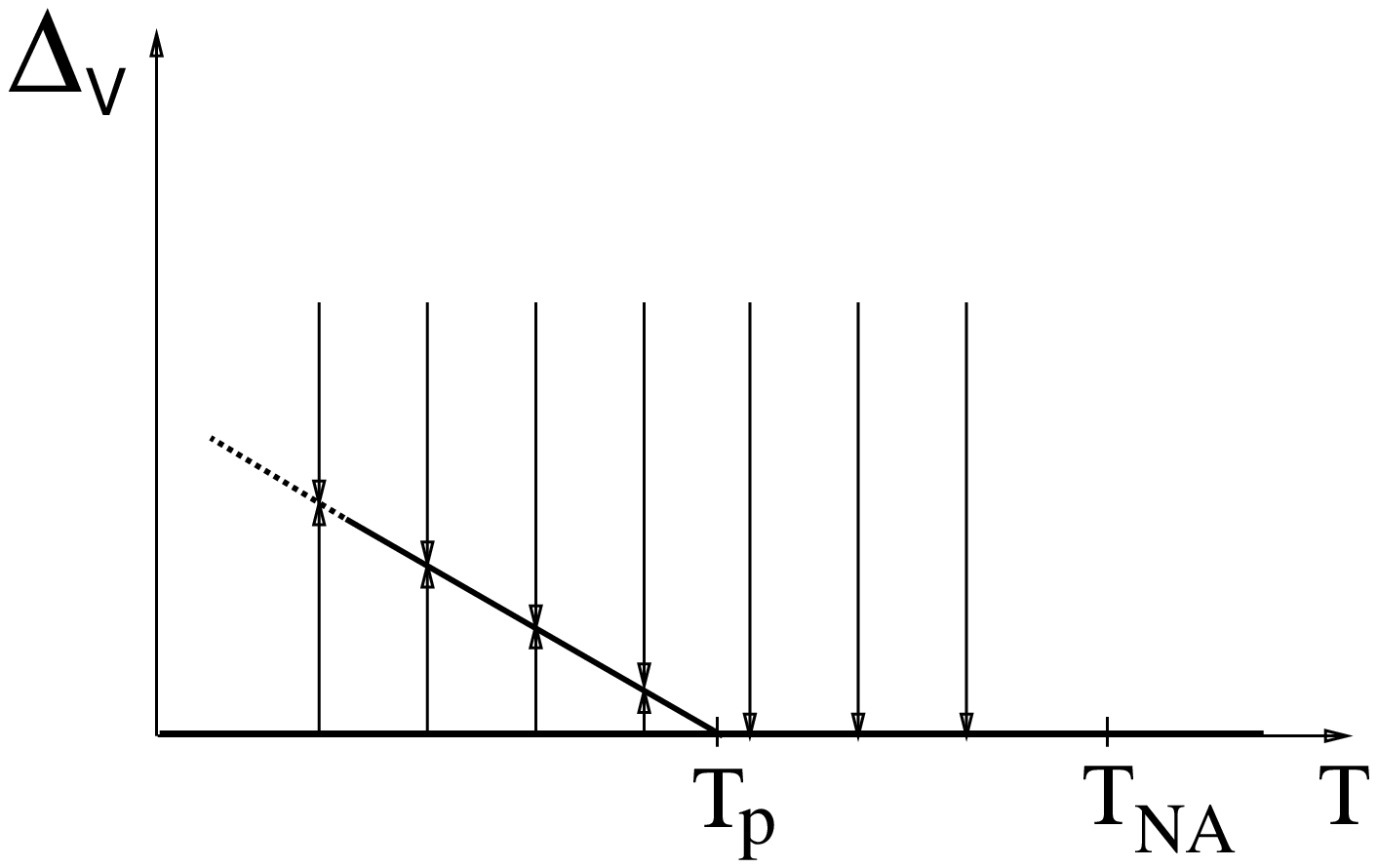}
\end{picture}}
\end{picture}}
Fig.6. A fixed line characterizing the pinned glassy phase of a three
dimensional smectic in aerogel. $T_p=16\pi\sqrt{K B}/q_0^2$ is the
pinning transition temperature, taking place {\em within} the 3d SBG
phase. This transition only survives within the harmonic elastic
approximation.
\label{fixedline}
\end{figure}
It is important to note that the transition at $T_p$, illustrated in
Fig.6, is taking place {\em within} the SBG phase. That is, $T_p$ is a
transition from the ($T>T_p$) SBG phase, within which the
translational random field disorder $\Delta_V$ is unimportant at long
scales, to the ($T<T_p$) SBG phase, in which $\Delta_V$ is relevant
and flows to a nontrivial fixed line. This $T_p$ pinning transition
must not be confused with the distinct higher temperature transition,
occuring just below $T_{NA}$, in which the putative unpinned ($T>T_p$)
SBG ``melts'' into the NEG by unbinding of dislocations. It is most
certain that the transition at $T_p$, derived here within the harmonic
elastic approximation, will be converted into a crossover by the
neglected elastic nonlinearities, analyzed in
Secs.\ref{anom_elasticity}-\ref{random_field_3-5}. However, it is
possible that a rounded ``ghost'' of the transition at $T_p$ will
persist in our full theory and will therefore be experimentally
observable.

It is enlightening to contrast these RG flows, with their perturbative
fixed line, Eq.\ref{g_fixed}, at which the relevant flows of
$\tilde{\Delta}_V$ terminate, with those of disorder-free {\em
thermal} Sine-Gordon models (e.g. {\em thermal} roughening
transition)\cite{SineGordon}, where the cosine coupling runs away to
strong coupling at low temperatures. In these latter systems the
cosine potential becomes relevant upon lowering the temperature, with
the system settling down in one of the minima of this {\em periodic}
pinning potential. As the cosine coupling continues to grow, it
further reduces thermal fluctuations, suppressing their ability to
average away its pinning effects, thereby further increasing its
pinning influence. Mathematically this manifests itself in the {\em
positive} nonlinear contribution to the flow equation for the cosine
coupling, which results in the absence of a stable {\em finite}
coupling fixed line. In contrast, for disordered problems of the type
considered here, as the disorder becomes relevant at low temperatures
and begins to grow, it leads to fluctuations (roughening) that are
larger than those from purely thermal fluctuations\cite{TD}. This
disorder-enhanced roughening subsequently leads to a more effective
averaging away of the pinning potential thereby suppressing its
effects beyond a certain strength. Mathematically, this is captured by
the {\em negative} sign of the nonlinear $\tilde{\Delta}_V^2$ term in
the flow Eq.\ref{gamma_tilde}, which results in the termination of the
flow of $\tilde{\Delta}_V$ and the glassy fixed line described by
Eq.\ref{g_fixed}.

The flow Eq.\ref{gamma_tilde} also implies that the random-field
disorder is irrelevant for $\eta>4$. Since the bulk modulus $B$
vanishes, while $K $ remains finite through $T_{NA}$, $\eta$ diverges
as $T \rightarrow T^-_{NA}$, and hence, sufficiently close to the NA
transition temperature T$_{NA}$, we are {\em guaranteed} to have a
range of temperatures (within the smectic phase) over which the
random-field disorder is irrelevant. However, as we will see below,
because tilt disorder $\Delta_h$ is a strongly relevant perturbation,
the three-dimensional quasi-long-range smectic order for $\eta>4$ will
be converted into short-range correlations, even when the random field
disorder $\Delta_V$ is irrelevant.

It is essential to stress that the renormalization group flow diagram
described above (i.e. relevance for $\eta<4$ and irrelevance for
$\eta>4$ of the random-field disorder) survives even despite the
strong relevance and runaway of the random tilt coupling
$\Delta_h$. This occurs because the dimension of the
$\cos\left[q_0(u_\alpha-u_\beta)\right]$ operator (the random field
disorder) is {\em independent} of $\Delta_h$, to all orders in
perturbation theory in $\Delta_V$. This can be easily seen to first
order in $\Delta_V$:
\begin{eqnarray}
\langle-H_{int}\rangle&=&{\Delta_V\over T}\int d^3
r\sum_{\alpha\beta}\langle\cos[q_0(u_\alpha-u_\beta)]\rangle_>\;,\nonumber\\
&=& {\cal\mbox{Re}}\;{\Delta_V\over T}\int d^3 r\sum_{\alpha\beta} e^{i
q_0(u^<_\alpha-u^<_\beta)}\langle e^{i
q_0(u^>_\alpha-u^>_\beta)}\rangle_>\;,\nonumber\\ 
&=&{\Delta_V\over T}\int d^3 r\sum_{\alpha\beta}
\cos[q_0(u^<_\alpha-u^<_\beta)]e^{-f_{\alpha\beta}}\;,\nonumber\\
&=&{\Delta_V\over T}\int d^3 r\sum_{\alpha\beta}
\cos[q_0(u^<_\alpha-u^<_\beta)]e^{-\eta\ell(1-\delta_{\alpha\beta})}
\;,\nonumber\\
\end{eqnarray}
where
\begin{equation}
f_{\alpha\beta}=q_0^2\big[G^>_{\alpha\alpha}(r=0)-
G^>_{\alpha\beta}(r=0)\big]\;,
\end{equation}
in $G^>_{\alpha\alpha}(0)$ there is no implied sum over $\alpha$, and
\begin{equation}
G_{\alpha\beta}^>({\bf r})=
\int_{\Lambda e^{-\ell}}^{\Lambda}{d^2
q_\perp\over(2\pi)^2}\int_{-\infty}^\infty{d q_z\over 2\pi}
G_{\alpha\beta}(q)e^{i{\bf q\cdot r}}\;.
\end{equation}
Using
\begin{equation}
G_{\alpha\alpha}(q)-G_{\alpha\beta}(q)=
{T\over K  q_\perp^4+B q_z^2}(1-\delta_{\alpha\beta})\;,
\end{equation}
which is obviously independent of the tilt-disorder coupling
$\Delta_h$, and Fourier transforming, we obtain the
$\Delta_h$-independent $\eta$ given in Eq.\ref{eta}. Performing length
rescalings, Eqs.\ref{rescale}, to restore the new cutoff $\Lambda
e^{-\ell}$ back to $\Lambda$ gives, to first order in $\Delta_V$, the
flow equation for $\Delta_V$, Eq.\ref{gamma_ell}.

To see that the recursion relation for $\Delta_V$,
Eq.\ref{gamma_tilde}, is independent of $\Delta_h$ to {\em all} orders
in $\Delta_V$, in this otherwise harmonic theory, one can return to
the non-replicated Hamiltonian, Eq.\ref{Huii} and completely eliminate
the random tilt field ${\bf h}({\bf r})$ via a change of variables
\begin{equation}
u({\bf r})\equiv u'({\bf r})+f({\bf r})\;,\label{change1}
\end{equation}
\top{-2.7cm}
with $f({\bf r})$ completely determined by the random tilt field ${\bf
h}({\bf r})$ via
\begin{equation}
(-B\partial_z^2 + K\nabla^4_\perp) f({\bf r})=\bbox{\nabla}_\perp
\cdot{\bf h}({\bf r}) \;,\label{change2}
\end{equation}
Inserting Eq.\ref{change1} into Eq.\ref{Huii}, and dropping terms that
only depend on the random tilt field ${\bf h}({\bf r})$, since, as
discussed earlier, these have no effect on {\em any} correlation
functions, we have
\end{multicols}
\begin{equation}
\hspace{-.3cm}H=\int \! d^d{\bf r}
\bigg[  {B\over 2}(\partial_z
{u}')^2 + {K\over 2}(\nabla^2_\perp {u}')^2
+ {\bf h}({\bf r})\cdot\bbox{\nabla_\perp} u'
+B\partial_z f\partial_z u'+K\nabla^2_\perp {f}\nabla^2_\perp {u}'
-|\psi_0|\left(V'({\bf r})e^{i q_0 u'({\bf r})}+
V^{'*}({\bf r})e^{-i q_0 u'({\bf r})}\right)
\bigg]\;,\label{Huii_prime}
\nonumber\\
\end{equation}
\bottom{-2.7cm}
\begin{multicols}{2}
\noindent
where we have defined $V'({\bf r})\equiv e^{i q_0 f({\bf r})}V({\bf
r})$. Note that the {\em statistics} of $V'({\bf r})$ are exactly the
same as the {\em statistics} of $V({\bf r})$, since all we have done
is added a new random phase factor to $V({\bf r})$. But since the
phase of $V({\bf r})$ was uniformly distributed between $0$ and
$2\pi$, the phase of $V'({\bf r})$ will also be. Hence, its statistics
are the same. Therefore, the model Eq.\ref{Huii_prime} is {\em only}
affected by the presence of the random tilt field ${\bf h}({\bf r})$
through the terms
\begin{equation}
\Delta H=\int \! d^d{\bf r}
\bigg[{\bf h}({\bf r})\cdot\bbox{\nabla_\perp} u'
+B\partial_z f\partial_z u'+K\nabla^2_\perp {f}\nabla^2_\perp {u}'\bigg]\;,
\label{DeltaHuii_prime}
\end{equation}
which, after integration by parts, vanish by virtue of the choice
Eq.\ref{change2}. Thus, the new Hamiltonian is unaffected by the
random field ${\bf h}({\bf r})$. Since the partition function is
obviously invariant under a change of variables, it and, therefore the
recursion relations for the parameters $B$, $K$, and most importantly
$\Delta_V$, cannot be affected by ${\bf h}({\bf r})$, and hence are
independent of $\Delta_h$.

The results described above for the disordered three-dimensional
smectic-A phase are quite closely analogous to those of Cardy and
Ostlund for random symmetry breaking fields in the two-dimensional
XY-model\cite{CardyOstlund}, except that for pinned smectics, the
random-field disorder (the cosine) is guaranteed to become irrelevant
as $T \rightarrow T^- _{NA}$ since $\eta(T\rightarrow
T_{NA})\rightarrow\infty$.

As can be seen from the detailed renormalization group analysis of
Appendix A and Eq.\ref{Deltah_ell}, even if the bare $\Delta_h=0$
(i.e., no initial tilt disorder), this type of disorder is generated
by the random-field disorder ($\Delta_V$) after the high-wavevector
degrees of freedom are integrated out. In contrast to the 2d
random-field XY-model, where the generated $\Delta_h$ disorder is only
marginally relevant and only weakly affects the QLRO order found for
$\Delta_h=0$ (converting $\ln(r)$ phase correlations to $\ln^2(r)$),
for the 3d smectic-A phase the $\Delta_h$ tilt disorder is strongly
relevant.  As we will see in the next section, the dimensionless
coupling that determines the effect of the tilt disorder is
\begin{equation}
\tilde{\Delta}_h\equiv A_1{\Delta_h}\;,\label{gh}
\end{equation}
From the recursion relations Eqs.\ref{gamma_ell}-\ref{Deltah_ell} we find
\begin{equation}
\frac{d \tilde{\Delta}_h}{d\ell}=2\tilde{\Delta}_h+{2c q_0^2\over\Lambda^2}\tilde{\Delta}_V^2
\label{g}
\end{equation}
For $\eta > 4$, $\tilde{\Delta}_V(\ell)\rightarrow 0$ (as we have seen),
and so $d \tilde{\Delta}_h/d\ell=2\tilde{\Delta}_h$, which is trivially solved to give
\begin{equation}
\tilde{\Delta}_h(\ell)=\tilde{\Delta}_h e^{2\ell}\;,
\end{equation}
where $\tilde{\Delta}_h$ is the non-universal dimensionless ``bare''
coupling, Eq.\ref{gh}.  Thus, we see that the tilt disorder is
strongly relevant, in contrast to the behavior of the two dimensional
random field XY model,\cite{CardyOstlund} for which the tilt disorder
was marginal, in the RG sense, in the phase in which the cosine was
irrelevant.

For $\eta < 4$, $\tilde{\Delta}_V\rightarrow\tilde{\Delta}_V^* > 0$,
with $\tilde{\Delta}_V^*$ given by Eq.\ref{g_fixed}.  Now, in the 2d
random field XY model, the existence of a non-zero
$\tilde{\Delta}_V^*$ in the low temperature phase implied completely
different behavior for the tilt coupling $\tilde{\Delta}_h(\ell)$ than
in the high temperature phase: a runaway of $\tilde{\Delta}_h(\ell)$
to infinity as $\ell\rightarrow \infty$ in the low temperature phase,
as opposed to a constant $\tilde{\Delta}_h(\ell \rightarrow \infty)$
in the high temperature phase.\cite{TD} In {\em our} problem, however,
$\tilde{\Delta}_h(\ell)$ runs away to infinity in {\em both} phases,
and {\em asymptotically in exactly the same way}!  One can see this by
solving Eq.\ref{g} with $\tilde{\Delta}_V(\ell)$ replaced by its
non-zero fixed point value $\tilde{\Delta}_V^*$ given by
Eq.\ref{g_fixed}, finding
\begin{equation}
\tilde{\Delta}_h(\ell)=\big[\tilde{\Delta}_h
+{c q_0^2\over\Lambda^2}({\tilde{\Delta}_V^*})^2\big] e^{2\ell}-{c
q_0^2\over\Lambda^2}({\tilde{\Delta}_V^*})^2\;,
\label{gsol}
\end{equation}
which has the same asymptotic behavior as $\ell\rightarrow \infty$
(namely $\tilde{\Delta}_{h0}e^{2\ell}$) as in the low temperature
phase.  Non-universal constants (like $\tilde{\Delta}_{h0}$) can be
different in the low temperature phase, but the scaling ($e^{2\ell}$)
is not.  We will see in the next section that this implies that {\em
equal time} correlation functions scale in exactly the same way in
both the pinned $(\eta < 4)$ and non-pinned $(\eta > 4)$ phase.  The only
difference between these two phases is in their dynamics, which are
divergently slower in the randomly pinned phase, as we shall show in a
future publication.\cite{RTunpublished}

Physically the strong relevance of $\Delta_h$ is expected, because in
the smectic-A phase the rotational invariance is {\em spontaneously}
broken. Since each realization of the random tilt disorder {\em
explicitly} breaks the rotational invariance (preserving it only
statistically), the smectic layer orientation has a divergent response
to $\Delta_h $ tilt disorder. At long scales where $\Delta_h\approx0$
is no longer valid, the {\em quasi-long-range} order, implied by the
nontrivial fixed line, Eq.\ref{g_fixed}, will undergo a rapid
crossover to short-range correlations of $\psi({\bf r})$, even in the
regime ($\eta>4$) where the random-field disorder (which always
generates random tilt $\Delta_h $) is irrelevant.

\section{Random Tilt-only Harmonic Elastic (Topologically Ordered) Model}
\label{tilt-only_model}
For finite $\Delta_h $ both the random field ($\Delta_V$) and random
tilt ($\Delta_h $) disorders must be treated simultaneously. Since, as
described above, our conclusions about the phase diagram for $\Delta_V$
are not affected by the tilt disorder $\Delta_h$, we can study the
effect of $\Delta_h $ on smectic-A order within the regime $\eta>4$,
where the random-field disorder is irrelevant. In this regime, at long
enough scales the smectic-A phase is effectively subjected to
tilt-disorder only. We can therefore safely set $\Delta_V=0$ and analyze
exactly the remaining quadratic theory for any value of $\Delta_h $.

As argued in the previous section, and to be shown in the next, the
scaling of the results we find here will also apply in the pinned
phase where $\eta < 4$ and the cosine disorder is relevant.

Since, within the harmonic-elastic approximation of this section, the
effective Hamiltonian is quadratic in the displacement fields $u$,
when $\Delta_V=0$ all the correlation functions can be computed
exactly.  In particular the quantity of interest is
\begin{equation}
C({\bf r}_\perp,z)=\overline
{\langle\left[u({\bf r_\perp},z)-u({\bf 0},0)\right]^2\rangle}\;.
\label{CTD}
\end{equation}
Fourier transforming the $u$-fields, we get
\begin{equation}
C({\bf r}_\perp,z)=2 \int\frac{d^2q_\perp dq_z}{(2\pi)^3}(1-e^{i{\bf
q}\cdot{\bf r}})\overline{\frac{\langle u({\bf q})u({\bf
q}')\rangle}{\delta^3({\bf q}+{\bf q}')}}\;.
\label{C1}
\end{equation}
We can write the quenched and thermal averaged $\overline{\langle
u({\bf q})u({\bf q}')\rangle}$ in terms of their replicated
correlation function:
\begin{eqnarray}
\overline{\langle u({\bf q})u({\bf q}')\rangle}
&=&\langle u_\alpha({\bf q})u_\alpha({\bf q}')\rangle\;,\nonumber\\
&=&\delta^3({\bf q}+{\bf q}')G_{\alpha\alpha}({\bf q})\;,\label{C2}
\end{eqnarray}
where no sum on $\alpha$ is implied, and the replica propagator
$G_{\alpha\beta}$ is given by Eq.\ref{Gab}.  Using that equation, and
Eq.\ref{C2} for the correlation function, gives
\begin{equation}
C({\bf r}_{\perp},z)=C_T({\bf r}_{\perp}, z) +C_\Delta({\bf r}_{\perp},z)
\end{equation}
where we have separated $C({\bf r}_{\perp},z)$ into ``thermal'' $C_T$
and ``disorder'' (frozen) $C_\Delta$ parts, given by
\begin{eqnarray}
C_T&=&2T\int{d^2{q_\perp}d
q_z\over(2\pi)^3}{1-e^{i{\bf q}\cdot{\bf r}}\over K  q_\perp^4 + B
q_z^2}\;,\nonumber\\
&=&{T\over2\pi\sqrt{K  B}}\;g_T\left({z\lambda\over r_\perp^2},
{r_\perp\over a}\right)\;,\nonumber\\
&=&{T\over2\pi\sqrt{K  B}}
\left[\ln\left({r_\perp\over a}\right)-
\frac{1}{2}{Ei}\left({-r_\perp^2\over 4\lambda|z|}\right)\right]\;,
\label{CT}
\end{eqnarray}
and
\begin{eqnarray}
C_\Delta&=&2\Delta_h\int{d^2{q_\perp}d
q_z\over(2\pi)^3}{q_\perp^2\left(1-e^{i{\bf q}\cdot{\bf r}}\right)
\over\left(K  q_\perp^4 + B q_z^2\right)^2}\;,\nonumber\\
&=&\frac{\Delta_h}{32\pi B^2\lambda^3}
\left\{4\lambda|z| e^{-r_\perp^2/4\lambda|z|}+\right.\nonumber\\
&+&\left.r_\perp^2\left[Ei\left({-4\lambda L_z\over L_\perp^2}\right)+
Ei\left({-r_\perp^2\over4\lambda|z|}\right)
+2\ln\left({L_\perp\over r_\perp}\right)\right]\right\}\;,\nonumber\\
\label{CD}
\end{eqnarray}
where $Ei(x)$ is the Exponential Integral function, and, as promised,
the effective dimensionless coupling $\tilde{\Delta}_h$, defined in
Eq.\ref{gh}, naturally appears above. In the above, we have considered
a finite system whose shape is a rectangular parallelepiped of linear
dimensions $L_\perp\times L_\perp\times L_z$, $L_z$ being the length
of the system along the ordering ($z$) direction.  Unless it has a
huge aspect ratio, such that $L_z \sim L_\perp^2/\lambda>> L_\perp$,
any large system ($L_\perp,L_z >>\lambda$) will have $\lambda L_z <<
L_\perp^2$.  In this limit, the asymptotic behaviors of $C_\Delta$ are
\begin{equation}
C_\Delta\approx\left\{\begin{array}{lr}
\frac{\Delta_h}{32\pi B^2\lambda^3}
\left[4 \lambda|z|+r_\perp^2\ln|\frac{L_z}{z}|\right]\;,
&\lambda|z|>>r_\perp^2\;,\\
&\lambda L_z<<L_\perp^2\label{limit1}\;,\\
\frac{\Delta_h}{16\pi B^2\lambda^3}
r_\perp^2\ln\left(\frac{2\sqrt{\lambda L_z}}{r_\perp}\right)\;,
&\lambda |z|<<r_\perp^2\;,\\
&\lambda L_z<<L_\perp^2\label{limit2}\;.
\end{array}\right.
\end{equation}
Note that, although in principle the second $(r_\perp^2\ln |L_z/z|)$
term in the $\lambda|z|>>r_\perp^2$ expansion of $C_\Delta$ dominates
the $\lambda|z|$ term in the thermodynamic limit $L_z \rightarrow
\infty$ (taking that limit at fixed $r_\perp$ and $z$), in practice,
for any reasonable system size $L_z$, the first $(\lambda|z|)$ term
actually dominates if $\lambda |z|/r_\perp^2$ gets appreciably
bigger than $1$.

An unusual feature of this result is that not only do the mean squared
fluctuations of $u$ at a given point in space diverge as a function of
system size, but even the {\em relative} displacement of two points
with {\em finite} separations $(r_\perp, z)$ diverge as the system
sizes $(L_\perp, L_z)$ go to infinity.  This is because the mean
squared real space {\em orientational} fluctuations
$\overline{\langle|\delta{\bf n}({\bf r})|^2\rangle}$ also diverge as
$L_{\perp,z}\rightarrow\infty$:
\begin{eqnarray}
\langle |\delta{\bf n}({\bf r})|^2\rangle &=&
\langle|\bbox{\nabla}_\perp u({\bf r})|^2\rangle\;,\nonumber\\
&=&\Delta_h\int{d^2{q_\perp}d q_z\over(2\pi)^3}
{q_\perp^4\over\left(K  q_\perp^4 + B q_z^2\right)^2}\;,\nonumber\\
&=&{\Delta_h\over16\pi^2B^2\lambda^3}
\int_{q_\perp>\mbox{max}\left[(\lambda
L_z)^{-1/2},L_\perp^{-1}\right]}
{d^2 q_\perp\over q_\perp^2}\;,\nonumber\\
&=&{\Delta_h\over8\pi B^2\lambda^3}
\ln\left({\mbox{min}}[\sqrt{\lambda L_z},L_\perp]\right)\;\nonumber\\
\label{orientation}
\end{eqnarray}

Although not experimentally relevant, it is instructive to generalize
this calculation to spatial dimensions $d>3$.  Keeping only the tilt
disorder (we will return to justify this later), and repeating the
calculation just presented, we find
\begin{equation}
C_\Delta({\bf r}_\perp,z)\propto\left\{\begin{array}{lr}
\Delta_h|z|^{(5-d)/2},&\lambda|z|>>r_\perp^2,\\
\Delta_h r_\perp^{5-d},&\lambda|z|<<r_\perp^2,
\end{array}\right.
\label{CDelta}
\end{equation}
which diverges at large distances for all $d < 5$.  This divergence
signals the destruction by tilt disorder of the (quasi-) long-ranged
smectic translational order for $d<5$.  Note that this divergence
occurs even for arbitrarily weak disorder (i.e., arbitrarily small
$\Delta_h$).  This agrees with the experimental observation that, even
when the aerogel density becomes very low, smectic translational order
is still destroyed, as manifested in the non-zero width of the X-ray
scattering peaks associated with the smectic layering. 

How are these results, Eq.\ref{CTD}, for the equal time correlation
functions modified by the presence of the cosine (random field)
disorder term in Eqs.\ref{Hu},\ref{Hr}?  Aside from a
temperature-dependent modification of the prefactor
$\Delta_h\rightarrow\Delta_h+{\rm const.}(T_p-T)^2$, below $d=5$, they
are not modified at all.  We will now demonstrate this fact in $d=3$,
and defer to section \ref{random_field_3-5} the demonstration for
$3<d<5$.

For $\eta > 4$, the cosine clearly does not affect the long distance
behavior, since its coefficient $\Delta_V$ is irrelevant (i.e., flows to
zero under renormalization). Furthermore, we need not worry that it
might be {\em dangerously} irrelevant, since we have just calculated
correlation functions with the $\Delta_V$ set equal to zero, and
encountered no difficulties.

What happens with $\eta < 4$?  For $\eta$ near $4$ ($4-\eta <<1$), the
fixed point value $\tilde \Delta_V^* = 4-\eta$ of $\tilde\Delta_V$
is also $<<1$, and our perturbative (in $\tilde{\Delta}_V$)
renormalization group remains valid.  We can therefore use this RG to
calculate the equal time correlations for $\eta < 4$ and $4-\eta <<1$.
This calculation must itself be done perturbatively in
$\tilde{\Delta}_V^*$, since the full Hamiltonian with $\tilde{\Delta}_V\neq 0$
is not quadratic.  However, such a perturbation theory diverges for
small wavevectors when $\eta < 4$.  Indeed, the growth under the RG of
small $\tilde\Delta_V$ for $\eta < 4$ is a signal and a consequence of
this divergence.  Fortunately, we can use the renormalization group
transformation to relate the correlation functions at small wavevector
to those at large wavevector, where perturbation theory should be
reliable.  Indeed, simply repeating for correlation functions the
rescalings done earlier for the partition function, we can show that:
\begin{eqnarray}
C_{\alpha\beta}({\bf q};B(0),K (0),\Delta_h(0),\Delta_V(0))
&\equiv&\overline{\frac{\langle u_\alpha({\bf q})u_\beta({\bf q}')\rangle}
{\delta^3({\bf q}+{\bf q}')}}\;\nonumber\\
&&\hspace{-2in}= e^{(2+\omega)\ell}
C_{\alpha\beta}(e^\ell{\bf q}_\perp,e^{\omega\ell}q_z;
B(\ell),K (\ell),\Delta_h(\ell),\Delta_V(\ell))\;.\nonumber\\
\label{match1}
\end{eqnarray}
Now, in order to insure that the rescaled correlation function on the
right hand side of this expression can be safely evaluated using
perturbation theory in $\tilde\Delta_V(\ell)$, we will choose $\ell$
such that $e^{\ell}q_\perp=\Lambda$, the ultraviolet cutoff.
With this choice, we have
\begin{eqnarray}
C_{\alpha\beta}({\bf q};B(0), K(0),\Delta_h(0),\Delta_V(0))&=&\nonumber\\
&&\hspace{-6.0cm}
\left({\Lambda\over q_\perp}\right)^{2+\omega}
\hspace{-0.6cm}
C_{\alpha\beta}(\Lambda,(\Lambda/q_\perp)^\omega q_z;
B(\ell^*),K (\ell^*),\Delta_h(\ell^*),\Delta_V(\ell^*))\;,\nonumber\\
\label{match2}
\end{eqnarray}
where $\ell^*\equiv\ln(\Lambda/q_\perp)$, and $B\equiv B(0)$, $K\equiv
K(0)$, $\Delta_h\equiv\Delta_h(0)$, $\Delta_V\equiv\Delta_V(0)$.  Note
that $\ell^*\rightarrow\infty$ as $q_\perp \rightarrow 0$.

To calculate the right hand side of Eq.\ref{match2} in perturbation
theory in $\Delta_V(\ell^*)$, we expand the cosine in the full
Hamiltonian Eq.\ref{Hr} to quadratic order, obtaining, in Fourier
space:
\begin{eqnarray}
H[u_\alpha]&=&{1\over2}\int_{\bf q}\bigg[\left(B q_z^2+K  q_\perp^4\right)
\sum_{\alpha=1}^n|u_\alpha({\bf q})|^2\nonumber\\
&+&{1\over2}\left(\Delta_h q_\perp^2+2\Delta_V q_0^2\right)
\sum_{\alpha,\beta=1}^n
|u_\alpha({\bf q})-u_\beta({\bf q})|^2\bigg]\;,\nonumber\\
\label{Hr2}
\end{eqnarray}
which we immediately recognize as {\em identical} to the tilt-only
Hamiltonian Eq.\ref{Hr0}, {\em except} for the replacement $\Delta_h
\rightarrow \Delta_h + 2\Delta_V q_0^2/q_\perp^2$.  Thus, we can immediately
calculate the correlation function on the right hand side of
Eq.\ref{match2} by simply making this replacement in the tilt
only propagator Eq.\ref{Gab}, obtaining
\begin{eqnarray}
&&C_{\alpha\beta}(\Lambda,(\Lambda/q_\perp)^\omega q_z;
B(\ell^*),K (\ell^*),\Delta_h(\ell^*),\Delta_V(\ell^*))=\nonumber\\
&&\hspace{.5in}={T\, \delta_{\alpha\beta}\over K (\ell^*)\Lambda^4+
B(\ell^*)(\Lambda/q_\perp)^{2\omega}q_z^2}+\nonumber\\
&&\hspace{.6in}{\Delta_h(\ell^*)\Lambda^2+2\Delta_V(\ell^*)q_0^2
\over\left[K (\ell^*)\Lambda^4+
B(\ell^*)(\Lambda/q_\perp)^{2\omega}q_z^2\right]^2}\;.\nonumber\\
\label{match3}
\end{eqnarray}
To finish writing this expression entirely in terms of $\bf q$, we
need to calculate the Hamiltonian parameters $B(\ell)$, $K(\ell)$,
$\Delta_h(\ell)$ and $\Delta_V(\ell)$ from the recursion relations
Eqs.\ref{gamma_ell}-\ref{Deltah_ell} and evaluate them at
$\ell^*=\ln(\Lambda/q_\perp)$.  We find
\begin{mathletters}
\begin{eqnarray}
B(\ell^*)&=&\left({\Lambda\over q_\perp}\right)^{2-\omega}B\;,
\label{Bmatch}\\
K (\ell^*)&=&\left({\Lambda\over
q_\perp}\right)^{\omega-2}K\;,
\label{Kmatch}\\
\Delta_h(\ell^*)&=&{\tilde{\Delta}_h(\ell^*)\over A_1(\ell^*)}\;,
\label{Deltah_match}\\
&=&\left({\Lambda\over q_\perp}\right)^{\omega}
\Big[\Delta_h+{c\,q_0^2\over\Lambda^2 A_1}
\tilde{\Delta}^{*2}_V\Big]\;,\nonumber\\
\Delta_V(\ell^*)&=&{\tilde{\Delta}_V(\ell^*)\over A_1(\ell^*)}\;,
\label{DeltaV_match}\\
&=&\left({\Lambda\over q_\perp}\right)^{\omega-2}
{\tilde{\Delta}_V^*\over A_1}\;,\nonumber
\end{eqnarray}
\label{coupling_relation}
\end{mathletters}
\noindent
where $A_1\equiv A_1(\ell=0)$, and in this last equation we have
assumed that $q_\perp$ is sufficiently small that $\ell^*
=\ln(\Lambda/q_\perp)$ is sufficiently {\em large} that
$\tilde{\Delta}_V(\ell^*)$ will have flown to very near its
$\eta$-dependent fixed point value $\tilde\Delta_V^*$, as given by
Eq.\ref{g_fixed}.  Likewise, in the expression for
$\tilde{\Delta}_h(\ell)$, we have used the solution Eq.\ref{gsol} for
$\tilde{\Delta}_h(\ell)$, the effective tilt coupling constant, and
again assumed that $\ell^* =\ln(\Lambda/q_\perp)$ is very large, so
that the second ($\ell$-independent) term in Eq.\ref{gsol} is
negligible compared to the first, exponentially growing term.  Both of
these approximations become asymptotically {\em exact} as $q_\perp
\rightarrow 0$, and $\ell^*$, as a result, $\rightarrow \infty$.
Inserting these results for the renormalized elastic and coupling
constants into the expression Eq.\ref{match3} for the rescaled
correlation function, and using {\em that} result in the matching
formula Eq.\ref{match2} for the original correlation function at small
wavevector yields, after a bit of algebra
\begin{eqnarray}
&&C_{\alpha\beta}({\bf q};B,K ,\Delta_h,\Delta_V)=\nonumber\\
&&{T\, \delta_{\alpha\beta}\over K  q_\perp^4+B q_z^2}
+{\Big(\Delta_h + 
{c\,q_0^2\tilde{\Delta}^{*2}_V\over\Lambda^2 A_1}\Big)q_\perp^2
+ {2 q_0^2\tilde{\Delta}_V^*\over A_1}(q_\perp/\Lambda)^4
\over\left(K  q_\perp^4+B q_z^2\right)^2}\;,\nonumber\\
\label{match4}
\end{eqnarray}
where all of the parameters ($K,B,\Delta_h, A_1$) in this expression
are {\em bare} parameters (i.e., evaluated at $\ell=0$). This
expression is {\em identical} to the result in the $\Delta_V$
irrelevant phase $(\eta > 4)$ except for: (i) the
$\tilde{\Delta}_V^*(q_\perp/\Lambda)^4$ term, which is clearly
negligible, (as $q_\perp\rightarrow 0$), relative to the $(\Delta_h +
c\,q_0^2\tilde{\Delta}^{*2}_V/\Lambda^2 A_1)q_\perp^2$ term, and (ii)
the enhancement of the strength of the tilt disorder according to
$\Delta_h\rightarrow\Delta_h + c\,q_0^2\tilde{\Delta}^{*2}_V/\Lambda^2
A_1$. The heretofore neglected effects of the {\em anharmonic} terms
coming from the cosine on the {\em rescaled} correlation function (and
hence on the original one, since they are related by the matching
formula), should be even smaller, since $\tilde{\Delta}_V^*$ is $<<1$
for $4-\eta <<1$.  As discussed earlier, this argument {\em cannot} be
invalidated by possible infra-red divergences in the perturbation
theory, since we are calculating the {\em rescaled} correlation
function at {\em large} wavevector.  The effects of these divergences
on the short wavelength correlation functions are {\em implicitly}
included in the matching expression Eq.\ref{match2} through our use of
the renormalized parameters on the right hand side, since these
renormalizations include anharmonic effects.

Hence, at long wavelengths, aside from the modification of the
prefactor $\Delta_h\rightarrow\Delta_h+{\rm const.}(T_p-T)^2$ (for
$T<T_p$), we recover the same asymptotic form for the {\em equal-time}
correlation functions in the randomly pinned $(\eta < 4)$ and
non-pinned $(\eta > 4)$ phases.  Although, strictly speaking, we have
only derived this result near $\eta = 4$, where our perturbation
theory is valid, it must apply throughout the entire low temperature
phase, since {\em only} a phase transition can change the asymptotic
scaling of the correlation functions.  Barring the existence of such a
phase transition controlled by some strong coupling fixed point, which
we of course cannot rule out with the perturbative analysis performed
here, our results for the scaling behavior of the equal-time $u-u$
correlation functions should persist all the way down to $\eta =0$;
i.e., throughout the smectic phase.

In any case, in this elastically harmonic phonon (no-dislocations)
model, there is {\em guaranteed} to be a pinned phase below $\eta=4$
whose {\em static} correlation functions have $r$ ($q$) dependence
{\em identical}, at long wavelengths, with those in the non-pinned
phase above $\eta =4$.  The distinction between these phases lies in
the temperature dependence of the parameters in these static
correlation functions (as we discuss below) and in their dynamical
properties.

That is, it is important to note, that in spite of the above finding
that static correlation functions {\em scale} (with position or
wavevector) identically in both the $\eta>4$ and $\eta<4$ phases, the
existence of the transition at $T_p$ to the low temperature
translationally pinned ($\eta<4$) phase {\em can} still in principle
be detected in a static experiment. This can be seen by noting that in
the presence of such a transition, all $\Delta_h$-dependent physical
quantities, which would otherwise be smooth functions of temperature,
are {\em non-analytic} in $T$ at $T_p$.  That is, because of the
additional contribution to $\Delta_h^{\rm eff}$ in Eq.\ref{match4},
proportional to $(\tilde{\Delta}_V^*)^2\propto|T_p-T|^2$, that arises
below $T_p$, $\Delta_h^{\rm eff}(T)=\Delta_h+{\rm
const.}(T_p-T)^2$. As a result, other physical quantities derived from
$\Delta_h$ will have a discontinuous second derivative. Although this
is a quite subtle effect, it is an unambiguous static experimental
signature of the transition to the pinned smectic glass phase at
$T=T_p$. However, as briefly discussed just below Fig.6, we
demonstrate in Secs.\ref{anom_elasticity}-\ref{random_field_3-5} that
this transition, derived here within the harmonic elastic theory, will
be rounded once nonlinear elastic effects are taken into account. It
is, however possible, that rounded remnants of the nonanalyticities at
$T_p$, discussed above, will still be experimentally observable.

The static correlation function derived in Eq.\ref{match4} shows that
long-ranged smectic translational order is destroyed in $d=3$.  To
make quantitative comparisons with experiments, it is useful to
calculate translational correlation lengths $\xi_\perp^X$ and
$\xi_z^X$, whose inverses will give the width of the broadened X-ray
diffraction peaks.  These correlation lengths are the distances
$r_\perp$ and $z$ at which the mean squared relative displacement
correlation function $C({\bf r}_\perp,z)$ is of order $a^2$, where
$a=2\pi/q_0$ is a lattice constant.

When there is no disorder $(\Delta_h=0)$, the above criterion leads to
\begin{mathletters}
\begin{eqnarray}
\xi_\perp^T&=&a e^{\pi a^2\sqrt{K B}/T}\;,\label{xiTperp}\\
&=&\sqrt{\lambda\xi_z^T}\;.\label{xiTz}
\end{eqnarray}
\label{xiT}
\end{mathletters}

As discussed in the extensive literature on the subject,\cite{dGP}
however, the slow (logarithmic) divergence of the thermal part $C_T$
of $C$ implies that, for $\eta < 2$, the X-ray scattering peaks do
{\em not}, in fact, become broad for $|{\bf q}_\perp|<1/\xi^T$;
instead, they become power-law divergences, rather than the
Lorentzians one might otherwise expect.

The more strongly divergent part $C_\Delta$ of $C$ {\em does} lead to
genuine broadening, for $|{\bf q}|<1/\xi_{\perp,z}^{X}$, where
$\xi_{\perp,z}^{X}$ are defined as those length scales (i.e., values
of $r_\perp$ and $z$) at which $C_\Delta(r_{\perp},z)=a^2$. Having
computed $C_\Delta$ within the elastically harmonic theory, we thereby
find
\begin{mathletters}
\begin{eqnarray}
\xi_z^{X}&=&a^2\frac{8\pi K B}{\Delta_h}\;,\label{xiDeltaz}\\
\xi_\perp^{X}&=&4a\left({\pi B^2\lambda^3\over\Delta_h
\ln(2\sqrt{\lambda L_z}/\xi_\perp^X)}\right)^{1/2}\;.\label{xiDeltap}
\end{eqnarray}
\label{xiDelta}
\end{mathletters}

In fact, we powder average the X-ray scattering, since, as we will
show in a moment, the smectic in aerogel lacks long-ranged
orientational order, as well as translational order. The broad {\em
ring} of X-ray scattering that results from this averaging will have
its width determined entirely by $\xi_z^{X}$:
\begin{equation}
\kappa_{powder}\cong(\xi_z^{X})^{-1}=\frac{\Delta_h}{8\pi K B a^2}\;,
\label{Kpowder}
\end{equation}
a gratifyingly simple prediction.

Although it is more difficult to measure experimentally, it is
nonetheless of interest to calculate the harmonic orientational
correlation lengths $\xi_{\perp,z}^{O}$, which are defined as the
values of $L_{\perp,z}$ beyond which the mean squared orientational
fluctuations $\overline{\langle|\delta{\bf n}|^2\rangle}$ get to be of
order $1$.  If we ignore dislocations and anharmonic elasticity, we
can obtain these lengths by simply equating the ``pure phonon'' result
Eq.\ref{orientation} for $\overline{\langle|\delta{\bf n}|^2\rangle}$
to $1$. This gives:
\begin{mathletters}
\begin{eqnarray}
\xi^{O}_\perp&=&a e^{4\pi B^2\lambda^3/\Delta_h}=
a e^{\lambda\xi^{X}_z/2a^2}\;,\label{xiOp}\\
\xi^{O}_z&=&{a^2\over\lambda}e^{8\pi
B^2\lambda^3/\Delta_h}={a^2\over\lambda}
e^{\lambda\xi^{X}_z/a^2}\;,\label{xiOz}
\end{eqnarray}
\end{mathletters}

In the next section, we will show that, unsurprisingly, these lengths
give the distance beyond which dislocations unbind, invalidating the
purely elastic phonon theory studied in this section.

Thus, orientational order persists out to {\em much} larger distances
than translational order, in the limit of weak disorder where {\em
all} the correlation lengths get large.  Indeed, in this limit the
orientational correlation lengths grow exponentially with the
translational ones.  This qualitatively agrees with experimental
determinations of the orientational correlation length, as indirectly
inferred from specific heat data and related measurements.\cite{Clark}

In Sec.\ref{anom_elasticity}, we will show that anharmonic elastic
effects, which we have ignored up to now, change the relation between
the X-ray correlation length $\xi^{X}_z$ and the orientational
correlation lengths $\xi_{\perp,z}^{O}$ from exponential to power
law. However, the fact that $\xi_{\perp,z}^{O}$ both remain
$>>\xi^X_z$ continues to hold, validating our use of elastic theory
(which was predicated on the assumption that orientational
fluctuations are small), to calculate the X-ray correlation lengths.
In any case, it would be very interesting to measure the relation
between translational and orientational correlation lengths as a
function of temperature.

All of the above results apply subject to our two initial assumptions:
(1) that dislocations were {\em not} generated by the disorder, and
(2) that anharmonic terms in the {\em elastic} Hamiltonian could be
neglected.

In the next section, we will show that, if we continue to assume (2),
assumption (1) is wrong: in the {\em harmonic} elastic approximation,
in 3d dislocations {\em are} created even by arbitrarily weak disorder.

However, the effects of these dislocations turn out, in the weak
disorder limit, to be felt only on length scales longer than the
smaller of the orientational length $\xi^O_{\perp,z}$ and dislocation
unbinding length $\xi_{\perp,z}^D$, both of which are {\em much}
longer than the translational correlation lengths.  Thus, our above
calculations of the correlation lengths remain valid.  However, in the
presence of these unbound dislocation loops, the smectic glass phase
is destroyed.  It may be replaced by a nematic glass phase,
however, as discussed in the Introduction. The static properties of
our system in the presence of dislocations are analytically
accessible, and we analyze them in the next section.

Furthermore, the behavior of the correlation functions found above is
modified by elastic nonlinearities (which lead to anomalous
elasticity), which also have a nontrivial effect on dislocations, as
we demonstrate in Sec.\ref{anom_elasticity_dislocations_correlations}.

\section{Dislocations in the Random Tilt-only Harmonic Model}
\label{dislocations}
In the previous sections, we have shown that there is no long-ranged
smectic translational order in the presence of the random pinning and
tilting fields. However, this by itself is {\it not} sufficient to
prove that there is no phase transition in this model. Indeed, we know
of many examples of transitions between two phases which {\it both}
lack long-ranged order. The Kosterlitz-Thouless transition in the
$d=2$ XY model\cite{KT} is perhaps the most famous example. In that
problem, the transition is associated not with the disappearance of
long-ranged order, but rather, with the unbinding of neutral pairs of
topological defects (vortices) with increasing temperature.

It is reasonable, therefore, to ask whether the same thing can happen
in our model: are topological defects (i.e., smectic dislocation
loops) still bound even in the translationally disordered phase we
have discussed? If they were, then an equilibrium phase transition
would be required to produce the expected high temperature phase, in
which dislocations are unbound.

Of course, the divergences that destroy long-ranged order are much
stronger (power-law) in our model for $d<5$ than in the 2d XY model,
where they are logarithmic. However, there are examples of phases much
more strongly disordered than the Kosterlitz-Thouless phase in which
topological defects nonetheless remain
bound.\cite{Rokhsar}. Furthermore, there has been considerable
speculation recently\cite{GL,GH,DSFisher} that a ``Bragg glass'' phase
might exist in pinned superconducting flux-line lattices. This ``Bragg
glass'' would be a phase in which the random pinning destroyed the
translational order of the flux lattice, but did {\it not} induce
dislocations in the lattice. It seems quite reasonable, therefore, to
ask whether an analogous phase occurs in smectics. This section
addresses this question, and shows analytically that smectic Bragg
glass does not occur in $d=3$ (within a model with {\em harmonic}
elasticity).  We will show later that this result maybe invalidated by
elastic anharmonic effects.

The starting point of our analytic theory is the ``tilt only'' model,
by which we mean Eq.\ref{Huii} with the random potential $V({\bf r})$
set to zero.  As discussed earlier, this theory correctly reproduces
all of the static correlation functions in {\em both} the {\em pinned}
($T<T_p$) and the non-pinned ($T>T_p$) regimes.  However, in view of
the very strong irrelevance of $\Delta_V$ to the static correlation
function in {\em both} phases in the ``phonon only'' approximation, it
seems quite plausible that it is also irrelevant when dislocations are
included. Indeed we will prove this in Sec.\ref{failure}.

Setting $V({\bf r})=0$ in the Hamiltonian Eq.\ref{Huii} reduces it to
a quadratic theory
\begin{equation}
H[{u}]=\int \! d^d{\bf r}
\bigg[  {B\over 2}(\partial_z
{u})^2 + {K\over 2}(\nabla^2_\perp {u})^2
+ {\bf h}({\bf r})\cdot\bbox{\nabla_\perp} u\bigg]\;.
\label{Hu_V=0}
\end{equation}
The effect of dislocations on such a theory can then be treated using
the techniques that many authors\cite{KT,Peskin,Pershan,NelsonToner} have
applied to a variety of disorder free systems, including smectics.

Most aspects of this procedure are already described quite well in the
literature; therefore, our explanation will be brief until we come to
those points that are affected by the presence of the disorder.

We begin by recalling the definition of dislocations in
smectics.\cite{Pershan} The simplest type is an edge dislocation,
illustrated in Fig.7.
\begin{figure}[bth]
{\centering
\setlength{\unitlength}{1mm}
\begin{picture}(150,60)(0,0)
\put(-20,-68){\begin{picture}(150,0)(0,0)
\includegraphics{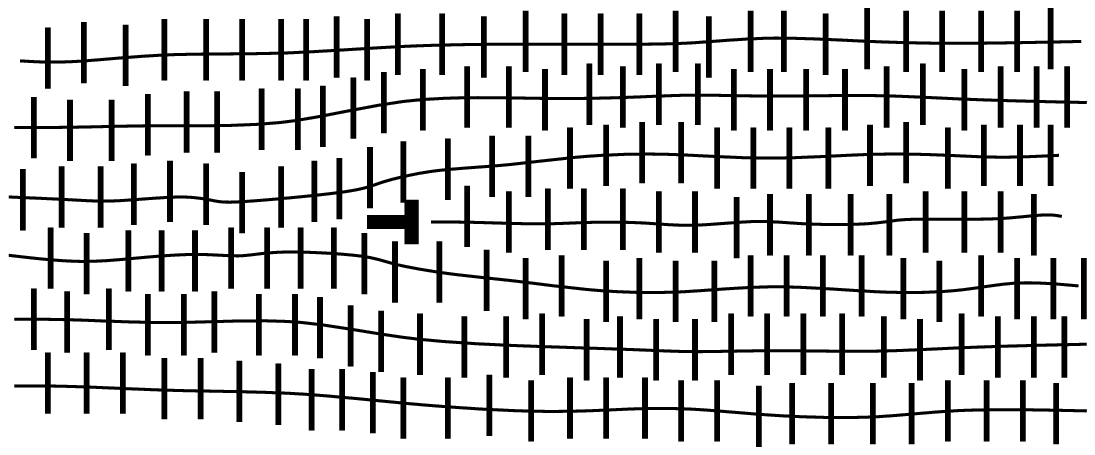}
\end{picture}}
\end{picture}}
Fig.7. {An {\it edge} dislocation in the smectic liquid crystal, with
the core coming out of the page at the position indicated by the lazy
``{\bf T}'' symbol.}
\label{edge_dislocation}
\end{figure}
This is the edge of an extra smectic layer inserted into the
smectic. Clearly, a pure edge dislocation must form a closed loop
lying in the smectic plane, as illustrated in Fig.8.  More generally,
dislocations can tip out of the plane of the smectic layers. An
extreme case is a screw dislocation, which runs {\it perpendicular} to
the layers.
\begin{figure}[bth]
{\centering
\setlength{\unitlength}{1mm}
\begin{picture}(150,80)(0,0)
\put(-28,-50){\begin{picture}(150,0)(0,0)
\includegraphics{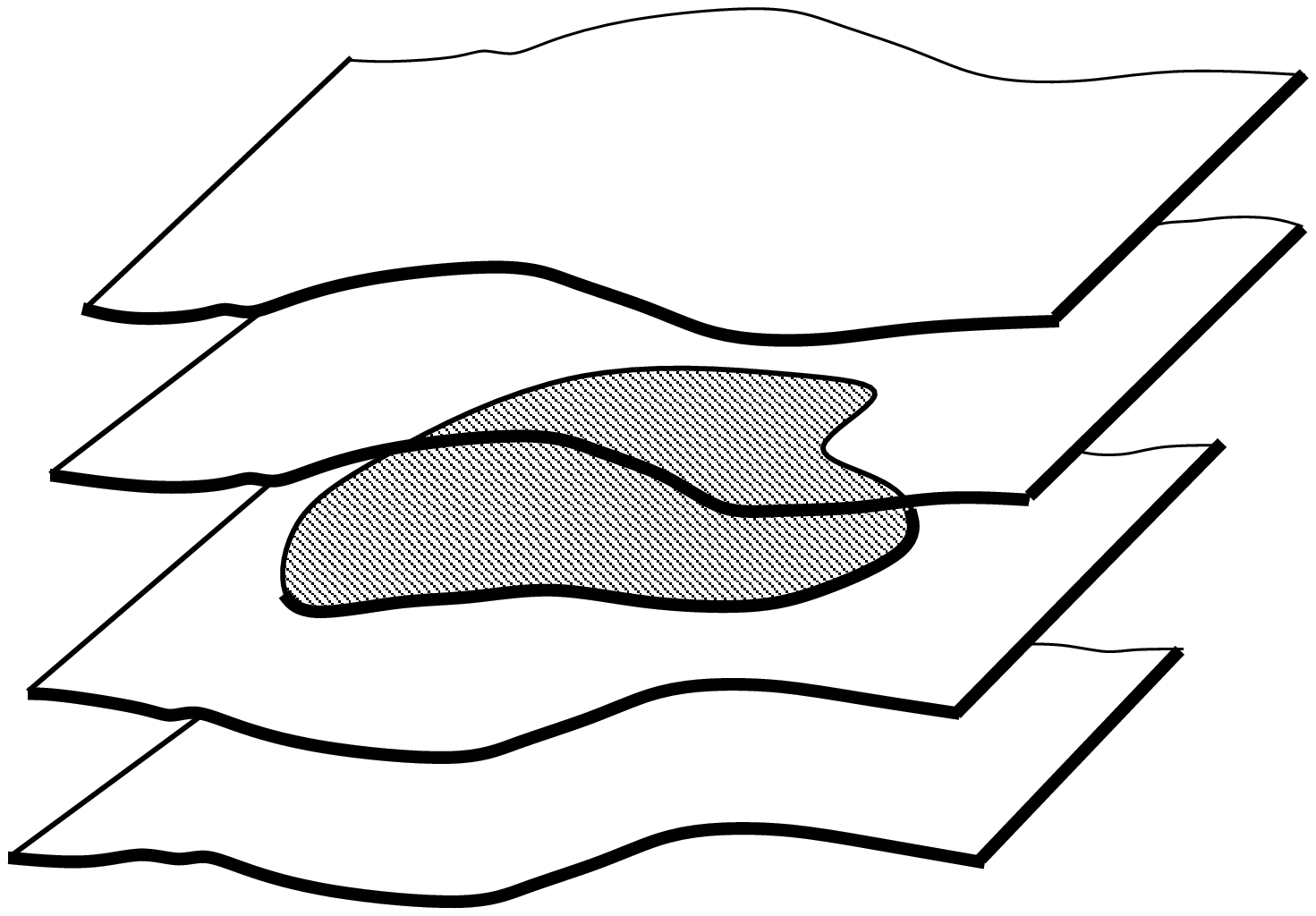}
\end{picture}}
\end{picture}}
Fig.8. {A three dimensional illustration of dislocation loop in the
smectic liquid crystal.}
\label{dislocation_loop}
\end{figure}

Mathematically a dislocation is a line, or, more generally, a curve,
with the property that when the gradient of the displacement $u$ is
integrated around any curve that encloses the dislocation, the result
is not zero, as it would be in a dislocation free system, but rather
an integral multiple $N$ of the layer spacing $a$. Mathematically,
this means
\begin{equation}
\oint\bbox{\nabla}u\cdot d \bbox{\ell}=a N\;,\label{dislocationEq}
\end{equation}
or, in differential form:
\begin{equation}
\bbox{\nabla}\times{\bf v}={\bf m}\;,\label{dislocationEqi}
\end{equation}
where we have defined
\begin{equation}
{\bf v}=\bbox{\nabla}u\;,\label{dislocationEqii}
\end{equation}
and
\begin{equation}
{\bf m}({\bf r})=\sum_i\int a N_i {\bf t}_i(s_i)
\delta^3({\bf r}-{\bf r}_i(s_i)) d s_i\;,
\label{m}
\end{equation}
where $s_i$ parameterizes the $i$'th dislocation loop, ${\bf
r}_i(s_i)$ is the position of that loop, ${\bf t}_i(s_i)$ is its local
unit tangent, and $N_i$ the ``charge'' or number of excess layers
associated with the dislocation. Note that $N_i$ is independent of
$s_i$, since the charge of a given line is constant along the line
defect. Furthermore, Eq.\ref{dislocationEqi} implies that
\begin{equation}
\bbox{\nabla}\cdot{\bf m}({\bf r})=0\;,
\label{continuity}
\end{equation}
which simply means that dislocation lines cannot end in the bulk of
the sample; they must either form closed loops or extend entirely
through the system.

Now, our procedure for adding dislocation lines to our previous pure
phonon model Eq.\ref{Hu_V=0} is the following standard one. We separate
the field ${\bf v}={\bbox\nabla} u$ into phonon (single-valued) and
dislocation (singular) parts
\begin{equation}
{\bf v}={\bf v}_p+{\bf v}_d\;,
\label{v}
\end{equation}
where the dislocation part ${\bf v}_d$ minimizes\cite{comment_vd} the
elastic Hamiltonian Eq.\ref{Hu_V=0}, subject to the constraint of
Eq.\ref{dislocationEqi} ($\bbox{\nabla}\times{\bf v}={\bf m}$). (For
an alternative derivation see Appendix C.)  This uniquely determines
${\bf v}_d({\bf r})$ given the dislocation configuration ${\bf m}({\bf
r})$. We then insert the decomposition Eq.\ref{v} back into the
elastic Hamiltonian Eq.\ref{Hu_V=0}. As a result of the construction
that ${\bf v}_d({\bf r})$ minimizes Eq.\ref{Hu_V=0}, all the cross
couplings between ${\bf v}_d({\bf r})$ and ${\bf v}_p({\bf r})$
vanish. We are thus left with a decoupled elastic Hamiltonian for
${\bf m}({\bf r})$, which we can use as a basis for studying the
statistical mechanics of dislocations.

Let us now implement this procedure. The Euler-Lagrange equation,
obtained by minimizing Eq.\ref{Hu_V=0}, is
\begin{equation}
(B\partial_z^2 - K\nabla_\perp^4)u_d({\bf r}) +
\bbox{\nabla}_\perp\cdot {\bf h}({\bf r})=0\;.
\label{EL}
\end{equation}
Rewriting this in terms of ${\bf v}_d({\bf
r})\equiv\bbox{\nabla}u_d({\bf r})$ gives
\begin{equation}
\partial_z v^z_{d} - \lambda^2\nabla^2_\perp\bbox{\nabla}_\perp\cdot
{\bf v}^\perp_{d}+{1\over B}\bbox{\nabla}_\perp\cdot{\bf h}({\bf r})=0\;,
\label{ELi}
\end{equation}
where $\lambda^2\equiv K/B$.

In Fourier space, this becomes
\begin{equation}
q_z v^z_{d}+ \lambda^2 q_\perp^2{\bf q}_\perp\cdot {\bf
v}^\perp_{d}+{1\over B}{\bf q}_\perp\cdot{\bf h}({\bf q})=0\;.
\label{ELii}
\end{equation}
The constraint Eq.\ref{dislocationEqi} becomes, in Fourier space,
\begin{equation}
i{\bf q}\times{\bf v}_d={\bf m}\;,\label{dislocationiF}
\end{equation}
which has the general solution
\begin{equation}
{\bf v}_d={i{\bf q}\times{\bf m}\over q^2}+{\bf q}\phi\;,
\label{phi}
\end{equation}
where $\phi$ is the smooth elastic distortion around the dislocation
line, to be determined by the Euler-Lagrange equation
Eq.\ref{EL}. Inserting the above expression for ${\bf v}_d$ into
Eq.\ref{ELii} and solving for $\phi$ gives
\begin{equation}
\phi=-{i q_z(1-\lambda^2 q_\perp^2)\epsilon_{z i j} q_i m_j\over \Gamma_q
q^2}-{{\bf q}_\perp\cdot{\bf h}\over B\Gamma_q}\;,
\label{phii}
\end{equation}
where we have defined the inverse of the smectic propagator
\begin{equation}
\Gamma_q\equiv q_z^2 + \lambda^2 q_\perp^4\;.
\label{Gamma}
\end{equation}

Inserting this into Eq.\ref{phi} and then substituting this final
expression for ${\bf v}_d$ into the original elastic Hamiltonian
Eq.\ref{Hu_V=0} gives the defect interaction Hamiltonian
\begin{equation}
H_d=\int_{\bf q}\left[{K q_\perp^2\over2\Gamma_q} P_{i j}^\perp
m_i({\bf q}) m_j(-{\bf q}) + {\bf m}({\bf q})\cdot{\bf a}(-{\bf q})\right]\;,
\label{Hd}
\end{equation}
where $P_{i j}^\perp({\bf q})=\delta_{i j}^\perp - q_i^\perp
q_j^\perp/q_\perp^2$, ${\bf a}({\bf q})$ (not to be confused with the
lattice spacing $a$) is a Fourier transform of the quenched field
related to the original random tilt field ${\bf h}({\bf q})$ via
\begin{equation}
{\bf a}({\bf q})=i \left [ \frac{{\bf q}\times {\bf h}}{q^2} -
\frac {\left (\hat{\bf z} \times {\bf q} \right ) {\bf
q} \cdot {\bf h}}{\Gamma_q q^2} q_z
\left (1-\lambda^2q^2_{\perp} \right ) \right ] \;,
\label{afield}
\end{equation}
and we have dropped unimportant terms that depend only on the quenched
random variables (and not on $\bf m$).  The first (${\bf
h}$-independent) term in this expression Eq.\ref{Hd} is just the usual
smectic dislocation energy for a pure system.\cite{NelsonToner}

To treat this model, we perform a duality transformation. We begin by
putting the model on a simple cubic lattice (to make the model well
defined at short distances); now, ${\bf m}({\bf r})$ is defined on the
sites ${\bf r}$ of the lattice, and takes on values
\begin{equation}
{\bf m}({\bf r})={a\over d^2}\big(n_x({\bf r}),n_y({\bf r}),n_z({\bf
r})\big)\;,
\label{mdescrete}
\end{equation}
where the $n_i$'s are integers, and $d$ is the cubic lattice constant
used in the discretization. The partition function for this model is
then
\begin{equation}
Z[{\bf h}({\bf q})]=\sum_{\{{\bf m}({\bf r})\}}^\prime e^{-S[{\bf m}]}\;,
\label{Z}
\end{equation}
where
\begin{equation}
S[{\bf m}]\equiv{1\over T}\left[H_d[{\bf m}] +
E_c {d^4\over a^2}\sum_{\bf r}|{\bf m}({\bf r})|^2\right]\;,
\label{S}
\end{equation}
and the sum is over all discrete configurations of ${\bf m}$'s given
by Eq.\ref{mdescrete}, satisfying the dislocation line continuity
constraint
\begin{equation}
\bbox{\nabla}\cdot{\bf m}=0\;,
\label{continuity2}
\end{equation}
where the divergence now represents a {\it lattice} divergence, $H_d$
is given by Eq.\ref{Hd}, and we have added a core energy term
$E_c\sum_{\bf r}|{\bf m}({\bf r})|^2$, to account for energies near
the core of the defect line that are not accurately treated by our
continuum elastic theory. We call the reader's attention to the fact
that the partition function still depends implicitly on the
configuration of the random tilt-disorder fields $\{\bf h\}$ through
$\bf a$ in Eq.\ref{Hd}.

To proceed, we enforce the constraint $\bbox{\nabla}\cdot{\bf m}=0$ by
introducing a new auxiliary field $\theta({\bf r})$, rewriting the
partition function Eq.\ref{Z} as
\begin{equation}
Z=\prod_{\bf r}\int d\theta({\bf r})
\sum_{\{{\bf m}({\bf r})\}} e^{-S[{\bf m}] + i\sum_{\bf
r}\theta({\bf r})\bbox{\nabla}\cdot{\bf m}({\bf r})d^2/a}\;,
\label{Z2}
\end{equation}
where the sum over $\{\bf m\}$ is now unconstrained. The constraint
$\bbox{\nabla}\cdot{\bf m}({\bf r})=0$ is enforced by
integration over $\theta({\bf r})$,\top{-2.7cm} 

\noindent
since
\begin{equation}
\delta(\bbox{\nabla}\cdot{\bf m}({\bf r}))
=\int_{0}^{2\pi} {d\theta({\bf r})\over 2\pi}
e^{i\theta({\bf r})\bbox{\nabla}\cdot{\bf m}({\bf r})d^2/a}\;,
\end{equation}
where the $\delta$ is a Kronecker delta, since ${\bf m}\, d^2/a$, and, hence,
$\bbox{\nabla}\cdot{\bf m}\,d^2/a$, are integer-valued.

Now we can ``integrate'' (actually sum) by parts, and rewrite
\begin{equation}
\sum_{\bf r}\theta({\bf r})\bbox{\nabla}\cdot{\bf m}({\bf r})=
-\sum_{\bf r}{\bf m}({\bf r})\cdot\bbox{\nabla}\theta({\bf
r})+\mbox{surface terms}\;.
\end{equation}
Our next step is to introduce a dummy gauge field $\bf A$ to mediate
the long-ranged interaction between defect loops in the Hamiltonian
Eq.\ref{Hd}. This is accomplished by rewriting the partition function
as
\begin{equation}
Z=\prod_{\bf r}\int d\theta({\bf r})d{\bf A}({\bf r})
\sum_{\{{\bf m}({\bf r})\}} e^{-S[{\bf m},\theta,{\bf A}]}
\delta(\bbox{\nabla}\cdot{\bf A})\delta(A_z)\;,
\label{Z3}
\end{equation}
with
\end{multicols}
\begin{equation}
S={1\over T}\sum_{\bf r}\bigg[{\bf m}
({\bf r})\cdot\big[-i{T d^2\over a}\bbox{\nabla}\theta({\bf r})
+d^3\big(i{\bf A}({\bf r})+{\bf a}({\bf r})\big)\big] 
+ E_c {d^4\over a^2}|{\bf m}|^2\bigg]
+{1\over2 T}\sum_{\bf q}{\Gamma_q\over K q_\perp^2}|{\bf A}|^2\;,
\label{Sq}
\end{equation}
\bottom{-2.7cm}
\begin{multicols}{2}
\noindent
where $\bf a$ is the quenched gauge field defined in Eq.\ref{afield}.

It is straightforward to check that, upon performing the Gaussian
integral over $\bf A$, subject to the indicated constraints
$\bbox{\nabla}\cdot{\bf A}=A_z=0$, we recover the original long-ranged
interaction between dislocation lines in Eq.\ref{Hd}.

The two goals of all of these manipulations have now\top{-2.7cm} been
achieved: the sum on $\{{\bf m}({\bf r})\}$ is now unconstrained, and
the sum on each site over ${\bf m}({\bf r})$ is now decoupled from
that on every other site. Furthermore, this sum is readily recognized
to be nothing more than the ``periodic Gaussian'' made famous by
Villain\cite{Villain}. The partition function Eq.\ref{Z3} can thus be
rewritten
\end{multicols}
\begin{equation}
Z=\prod_{\bf r}\int d\theta({\bf r})d{\bf A}({\bf r})
\delta(\bbox{\nabla}\cdot{\bf A}({\bf r}))\delta(A_z)
\exp\bigg[-\sum_{{\bf r},i} V_p[\theta({\bf r + \hat{x}_i})-
\theta({\bf r})-{a d\over T}\big(A_i({\bf r})-i a_i({\bf r})\big)]
-{1\over2T}\sum_{\bf q}{\Gamma_q\over K q_\perp^2}|{\bf A}|^2\bigg]\;,
\label{Z4}
\end{equation}
\bottom{-2.7cm}
\begin{multicols}{2}
\noindent
where the well-known $2\pi$-periodic Villain potential $V_p(x)$,
defined by
\begin{equation}
e^{-V_p(x)}\equiv\sum_{n=-\infty}^\infty e^{-n^2 E_c/T + i x n}\;
\label{Vp}
\end{equation}
has the usual property that the {\it smaller} $E_c/T$ is (i.e., the
{\it higher} the temperature in the original random-tilt smectic
model), the sharper the potential minima. Thus {\it raising} the
temperature in the original model is like {\it lowering} the
temperature in the dual model Eq.\ref{Z4}. It is precisely this
familiar temperature inversion associated with duality that leads to
an {\it inverted} XY transition for three-dimensional disorder-free
superconductors\cite{HLM} and bulk smectics\cite{Toner}. It also plays
an important role here, as we shall see in a moment.

Standard universality arguments imply that replacing the periodic
potential $V_p(x)$ in Eq.\ref{Z4} by {\it any} other non-singular
periodic function should not change the universality class of the
transition. In particular, we could replace $V_p(x)$ by $\cos(x)$. The
resultant model would be precisely the ``fixed length'' version of the
``soft spin'', or Landau-Ginsburg-Wilson model, with the {\em
complex}\top{-2.7cm} 

\noindent
``action''
\end{multicols}
\begin{equation}
S=\sum_{\bf r}\bigg[{c\over2}\big[\bbox{\nabla}+
{a d\over T}(i{\bf A}+{\bf a})\big]\psi^*\cdot
\big[\bbox{\nabla}-{a d\over T}(i{\bf A}+{\bf a})\big]\psi
+t|\psi|^2+u |\psi|^4\bigg]
+\sum_{\bf q}{\Gamma_q\over 2T K q_\perp^2} |{\bf A}({\bf q})|^2\;,
\label{S3}
\end{equation}
\bottom{-2.7cm}
\begin{multicols}{2}
\noindent
where $\psi({\bf r})$ is a complex ``disorder'' parameter field whose
phase is $\theta({\bf r})$; the reduced temperature $t$, quartic
coupling $u$, and $c(T)$ are parameters of the model with
\begin{mathletters}
\begin{eqnarray}
c(T)&=&{d^2V_p(x)\over d^2x}\Big|_{x=0}\times O(1)\label{c(T)a}\;,\\
&\approx&\left\{\begin{array}{lr}
2e^{-E_c/T}, & T<<E_c\\
{T/2E_c}, & T>>E_c\\
\end{array} \right.
\label{c(T)b}
\end{eqnarray}
\label{c(T)}
\end{mathletters}

Because of the duality transformation's inversion of the temperature
axis, the reduced temperature $t$ is a monotonically {\it decreasing}
function of the temperature $T$ (of the original dislocation loop
model), which vanishes at the mean-field transition temperature
$T_{MF}$ of the fixed length model Eq.\ref{Z4}.

Universality also implies that this ``soft-spin'' model should be in
the same universality class as the fixed length model Eq.\ref{Z4}. We
shall, therefore, henceforth work with model Eq.\ref{S3}, because it
is more straightforward to analyze perturbatively.

As we undertake that analysis, it is important to keep in mind that,
as a consequence of the duality inversion of the temperature axis, the
{\it ordered} phase of the dual model Eq.\ref{S3} corresponds to the
{\it disordered} (i.e., dislocation loops unbound) phase of the
original dislocation loop gas model. That is, the low dual-temperature
phase described by
\top{-2.7cm}
\begin{equation}
\langle\psi({\bf r})\rangle\neq 0\;,
\end{equation}
corresponds to the {\em disordered} dislocation-unbound phase of the
smectic liquid crystal.

In the absence of disorder (${\bf a}=0$) this model is exactly the
dual version of the NA dislocation loop model derived and studied by
one of us.\cite{Toner}

Disorder is included in Eq.\ref{S3} through the quenched gauge-field
${\bf a}({\bf r})$, which is related to the random tilt field ${\bf
h}({\bf r})$ by Eq.\ref{afield}. The partition function
\begin{equation}
Z[{\bf h}]=\int [d\psi][d{\bf A}]
e^{-S[\psi,{\bf A},{\bf h}]}
\delta(\bbox{\nabla}\cdot{\bf A})\delta(A_z)\;,
\label{Z5}
\end{equation}
with $S$ given by Eq.\ref{S3}, is thus an implicit function of the
random tilt field configuration ${\bf h}({\bf r})$.

As in the previous section, we will cope with this dependence of $Z$
on the quenched field ${\bf h}({\bf r})$ using the replica trick.
Doing so leads us to calculate
\begin{eqnarray}
\overline{Z^n} = &&\int [d {\bf a}] \prod_{\alpha=1}^n
\left[d\psi_\alpha\right]\left[d{\bf A}_\alpha\right] 
e^{-S_r\left[\psi_\alpha,{\bf A}_\alpha,{\bf a}\right]} \nonumber \\
&&\times P[{\bf a}] \delta(\bbox{\nabla}\cdot{\bf
A}_\alpha)\delta(A_z^\alpha) \; ,
\label{Zr}
\end{eqnarray}
with
\end{multicols}
\begin{equation}
S_r=\sum_{{\bf r},\alpha}\bigg[{c\over2}\big[\bbox{\nabla}
+{a d\over T}(i{\bf A}_\alpha+{\bf a})\big]\psi^*_\alpha\cdot
\big[\bbox{\nabla}-
{a d\over T}(i{\bf A}_\alpha+{\bf a})\big]\psi_\alpha
+t|\psi_\alpha|^2+u |\psi_\alpha|^4\bigg] 
+\sum_{{\bf q},\alpha}{\Gamma_q\over 2T K q_\perp^2} 
|{\bf A}_\alpha({\bf q})|^2\;,
\label{Sr}
\end{equation}
\bottom{-2.7cm}
\begin{multicols}{2}
\noindent

The probability distribution $P[{\bf a}]$ of the field $\bf a$ in
Eq.\ref{Zr} is Gaussian, since $\bf a$ is linear in $\bf h$ and the
distribution of $\bf h$ is Gaussian, defined by Eq.\ref{h_corr}.
Thus, the distribution $P[{\bf a}]$ is completely specified by the
average $\overline{a_i({\bf q})a_j(-{\bf q})}$.  This is easily
evaluated, using the relation Eq.\ref{afield} between ${\bf a}({\bf
q})$ and ${\bf h}({\bf q})$. We find
\begin{mathletters}
\begin{eqnarray}
\overline{a_i({\bf q})a_j(-{\bf q})} 
&=&\overline{h_l({\bf q})h_n(-{\bf q})}\label{aa1}\\
&\times&\left(\frac{\epsilon_{ikl}q_k}{q^2}-\frac{\epsilon_{zki}q_k}
{\Gamma_q q^2}q_l q_z(1-\lambda^2q_\perp^2)\right)\nonumber\\
&\times&\left(\frac{\epsilon_{jmn} q_m}{q^2}-
\frac{\epsilon_{zmj}q_m}{\Gamma_qq^2}q_n q_z(1-\lambda^2q_\perp^2)\right)
\;,\nonumber\\
&=&\Delta_h\left(\frac{P_{ij}}{q^2}+
\frac{q_z^2q^2_{\perp}(1-\lambda^2q_\perp^2)^2P^{\perp}_{ij}}{\Gamma^2_q q^2}
\right)\;,
\label{aa2}
\end{eqnarray}
\end{mathletters}
where $\epsilon_{i j k}$ is the usual fully antisymmetric third rank
unit tensor. To obtain the second equality we have used
$\overline{h_l({\bf q})h_n({\bf -q})} = \Delta_h \delta_{l n}$, and
defined the standard transverse projection operators
\begin{mathletters}
\begin{eqnarray}
P_{ij}&=&\delta_{ij}-q_i q_j/q^2\;,\label{Pij}\\
P^\perp_{ij}&=&(1-\delta_{iz})(1-\delta_{jz})
(\delta_{ij}-q^\perp_i q^\perp_j/q_\perp^2)\;.\label{Pijperp}
\end{eqnarray}
\end{mathletters}

We now need merely to consider the statistical mechanics of the model
defined by Eqs.\ref{Zr},\ref{Sr}, in the limit $n\rightarrow 0$. 

From previous classic works\cite{HLM,BC,Toner} we have considerable
experience analyzing the critical properties of gauge theories of the
type defined by the action in Eq.\ref{Sr}. As we know from work on the
zero field Normal-to-Superconductor transition\cite{HLM}, and the
analogous analysis of the bulk, clean Nematic-to-Smectic-A
transitions\cite{Toner}, there are two possible regimes: (i) the
extreme type I regime, in which $u<<1$, and, as a result, the $\psi$
fluctuations are subdominant to those of the gauge field $\bf A$,
allowing a mean-field treatment of $\psi$. In this case the remaining
gauge field fluctuations can be treated {\em exactly}. (ii) the type
II regime, in which $u>1$ and {\em both} $\psi$ and $\bf A$
fluctuations must be treated within a considerably more involved RG
analysis.

A complete analysis of the critical properties of this smectic
dislocation loop unbinding transition, described by the action in
Eq.\ref{Sr}, is beyond the scope of the present
paper.\cite{RTunpublished} Here our goal is to determine the stability
of the topologically ordered smectic Bragg glass phase to
disorder-induced dislocation loop unbinding. To answer this question,
it is sufficient to study the effect of diagrammatic corrections on
the reduced dual temperature $t$, as we demonstrate in detail in
Appendix D for the type I limit, delaying the analysis of type II
limit to a future publication.\cite{RTunpublished} By computing the
disorder-averaged free energy, we find that in {\em either} regime the
lowest order contribution to the renormalized {\em dual} temperature
$t_R$ comes from the average of the ``diamagnetic'' terms
\begin{equation}
\delta S={c a^2 d^2\over2 T^2}\sum_{\bf r}\big(\langle|{\bf A}|^2\rangle 
- \overline{|{\bf a}|^2}\big)|\psi|^2\;,\label{deltaS}
\end{equation}
graphically illustrated in Fig.9.
\begin{figure}[bth]
{\centering
\setlength{\unitlength}{1mm}
\begin{picture}(150,30)(0,0)
\put(-20,-68){\begin{picture}(150,0)(0,0)
\includegraphics{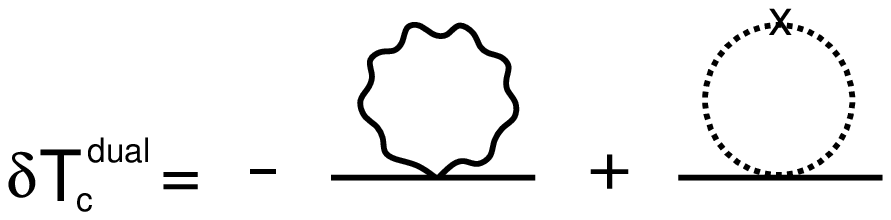}
\end{picture}}
\end{picture}}
Fig.9. {Feynman diagrams that dominate the renormalization of the
reduced {\em dual} temperature $t$, leading to an {\it upward} shift
in the {\it dual} $T_c$. In $d\leq3$ these drive the transition
temperature in the original smectic model to $0$.}
\label{dualTc}
\end{figure}
Generalizing to $d$ dimensions these give
\begin{equation}
t_R=t_0 + {(d-2)c a^2 d^2\over T^2}\int{d^d q\over(2\pi)^d}\left[{K T
q_\perp^2\over\Gamma_q} - {\Delta_h q_z^2 q_\perp^2\over q^2
\Gamma_q^2}\right],
\label{tR}
\end{equation}
where the first term in the square brackets comes from the first graph
in Fig.9, with the internal wiggly line representing ${\bf A}_\alpha$
fluctuations, while the second comes from the second graph with the
internal dotted line representing the quenched gauge field $\bf
a$. The negative sign of the disorder contribution leads to an {\em
increase} in the {\em dual} $T_c$, and can be traced back to the fact
that the action $S_r$ Eq.\ref{Sr}, is complex.

The second, disorder, term in this integral dominates the first as
${\bf q}\rightarrow 0$. Indeed, this integral diverges in the infra-red for
$d\leq 3$
\begin{eqnarray}
\int d^d q {q_z^2 q_\perp^2\over q^2(q_z^2 + \lambda^2
q_\perp^4)^2}&\propto&\int d^{d-1} q_\perp {q_\perp^8\over
q_\perp^{10}}\nonumber\\
&\propto&\int {d^{d-1}\over q_\perp^2}\;,
\label{diverge}
\end{eqnarray}
where the first proportionality follows from the fact that the
dominant regime of the integral over $q_z$ is $q_z\sim\lambda
q_\perp^2$, as can be easily seen by contour integration. This
divergence implies that the renormalized {\em dual} temperature $t_R$
is driven to $-\infty$ (note the sign) by the disorder in
$d\leq 3$. Indeed, we find in $d=3$
\begin{equation}
t_R=t_0 - {c a^2 d^2\Delta_h\over \lambda T^2}\ln(L/a)\times O(1)\;,
\label{tRi}
\end{equation}
where $L$ is an infra-red cutoff (e.g., the lateral extent of the
smectic layers) and $a$ is the ultra-violet cutoff (e.g., the size of
the liquid crystal molecules $\approx 10$\AA).

The unusual {\em negative} ($d=3$) divergent correction to $t$ in
Eq.\ref{tRi} implies that the {\it dual} order parameter is always in
its {\em ordered} phase, which, in turn, implies that the dislocation
loops of the original smectic model are always {\it unbound}, thereby
destroying the topological order of the smectic Bragg glass, even at
$T=0$, for {\em any} amount of disorder, no matter how small. This
conclusion, of course, only holds within the {\em harmonic} elastic
approximation made in this section, with the treatment that
approximately includes these nonlinear elastic effects given in
Sec.\ref{anom_elasticity_dislocations_correlations}.

It is important to note that although the dual disordered gauge model,
Eq.\ref{S3}, can be extended to any dimension, it rigorously only
describes the original disordered smectic in $d=3$, the dimension in
which the duality correspondence was established between the two
models.  However, if one can entertain the idea that the relation
between the two models extends to $d>3$, then we have found a quite
nontrivial result for this higher dimensional disordered smectic. For
$3<d\leq5$ the disordered smectic will then have strongly divergent
displacement fluctuations Eq.\ref{CDelta}, but will maintain
topological order, i.e. no condensation of dislocation loops takes
place in this range of dimension (for $d>3$). From this point of view
this higher dimensional disordered smectic is in fact a true ``Bragg''
glass.\cite{GL,GH,DSFisher}

\section{Anomalous Elasticity in the Tilt-only Model of Disordered Smectics}
\label{anom_elasticity}

Up to now we have analyzed disordered smectics ignoring {\em elastic
anharmonic} effects. These effects are important even in pure
smectics, as was discovered some time ago by Grinstein and Pelcovits
(GP)\cite{GP}. As reported in a recent Letter\cite{anomPRL}, we have
discovered that in {\em disordered} smectics these elastic
anharmonicities play an even more important role; specifically, we
find disorder-driven {\em power law} divergences, in contrast to the
weak, thermally-driven {\em logarithmic} divergences that occur in
pure smectics.\cite{GP} Here we provide details of these calculations,
describe their effect on the analysis of previous sections, and
discuss them in the context of recent
experiments.\cite{review,Bellini,Clark,Wu,Kutnjak}

The anomalous elasticity of {\em pure} (bulk) smectics\cite{GP} is
characterized by layer compressional and tilt moduli $B({\bf k})$ and
$K({\bf k})$ which vanish and diverge, respectively, at long
wavelengths (${\bf k}\rightarrow0$).  This beautiful result is
actually a general property of all of one-dimensional crystals, in
which the direction of the 1d ordering wavevector is chosen {\em
spontaneously}. As a consequence of this {\em spontaneous} breaking of
rotational symmetry (a property possessed by smectics but not, e.g.,
charge density waves), in the presence of fluctuations a compression
can be relieved by smoothing out these fluctuations, thereby leading
to an effective compressional modulus $B({\bf k})$ that vanishes at
long wavelengths. This is very similar to the ability of free
dislocations in a solid to relieve an imposed shear. Similarly, in the
presence of fluctuations, a bending of smectic layers necessarily
leads to a compression, which implies an effective tilt modulus
$K({\bf k})$ which diverges at long wavelengths.  This phenomenon is
also analogous to the thermally-driven anomalous elasticity of
polymerized membranes, in which the bending rigidity modulus diverges,
and the shear and bulk moduli vanish at long length
scales.\cite{NP,AL,LeDR}

Although quite fascinating, the anomalous elasticity of GP for pure 3d
smectics is somewhat academic, because for $d=3$ the nonlinearities
responsible for the anomalous elasticity are marginally irrelevant,
and therefore lead to weakly (logarithmically) $k$-dependent elastic
moduli. These, in practice, are quite difficult to detect
experimentally, and have never been observed in bulk (pure) smectics.

As mentioned above, the main ingredients necessary for the existence
of anomalous elasticity are {\em spontaneously} broken rotational
invariance, and the presence of fluctuations--i.e. wrinkles in the
smectic layers--both of which exist (even at zero temperature) in
smectics in random environments. In this section, we demonstrate the
existence of anomalous elasticity that is significantly stronger, in
all spatial dimensions $d<5$, than the marginal anomalous elasticity
of thermal smectics. This new, stronger anomalous elasticity is driven
by quenched disorder, and controlled by a new, zero-temperature fixed
point that is perturbatively accessible in $d=5-\epsilon$.  One can
physically appreciate why this elastic anomaly is so much stronger in
smectics with quenched disorder by realizing that the quenched
disordered fluctuations are more divergent (at long wavelengths) than
their thermal counterparts (see e.g., Eq.\ref{CT} and Eq.\ref{CD}).

As discussed at the end of Sec.\ref{stability_harmonic_elastic}, in
dimensions $d<5$ we expect that the random field disorder to be
significantly less important than the random tilt disorder for the
properties of smectics with quenched disorder. We therefore expect the
effective elastic Hamiltonian in Eq.\ref{Hu} with $U({\bf r})=0$ to be
a good starting point for the description of smectics confined inside
low-density aerogels. However, such a model misses an important
ingredient: nonlinear elasticity which takes into account the
underlying rotational invariance of the smectic phase,\cite{GP} hidden
by the {\em spontaneous} choice of the layers to stack along the
$\hat{\bf z}$-direction.  A careful analysis, starting with the de
Gennes model, that keeps track of such elastic nonlinearities, leads
to\cite{GP}
\begin{eqnarray}
H[u]&=&\int_{\bbox r}\bigg[{K\over 2}(\nabla^2_\perp {u})^2 + {B\over
2}\big(\partial_z u - {1\over2}(\bbox{\nabla}_\perp u)^2\big)^2
\nonumber\\ 
&+&{\bf h}({\bf r})\cdot\bbox{\nabla_\perp} u\bigg]\;,
\label{Hanharmonic}
\end{eqnarray}
as the proper rotationally invariant elastic Hamiltonian for the
disordered smectic. This form guarantees that a uniform rotation of
smectic layers costs zero energy in the absence of the random tilt
field.\cite{comment_nonlinear_elasticity} The field ${\bf h}({\bf
r})\equiv g_z({\bf r}){\bf g}({\bf r})$ is the quenched random tilt
disorder, defined in Sec.\ref{model}. As discussed in that section, it
has a vanishing average and Gaussian statistics, completely
characterized by the two-point correlation function
\begin{equation}
\overline{h_i({\bf r})h_j({\bf r'})}=
\Delta_h\delta^d({\bf r}-{\bf r'})\delta_{ij}\;,
\label{Delta}
\end{equation}
which we take to be short-ranged for the reasons discussed in detail
in Sec.\ref{model}.

To compute self-averaging quantities, (e.g. the disorder averaged free
energy) and to assess the importance of the nonlinearities at long
scales it is convenient (but not necessary) to employ the replica
``trick''\cite{Anderson} that relies on the identity $\overline{\ln
Z}=\lim_{n\rightarrow 0}(\overline{Z^n}-1)/n$. As discussed in
Sec.\ref{stability_harmonic_elastic}, this allows us to work with a
translationally invariant field theory at the expense of introducing
$n$ replica fields (with the $n\rightarrow 0$ limit to be taken at the
end of the calculation). After replicating and integrating over the
disorder ${\bf h}({\bf r})$, and using Eq.\ref{Delta}, we obtain an
effective translationally invariant replicated Hamiltonian
\begin{eqnarray}
H[u_\alpha]&=&{1\over2}\int_{\bf r}
\sum_{\alpha=1}^n\bigg[{K}(\nabla_\perp^2
u_\alpha)^2+{B}\big(\partial_z u_\alpha-
{1\over2}(\bbox{\nabla}_\perp u_\alpha)^2\big)^2\bigg]\nonumber\\
&-&{\Delta_h\over 2T}\int_{\bf r}\sum_{\alpha,\beta=1}^n
\bbox{\nabla_\perp}u_\alpha\cdot\bbox{\nabla_\perp}u_\beta
\label{Hr_anharmonic}
\end{eqnarray}
from which the noninteracting propagator $G_{\alpha\beta}({\bf
q})\equiv V^{-1}\langle u_\alpha({\bf q})u_\beta(-{\bf q})\rangle_0$
(where $V$ is the system volume) can be easily obtained (see
Sec.\ref{stability_harmonic_elastic}),
\begin{equation}
G_{\alpha\beta}(q)=T G({\bf q})\delta_{\alpha\beta}+
\Delta_h q_\perp^2 G({\bf q})^2\;,
\label{G}
\end{equation}
with $G({\bf q})=1/(K q_\perp^4+B q_z^2)$.

We first attempt to assess the effects of the anharmonicities,
disorder and thermal fluctuations by performing a simple perturbation
expansion in the nonlinearities of $H[u_\alpha]$. The lowest order
correction $\delta B$ to the bare elastic compressional modulus $B$
comes from a part of the diagram in Fig.10. A standard analysis gives
\begin{mathletters}
\begin{eqnarray}
\delta B&=&-{B^2\over2}\int_{\bf q}^>\left[T G({\bf q})^2
+2\Delta_h q_\perp^2 G({\bf q})^3\right]q_\perp^4\;,\label{deltaBa}\\
&\approx&-B^2\Delta_h\int_{-\infty}^{\infty}{d q_z\over2\pi}
\int^>{d^{d-1}q_\perp\over(2\pi)^{d-1}}
{q_\perp^6\over(K q_\perp^4+B q_z^2)^3}\;,\label{deltaBb}\\
&\approx&-{3\over16} {C_{d-1}\over 5-d}\;
B\Delta_h\left({B\over K^5}\right)^{1/2} L^{5-d}\;,\label{deltaBc}
\end{eqnarray}
\end{mathletters}
where we have dropped the subdominant thermal part, focused on $d<5$,
which allows us to drop the uv-cutoff ($\Lambda$) dependent part which
vanishes for $\Lambda\rightarrow\infty$, and cutoff the divergent
contribution of the long wavelength modes via the infra-red cutoff
restriction $q_\perp>1/L$, where $L$ is the linear extent of the
system. The constant $C_d=2\pi^{d/2}/[(2\pi)^d\Gamma(d/2)]$ is the
surface area of a $d$-dimensional sphere divided by
$(2\pi)^d$. Clearly the anhamonicities become important when the
fluctuation corrections to the elastic constants (e.g., $\delta B$
above) become comparable to the bare values. The divergence of this
correction as $L\rightarrow\infty$ signals the breakdown of
conventional harmonic elastic theory on length scales longer than a
crossover scale $\xi^{NL}_\perp$, which we define as the value of $L$
at which $|\delta B(\xi^{NL}_\perp)|=B$. In $d$ dimensions, this
definition gives:
\begin{equation}
\xi^{NL}_\perp=\left({16(5-d)K^{5/2}\over3
C_{d-1}B^{1/2}\Delta_h}\right)^{1/(5-d)}\;,\label{xiNLperp}
\end{equation}
which for physical 3d smectics is given by 
\begin{equation}
\xi^{NL}_\perp=\left[{64\pi\over3}{K^{5/2}\over
B^{1/2}\Delta_h}\right]^{1/2}\;,\;\;\mbox{for}\;\;d=3\;.\label{xiNLperp3d}
\end{equation}
We can also obtain a non-linear crossover length in the $z$-direction
\begin{mathletters}
\begin{eqnarray}
\xi^{NL}_z&=&(\xi^{NL}_\perp)^2/\lambda\;,\label{z-perp_relation}\\ 
&=&{64\pi K^2\over3\Delta_h}\;,\;\;\mbox{for}\;\;d=3\;,\label{xiNLz3d}
\end{eqnarray}
\end{mathletters}
where $\lambda\equiv(K/B)^{1/2}$, by imposing the infra-red cutoff in
the $z$-direction.

To understand the physics beyond this crossover scale, -- i.e., to make
sense of the apparent infra-red divergences found in
Eq.\ref{deltaBc}\ -- we turn to the renormalization group. As discussed
in Sec.\ref{stability_harmonic_elastic}, we employ the standard
momentum shell renormalization group transformation\cite{Wilson}, by
writing the displacement field as $u_\alpha({\bf r})=u^<_\alpha({\bf
r})+u^>_\alpha({\bf r})$, integrating perturbatively in the nonlinear
coupling $B$ the high wavevector part $u^>_\alpha({\bf r})$ with
support in Fourier space in an infinitesimal cylindrical shell
($\Lambda e^{-\ell}<q_\perp<\Lambda$ and $-\infty<q_z<\infty$), and
rescaling the lengths and long wavelength part of the fields as in
Eqs.\ref{rescale} with $r_\perp=r_\perp'e^{\ell}$,
$z=z'e^{\omega\ell}$ and $u^<_\alpha({\bf r})=
e^{\phi\ell}u_\alpha'({\bf r'})$, so as to restore the ultraviolet
cutoff back to $\Lambda$. The underlying rotational invariance insures
that the graphical corrections preserve the rotationally invariant
operator $\big(\partial_z u - {1\over2}(\bbox{\nabla}_\perp
u)^2\big)$, renormalizing it as a whole. It is therefore convenient
(but not necessary) to choose the dimensional rescaling that also
preserves this operator. It is easy to see that this choice leads to
\begin{equation}
\phi=2-\omega\;.\label{chi_choice}
\end{equation}
This rescaling then leads to the zeroth order RG flow of the effective
couplings
\begin{mathletters}
\begin{eqnarray}
K(\ell)&=&K e^{(d-1-\omega)\ell}\;,\label{Kflow}\\
B(\ell)&=&B e^{(d+3-3\omega)\ell}\;,\label{Bflow}\\
\left(\Delta_h\over T\right)(\ell)&=&\left(\Delta_h\over T\right)
e^{(d+1-\omega)\ell}\;.\label{Deltaflow}
\end{eqnarray}
\label{KBDflow}
\end{mathletters}
From these dimensional couplings one can construct two other effective
dimensional couplings,
\begin{mathletters}
\begin{eqnarray}
\tilde{g}_1&\equiv&\left(B\over K^3\right)^{1/2}\;,\label{g1define}\\
\tilde{g}_2&\equiv&\Delta_h\left(B\over K^5\right)^{1/2}\;,\label{g2define}
\end{eqnarray}
\label{g1g2define}
\end{mathletters}
whose flows are independent of the arbitrary rescaling exponents
$\omega$ and $\phi$, and are given by
\begin{mathletters}
\begin{eqnarray}
\tilde{g}_1(\ell)=\tilde{g}_1e^{(3-d)\ell}\;,\label{g1rescale}\\
\tilde{g}_2(\ell)=\tilde{g}_2 e^{(5-d)\ell}\;.\label{g2rescale}
\end{eqnarray}
\label{g1g2rescale}
\end{mathletters}
$\tilde{g}_1$ is just the coupling that becomes relevant in $d<3$ and
was discovered by Grinstein and Pelcovits to lead to anomalous
elasticity in pure thermal smectics. It is, however, only marginally
irrelevant in $d=3$ and therefore only leads to weak anomalous
elasticity in physical 3d smectics.\cite{GP} In contrast, the upper
critical dimension $d_{uc}$ below which $\tilde{g}_2$ becomes relevant
is $d_{uc}=5$, and we therefore expect a significantly stronger
anomalous elasticity, that should be experimentally observable in {\em
disordered} 3d smectics. These observations imply that temperature is
a strongly irrelevant variable near the disorder dominated fixed point
and does not feed back in a dangerous way. We will therefore set $T=0$
in all subsequent calculations.

We now turn to the RG computation of the one-loop graphical
corrections to the flow of the couplings defined above. The required
integration over the high wavevector components of $u_\alpha$ can only
be accomplished perturbatively in nonlinearities of $H[u]$. Since the
most relevant coupling $\tilde{g}_2$ becomes important for $d<5$, we
will control the infrared divergences by performing an expansion in
$\epsilon=5-d$.

The change in the Hamiltonian due to integrating out these short
length modes is
\begin{equation}
\delta H[u_\alpha^<]=\langle H_{i}[u_\alpha^<+u_\alpha^>]\rangle_>
-{1\over 2 T}\langle H_{i}^2[u_\alpha^<+u_\alpha^>]\rangle_>^c\ldots\;,
\label{deltaH}
\end{equation}
where $H_{i}[u_\alpha^<+u_\alpha^>]$ is the nonlinear part of the
Hamiltonian $H$, Eq.\ref{Hr}, which contains three- and four-point
vertices. The averages above are performed with the quadratic part of
$H$, Eq.\ref{Hr}, using the propagator $G_{\alpha\beta}({\bf q})$,
Eq.\ref{G}, with only an infinitesimal cylindrical shell (i.e. no
cutoff on $q_z$) of modes $u_\alpha^>$ integrated out. The superscript
$c$ denotes a cumulant average. It is easy to verify that the first
term in $\delta H$ (first order in $H_i$) does not lead to corrections
of the elastic constants. Aside from correcting the free energy, it
generates an operator {\em linear} in $\big(\partial_z u -
{1\over2}(\bbox{\nabla}_\perp u)^2\big)$ which corresponds to a
renormalization of the smectic wavevector $q_0$. These corrections
turn out to be finite (irrelevant) near $d=5$, and we will therefore
not keep track of them here.

The renormalization of $K$, $B$ and $\Delta_h$ comes from the second
term in Eq.\ref{deltaH}, for example from parts which are second order
in the three-point vertex, with $4$ of the $6$ fields contracted. The
generic diagram that corrects the elastic moduli and the disorder
variance is illustrated in Fig.10, with the part diagonal in
the replica indices $\alpha,\beta$ (i.e. part proportional to
$\delta_{\alpha\beta}$) renormalizing $K$ and $B$, and the part
independent of $\alpha,\beta$ correcting $\Delta_h$.
\begin{figure}[bth]
\begin{center}
\setlength{\unitlength}{1mm}
\begin{picture}(30,30)(0,0)
\put(-52,-77){\begin{picture}(30,30)(0,0)
\includegraphics{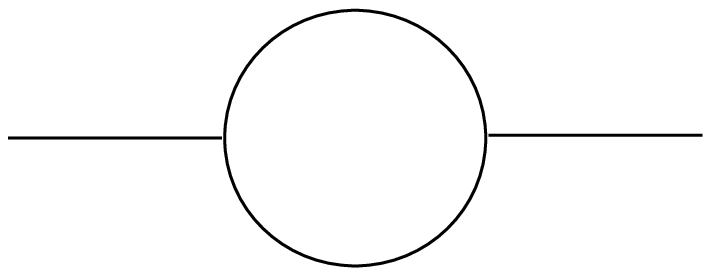}
\end{picture}}
\end{picture}
\end{center}
Fig.10. {Feynman graph that renormalizes the elastic moduli $K$, $B$
and the tilt-disorder variance $\Delta_h$.}
\label{deltaBfig}
\end{figure}
For the calculation of $\delta B$, the loop integrals can be performed
at vanishing external momentum, and after a short calculation one
finds
\begin{eqnarray}
\delta B&=&-{B^2\over2}\int_{\bf q}^>\left[T G({\bf q})^2
+2\Delta_h q_\perp^2 G({\bf q})^3\right]q_\perp^4\;,\nonumber\\
&\approx&-B^2\Delta_h\int_{-\infty}^{\infty}{d q_z\over2\pi}
\int^>{d^{d-1}q_\perp\over(2\pi)^{d-1}}
{q_\perp^6\over(K q_\perp^4+B q_z^2)^3}\;,\nonumber\\
&\approx&-{3\over16}g_2 B d\ell\;,\label{deltaB3}
\end{eqnarray}
where going from the first line to second line above we dropped the
irrelevant finite temperature part, and defined a dimensionless
coupling constant
\begin{equation}
g_2\equiv\Delta_h\left(B\over K^5\right)^{1/2}
C_{d-1}\Lambda^{d-5}\;.\label{g2dimensionless}
\end{equation}
Similar, but more involved calculations for $K$ and $\Delta_h$ give
\begin{mathletters}
\begin{eqnarray}
\delta K&\approx&{1\over32}g_2 K d\ell\;,\label{deltaK}\\
\delta\left({\Delta_h\over T}\right)&\approx&{1\over64}g_2
\left({\Delta_h\over T}\right) d\ell\;,
\label{deltaDelta}
\end{eqnarray}
\end{mathletters}
which lead to the following RG flow equations
\begin{mathletters}
\begin{eqnarray}
\frac{d B(\ell)}{d\ell}&=&(d+3-3\omega-{3\over16}g_2(\ell))B(\ell)
\;,\label{Banom_flow}\\
\frac{d K(\ell)}{d\ell}&=&(d-1-\omega+{1\over32}g_2(\ell))K(\ell)
\;,\label{Kanom_flow}\\
\frac{d\Delta_h(\ell)}{d\ell}
&=&(d+1-\omega+{1\over64}g_2(\ell))\Delta_h(\ell)\;,
\label{DeltaTanom_flow}
\end{eqnarray}
\end{mathletters}
These, together, lead to the RG flow of the dimensionless coupling
$g_2$ defined in Eq.\ref{g2dimensionless}
\begin{equation}
\frac{d g_2(\ell)}{d\ell}=\epsilon g_2(\ell) - {5\over32} g_2(\ell)^2\;,
\label{g2}
\end{equation}
where we remind the reader that in this section $\epsilon=5-d$. As
required, the flow of $g_2$ is independent of the arbitrary choice of
the anisotropy rescaling exponent $\omega$. The RG flow Eq.\ref{g2}
shows that the Gaussian $g_2^*=0$ fixed point becomes unstable for
$d<5$, and the low temperature phase is controlled by a stable,
nontrivial, glassy $T=0$ fixed point at
\begin{equation}
g_2^*={32\over5}\epsilon\;.\label{anomalous_fp}
\end{equation}
The existence of this nontrivial fixed point leads to anomalous
elasticity.  To see this it is convenient to use our RG results to
evaluate the connected disordered averaged 2-point $u({\bf k})$
correlation function $G({\bf k})\propto\overline{\langle|u({\bf
k})|^2\rangle}- \overline{\langle u({\bf k})\rangle\langle u(-{\bf
k})\rangle}$. As discussed (and utilized) in
Sec.\ref{stability_harmonic_elastic}, the power of the renormalization
group is that it establishes a connection between a correlation
function at a small wavevector (which is impossible to calculate in
perturbation theory due to the infra-red divergences) to the same
correlation function at large wavevectors, which can be easily
calculated in a controlled perturbation theory. This relation for
$G({\bf k})$ is
\begin{eqnarray}
G({\bf k}_\perp, k_z, K, B, g_2)&=&e^{(3+d-\omega)\ell}\times\nonumber\\
&&G({\bf k}_\perp e^\ell, k_z e^{\omega\ell}, K(\ell), B(\ell),
g_2(\ell))\;,\nonumber\\
\label{relationG}
\end{eqnarray}
where the prefactor on the right hand side comes from the dimensional
rescaling (remembering the momentum conserving $\delta$-function),
after using the exact rotational Ward identity $\phi=2-\omega$, and we
have ``traded-in'' the disorder variable $\Delta_h$ for the
dimensionless coupling $g_2$. To establish the anomalous behavior of
$K$, we look at $k_z=0$. We then choose the rescaling variable
$\ell^{\ast}$ such that
\begin{equation}
k_\perp e^{\ell^{\ast}}=\Lambda\;.\label{matching_anom}
\end{equation}
We also choose $k_\perp$ sufficiently small such that
$g_2(\ell^{\ast})$ has reached our nontrivial fixed point
$g_2^{\ast}$. Eliminating $\ell^{\ast}$ in favor of $k_\perp$, we then
obtain
\begin{eqnarray}
G({\bf k}_\perp, 0, K, B, g_2)&=&
\left(\Lambda\over k_\perp\right)^{3+d-\omega}\times\nonumber\\
&&G\left(\Lambda, 0,
K(\ell^{\ast}),B(\ell^{\ast}),g_2^{\ast}\right)\;,\nonumber\\
\label{relationG2}
\end{eqnarray}
Since the right hand side is evaluated with the transverse wavevector
at the Brillouin zone boundary, it is easily evaluated perturbatively
for a small fixed point coupling $g_2^{\ast}$. To lowest order we
obtain
\begin{mathletters}
\begin{eqnarray}
G({\bf k}_\perp, 0, K, B, g_2)&\approx&
{(\Lambda/k_\perp)^{3+d-\omega}\over\Lambda^4
K\left(\Lambda/k_\perp\right)^{(d-1-\omega + g_2^{\ast}/32)}}
\;,\label{relationG3}\\
&\equiv&{1\over K(k_\perp) k_\perp^4}
\end{eqnarray}
\end{mathletters}
where we integrated Eq.\ref{Kanom_flow} to obtain $K(\ell^{\ast})$,
and defined the anomalous tilt modulus which diverges at long length
scales
\begin{equation}
K(k_\perp)=K\left(k_\perp/\Lambda\right)^{-\eta_K}\;,\label{Kanom}
\end{equation}
with an anomalous exponent
\begin{mathletters}
\begin{eqnarray}
\eta_K&=&{1\over32}g_2^{\ast}\;,\label{etaK1}\\
&=&{1\over 5}\;\epsilon\;,\label{etaK2}\\
&=&{2\over5}\;,\;\;\;\mbox{for}\; d=3\;.\label{etaK3}
\end{eqnarray}
\end{mathletters}

Similar calculations for the other coupling constants and other
directions of $\bf k$ show that, in general, 
\begin{mathletters}
\begin{eqnarray}
K({\bf k})&=&K\left(k_\perp\xi^{NL}_\perp\right)^{-\eta_K}
f_K(k_z\xi^{NL}_{z}/(k_\perp\xi^{NL}_\perp)^\zeta)\;,\label{Kg}\\
B({\bf k})&=&B\left(k_\perp\xi^{NL}_\perp\right)^{\eta_B}
f_B(k_z\xi^{NL}_{z}/(k_\perp\xi^{NL}_\perp)^\zeta)\;,\label{Bg}\\
\Delta_h({\bf k})&=&\Delta_h\left(k_\perp\xi^{NL}_\perp\right)^{-\eta_\Delta}
f_\Delta(k_z\xi^{NL}_{z}/(k_\perp\xi^{NL}_\perp)^\zeta)\;,\label{Deltag}
\end{eqnarray}
\label{KgBgDg}
\end{mathletters}
with the anisotropy exponent
\begin{equation}
\zeta\equiv 2-(\eta_B+\eta_K)/2\;,\label{zeta}
\end{equation}
which would $=2$ in the absence of anharmonic effects,
\begin{mathletters}
\begin{eqnarray}
\eta_B&=&{3\over16}g_2^{\ast}\;,\label{etaB1}\\
&=&{6\over 5}\;\epsilon\;,\label{etaB2}\\
&=&{12\over5}\;,\;\;\;\mbox{for}\; d=3\;,\label{etaB3}
\end{eqnarray}
\end{mathletters}
and 
\begin{mathletters}
\begin{eqnarray}
\eta_\Delta&=&{1\over64}g_2^{\ast}\;,\label{etaDelta1}\\
&=&{1\over 10}\;\epsilon\;,\label{etaDelta2}\\
&=&{1\over5}\;,\;\;\;\mbox{for}\; d=3\;.\label{etaDelta3}
\end{eqnarray}
\end{mathletters}

Of course, we do not completely trust the extrapolation of these small
$\epsilon$ results down to $\epsilon=2$ ($d=3$). However, since by
definition $d g_2/d\ell = 0$ at the nontrivial fixed point, this
condition implies an {\em exact} relation between the anomalous
exponents
\begin{equation}
5-d + \eta_\Delta = {\eta_B\over 2} + {5\over 2}\eta_K\;,
\label{WI}
\end{equation}
which is obviously satisfied by the anomalous exponents,
Eqs.\ref{etaK2},\ref{etaB2},\ref{etaDelta2}, computed here to first
order in $\epsilon$. This Ward identity between the anomalous
exponents can be equally easily obtained from a self-consistent
integral equation for the $u-u$ correlations functions, using
renormalized wavevector dependent elastic moduli and disorder
variance.\cite{RTunpublished}

At length scales beyond $\xi^{NL}_\perp$ and $\xi^{NL}_z$, the
elasticity and fluctuations of the disordered smectic are controlled
by our new glassy fixed point. One of the important consequences can
be seen in the layer fluctuations, which can be inferred from X-ray
scattering experiments.  For instance, layer displacement fluctuations
along $z$ are described by
\begin{eqnarray}
C(z)&\equiv&\overline{\langle\left(u(0_\perp,z)-
u(0_\perp,0)\right)^2\rangle}\;,\nonumber\\
&\approx&\int{d^{d}k\over(2\pi)^{d}} {2(1-\cos(k_z z))\Delta_h({\bf k})
k_\perp^2\over\big[K({\bf k}) k_\perp^4+B({\bf k}) k_z^2\big]^2}
\;,\label{Cz}
\end{eqnarray}
One can then naturally define the X-ray translational correlation
length $\xi_z^{X}$ as the length along $z$ at which
$C(z=\xi_z^{X})\equiv a^2$, where $a$ is the smectic layer spacing. A
simple calculation, using Eqs.\ref{Kg}-\ref{Deltag} leads in 3d to
\begin{mathletters}
\begin{eqnarray}
\xi_z^{X}&=&\left(a/\lambda\right)^{\zeta/\chi}K^2/\Delta_h\;,\\
&=&\left(a/\lambda\right)^{\zeta/\chi}\xi_z^{NL}\;,\label{xi_z}
\end{eqnarray}
\label{xi_z_ab}
\end{mathletters}
where 
\begin{equation}
\chi\equiv(\eta_B+\eta_K)/2\;,\label{anisotropy_gamma}
\end{equation}
and this result should be contrasted with the expression for the X-ray
correlation length $\xi_z^{X}$, Eq.\ref{xiDeltaz}, calculated within
the {\em harmonic} theory, i.e., ignoring the anomalous
elasticity. Note that this X-ray correlation length is finite even as
$T\rightarrow 0$, as illustrated in Fig.3. This result is consistent
with the experimental observation\cite{review,Clark} that the X-ray
correlation length for smectics in aerogel saturates at some finite
value at low temperatures. Note also that this length should be
different for different smectics in the same aerogel, since $ B, K$,
and $\Delta_h$ will change from smectic to smectic. Since, as
discussed in Sec.\ref{model}, we expect $\Delta_h$ to be a
monotonically increasing function of the aerogel density $\rho_A$,
with a simplest aerogel model giving $\Delta_h\propto\rho_A$,
Eq.\ref{rhorho_b}, the aerogel density dependence of $\xi_z^X$ could
test the prediction of Eq.\ref{xi_z_ab}. Likewise, the {\em
temperature} dependence of $\xi_z^X$ could be used to determine the
ratio $\zeta/\chi$, since the {\em bulk} $K(T)$ and $B(T)$ that
implicitly appear in Eq.\ref{xi_z_ab} have temperature dependences
that can be extracted from measurements on {\em bulk} materials.

Note also that this correlation length is longer than the non-linear
crossover length for $\lambda<a$ (i.e., for large $B$). For
$\lambda>>a$ (small $B$), $C(z)$ reaches $a^2$ before $z$ reaches
$\xi^{NL}_z$, and hence anharmonic effects are unimportant. In this
case, the correlation length $\xi_z^X$ can be determined in the
harmonic theory (which amounts to evaluating the integral in
Eq.\ref{Cz} with $K({\bf k})$, $B({\bf k})$, and $\Delta_h({\bf k})$
replaced by their constant (bare) values $K$, $B$, and $\Delta_h$). As
shown in Sec.\ref{tilt-only_model}, this gives $\xi_z^X=a^2 B
K/\Delta_h=(a/\lambda)^2\xi_z^{NL}$, which is, reassuringly, much less
than $\xi_z^{NL}$ in the limit $a<<\lambda$ in which we have asserted
it applies.

In this section, in our treatment of elastic nonlinearities we have
ignored dislocations, which in Sec.\ref{dislocations} we have shown to
unbind in a {\em harmonic} theory, for arbitrarily weak disorder.
However, as we have shown in that section, their effects are only felt
on length scales greater than $\xi_z^{D}=
(a^2/\lambda)e^{\lambda\xi_z^X/a^2}$, which is much longer, in the
weak disorder ($\Delta_h\rightarrow0$, $\xi_z^X\rightarrow\infty$)
limit, than the translational correlation length $\xi_z^X$ found here.
However, it is, nevertheless important to study the simultaneous
effects of anomalous elasticity and dislocations. We do this in
Sec.\ref{anom_elasticity_dislocations_correlations} by reconsidering
the criterion for dislocation unbinding, studied in
Sec.\ref{dislocations}, now in the presence of anomalous elasticity.
We will find that anomalous elasticity leads to shortening of
$\xi_z^{D}$, which nonetheless remains much longer than $\xi_z^X$.
Hence, for low density aerogels, our predictions for $\xi_z^X$ and the
anomalous elasticity studied here remain valid.

Finally, our entire discussion so far has focused only on the effects
of {\it orientational} tilt disorder. In Sec.\ref{random_field_3-5} we
will consider the {\it translational} disorder (i.e., random pinning
of the positions of the layers) and will demonstrate that it is {\it
less} important, at long wavelengths (in $d<5$), than the
orientational disorder. Thus, the results described herein are
directly applicable to real smectics, where both kinds of disorder are
present.

\section{Effects of anomalous elasticity on dislocation unbinding 
and smectic correlations}
\label{anom_elasticity_dislocations_correlations}

In Section \ref{anom_elasticity} we established that, at least at the
``phonon-only'' level, the anharmonic terms in the elastic energy are
extremely important below $d = 5$. We therefore need to also consider
the effects of anharmonic elasticity (and the resulting anomalous
elasticity) on dislocations and smectic correlation functions.

\subsection{Dislocations and orientational order}
\label{anom_elasticity_dislocations}
In this subsection we incorporate anomalous elasticity into our
duality model and use it to derive a criterion, in terms of the
anomalous exponents $\eta_B$ and $\eta_K$, for the dislocation
unbinding to take place in the presence of anomalous
elasticity. Combining these results with an analysis of orientational
order we establish bounds on $\eta_B$ and $\eta_K$ for the stability
of the smectic Bragg glass phase.  However, since we do not know the
numerical values of $\eta_B$ and $\eta_K$ in $d = 3$, we cannot say
whether or not, in the full, anharmonic theory, dislocations are, in
fact, unbound.

Unfortunately, a full {\em anharmonic} theory of dislocations is
simply intractable.  In particular, the fact that, in an anharmonic
theory, the interaction energy between dislocations {\em cannot} be
written as a sum of pairwise interactions (since their fields do not
simply add) makes it impossible to even write down a general
expression for the energy of an arbitrary dislocation configuration.
At best, one might hope to be able to write down the energy for a few
simple, high symmetry configurations (e.g., a single, straight
dislocation line).  Such specialized results would be of no use in a
full statistical theory of defect unbinding, which requires
consideration of very complicated, tangled configurations of
dislocations, which, for entropic reasons, dominate the free energy
near the dislocation binding transition.

Furthermore, even {\em if} one {\em could} write down the anharmonic
energy for an arbitrary dislocation configuration, it would presumably
be anharmonic in the dislocation fields ${\bf m}({\bf r})$, and hence,
those fields could {\em not} be decoupled by a simple
Hubbard-Stratonovich trick, as they can in the harmonic approximation.

For all of these reasons, a completely honest treatment of
dislocations in the full, anharmonic model is impossible, at least for
mere mortals like us.  Rather than despair, however, we will use a
slightly dishonest (i.e., uncontrolled) approximation: we will simply
replace the elastic {\em constants} $K$ and $B$, and tilt disorder
variance $\Delta_h$ in our expression Eq.\ref{tR} for the renormalized
dislocation unbinding transition temperature, with the renormalized,
wavevector dependent moduli $K({\bf q})$, $B({\bf q})$, and
$\Delta_h({\bf q})$, derived from the anharmonic, tilt-only elastic
theory of Sec.\ref{anom_elasticity}.

Doing so, we obtain for the downward renormalization of the {\em dual}
transition temperature
\begin{equation}
\delta t = - {\rm O}(1)\times{c a^2 d^2\over T^2}
\int{\Delta_h({\bf q}) q^2_z d^dq\over
\big[q^2_z + \lambda^2({\bf q})q^4_{\perp}\big]^2}\;,
\label{delta_t_anom}
\end{equation}
where $\Delta_h({\bf q})$ is given by Eq.\ref{Deltag}, $c$ is given by
Eq.\ref{c(T)}, and $\lambda^2({\bf q})\equiv K({\bf q})/B({\bf q})$,
with $K({\bf q})$ and $B({\bf q})$ given by Eqs.\ref{Kg} and \ref{Bg},
respectively. In writing Eq.\ref{delta_t_anom} we have anticipated
that the integral is dominated by region $q_z<<q_\perp$, as can be
seen from Eq.\ref{delta_t_anom2}, below.

Imposing an infra-red cutoff $q_{\perp} > L^{-1}$ on the wavevector
integral in Eq.\ref{delta_t_anom}, where $L$ is the lateral $(\perp)$
real-space linear extent of the system, and changing variables in that
integral as follows:
\begin{mathletters}
\begin{eqnarray}
{\bf q}_{\perp}&\equiv&{{\bf Q}_{\perp} \over L}
\label{rescale_perp}\;,\\
q_z&\equiv&{Q_z \over L^{\zeta}}{\lambda\over(\xi_\perp^{NL})^\chi}
\label{rescale_z}\;,
\end{eqnarray}
\end{mathletters}
with $\chi\equiv(\eta_B+\eta_K)/2$, we find that
\end{multicols}
\begin{equation}
\delta t\approx -\left({L\over\xi_\perp^{NL}}\right)^{\gamma}
{c a^2 d^2\Delta_h\over\lambda T^2}\int_{L\Lambda > Q_{\perp} > 1}
 {Q^{-\eta_{\Delta}}_{\perp}\, Q_z^2\, f_{\Delta} 
\left(Q_z/Q^{\zeta}_{\perp}\right)\, d^dQ\over 
\left[Q^2_z + {f_K\left(Q_z/Q^{\zeta}_{\perp}\right)
\over f_B\left(Q_z/Q^{\zeta}_{\perp}\right)}\, Q^{2\zeta}_{\perp}\right]^2}
\;,\label{delta_t_anom2}
\end{equation}
\bottom{-2.7cm}
\begin{multicols}{2}
\noindent
with
\begin{equation}
\gamma = \zeta + \eta_{\Delta} + 1 - d\;.
\label{gamma1}
\end{equation}
In deriving Eq.\ref{gamma1}, we have used the scaling relation $\zeta
= 2 - (\eta_B + \eta_K)/2$, and Eq.\ref{WI} defined in
Sec.\ref{anom_elasticity}.

Now, it is equally straightforward to show, by anisotropic power
counting, that the integral over $\bf Q$ in Eq.\ref{delta_t_anom2} is
convergent in the ultra-violet when $\gamma > 0$, and is therefore
independent of its upper cutoff $L\Lambda$, for $L\rightarrow\infty$.
As a result, for $\gamma > 0$, the correction to the reduced {\em
dual} temperature $t$ depends on $L$ only through the overall
prefactor $L^\gamma$, i.e.
\begin{equation}
\delta t\approx -\left({L\over\xi_\perp^{NL}}\right)^\gamma
{c a^2 d^2\Delta_h\over\lambda T^2}\times{\rm O}(1)\;,
\label{delta_t_final}
\end{equation}
and therefore, for $\gamma > 0$, diverges with system size.

As discussed in our original treatment of duality, such an infinite
{\em downward} renormalization of $t$ in the {\em dual} model implies
that dislocations are {\em always} unbound, even at zero temperature,
even for arbitrarily weak disorder.  Thus, our criterion for
determining whether or not dislocations are always unbound in $d = 3$
is very simple: if $\gamma$, as given by Eq.\ref{gamma1}, is $> 0$,
dislocations {\em are} always unbound, and {\em no} stable Bragg glass
phase exists, while if $\gamma < 0$, dislocations will be bound, and a
stable Bragg glass phase {\em will} exist, at least at sufficiently
low temperatures and weak disorder.

So it behooves us to determine $\gamma$ in $d = 3$.  Using $\zeta = 2-
(\eta_B + \eta_K)/2$, and the exact scaling relation Eq.\ref{WI}
between $\eta_{\Delta}, \eta_B,$ and $\eta_K$, we can rewrite
Eq.\ref{gamma1} in terms of $\eta_{\Delta}$ and $\eta_K$ as:
\begin{equation}
\gamma = 2 \left( \eta_K -1 \right)\;.
\label{gamma2}
\end{equation}
Using our $\epsilon$-expansion result Eq.\ref{etaK} for $\eta_K$, we
obtain
\begin{mathletters}
\begin{eqnarray}
\gamma&=&2 \left( {\epsilon \over 5} -1 \right) + O(\epsilon^2)\;,\\
&=&-{6\over5}<0\;,
\end{eqnarray}
\label{gamma3}
\end{mathletters}
the last equality holding in $d = 3$ ($\epsilon = 2$), if we drop the
unknown $O(\epsilon ^2)$ term. This result {\em appears} to imply that
dislocations are bound, and the smectic Bragg glass is {\em stable} in
three dimensions.  Of course, it is always risky to quantitatively
extrapolate an expansion for small $\epsilon$ out to $\epsilon = 2$,
but this is the best we can do in $d = 3$.  And it seems to suggest
that the smectic Bragg glass is {\em stable} in $d = 3$.

Independent of the $\epsilon$-expansion, our result implies the strict
{\em necessary, but not sufficient} condition for the stability of the
smectic Bragg glass phase
\begin{equation}
\eta_K  < 1\;.
\label{stable_disl}
\end{equation}

If, on the other hand, this condition is violated, then dislocations
will unbind on length scales longer than $\xi_\perp^D$, where
$\xi_\perp^D$ is the value of $L$ in Eq.\ref{delta_t_final} such that
the correction $\delta t$ becomes of $O(t)$. This gives the value of
$\xi^D_\perp$ quoted in the Introduction, with $c(T)$ given by
Eq.\ref{c(T)}. We note that, although $\lambda(T)\rightarrow\infty$ as
$T\rightarrow T_{NA}$ from below, so does $c(T)$, since the
renormalized dislocation core energy vanishes at $T_{NA}$. We expect
the divergence of $c(T)$ to overwhelm that of $\lambda(T)$, leading to
a dislocation length that gets ever shorter as $T\rightarrow T_{NA}$
from below. Eventually, as we approach $T_{NA}$ from below,
$\xi_\perp^D$, as given by Eq.\ref{xi_perpD}, will get smaller than
$\xi_\perp^{NL}$; at this point, we are in the strong coupling regime,
and our weak-disorder results no longer apply.

There is another criterion in addition to Eq.\ref{stable_disl}, that
must be satisfied for the smectic Bragg glass to be stable in $d = 3$:
{\em long-ranged orientational} order must exist.  This is clear,
since we have implicitly assumed the existence of such order
throughout our calculation (e.g., in writing $\hat{\bf n} = \hat{\bf
z} + \delta{\bf n}$ and {\em assuming} $|\delta{\bf n}|\ll 1$).
Surely, if the layer orientations are not ordered, the layers
themselves cannot retain their integrity (i.e., remain undislocated).
So long-ranged orientational order must be preserved for the smectic
Bragg glass phase to be stable.

An obvious corollary of this conclusion is that orientational {\em
fluctuations} $\overline{\langle|\delta{\bf n}({\bf r})|^2\rangle}$
must remain {\em finite} (as the system size $L \rightarrow \infty$)
for the smectic Bragg glass to be stable.

We will now show, that, within a $5-d=\epsilon$-expansion, this
condition is definitely {\em not} satisfied: orientational
fluctuations {\em do} diverge in $d = 3$.  Hence, the
$\epsilon$-expansion implies that (notwithstanding the result
$\gamma<0$, Eq.\ref{gamma3}) the smectic Bragg glass phase is {\em
not} stable in three dimensions.

To show this, we use the fact that $\delta{\bf n}({\bf r})=
{\bbox\nabla}_\perp u({\bf r})$ to write
\begin{mathletters}
\begin{eqnarray}
\overline{\langle|\delta{\bf n}({\bf r})|^2\rangle} &=&
\int{d^dq\over(2\pi)^d}
\overline{\langle|\delta{\bf n}({\bf q})|^2\rangle}\;,\\
&=&\int{d^dq\over(2\pi)^d} q^2_{\perp}
\overline{\langle|{\bf u}({\bf q})|^2\rangle}\;,\\
&=&\int{d^dq\over(2\pi)^d}{\Delta_h({\bf q})q^4_{\perp}
\over\left[B({\bf q})q^2_z + K({\bf q})q^4_{\perp}\right]^2}\;,
\end{eqnarray}
\label{delta_n}
\end{mathletters}
where we have kept only the dominant, disorder induced term in
$\overline{\langle|u({\bf q})|^2\rangle}$, Eq.\ref{Gab}, in
writing the last equality.  Imposing an infrared cutoff $q_{\perp} >
L^{-1}$, and power-counting on this integral exactly as we just did in
evaluating Eq.\ref{delta_t_anom2}, we find that
\begin{equation}
\overline{\langle|\delta{\bf n}({\bf r})|^2\rangle}\propto L^{\delta}\;,
\label{delta_n2}
\end{equation}
with
\begin{equation}
\delta = -d - 3 + \eta_{\Delta} + 2\eta_B + 3\zeta\;.
\label{delta1}
\end{equation}
Using $\zeta = 2 - (\eta_B + \eta_K)/2$ and the exact scaling relation
Eq.\ref{WI}, we can rewrite this as
\begin{mathletters}
\begin{eqnarray}
\delta &=& \eta_B + \eta_K - 2\;,\label{delta2}\\
&=&2(\chi-1)\;.
\end{eqnarray}
\end{mathletters}
Thus, stability of long-ranged {\em orientational} order, and hence,
stability of the smectic Bragg glass phase itself, requires $\delta <
0$, or
\begin{equation}
\eta_B + \eta_K < 2\;,
\label{stable_orient}
\end{equation}
which, {\em not} coincidentally, is equivalent to the condition that
the anisotropy exponent $\zeta > 1$.  The reason that these two
conditions are equivalent is clear. As we go down in dimension,
$\eta_B$ and $\eta_K$ grow, and the anisotropy exponent $\zeta$
decreases below $2$, with the system becoming more isotropic. The
dimension at which the orientational order disappears must be {\em
exactly} the dimension at which the system becomes completely
isotropic, i.e. $\zeta=1$, since, without orientational order, there
can be no distinction between different directions.  Thus, the lower
critical dimension for orientational order is given by $\zeta = 2 -
(\eta_B + \eta_K)/2 = 1$, which implies $\eta_B + \eta_K = 2$, the
borderline of the inequality Eq.\ref{stable_orient}.

If the condition Eq.\ref{stable_orient} is violated, then for length
scales beyond the orientational correlation length, $\xi_O$,
\begin{equation}
\xi_O=\xi_\perp^{NL}\left({\xi_z^{NL}
\over\lambda}\right)^{1\over2(\chi-1)}\;,
\label{xi_Ocalcs}
\end{equation}
obtained by equating $\overline{\langle|\delta{\bf n}({\bf
r})|^2\rangle}$ to $1$, the orientational order is destroyed, our
smectic description breaks down, no stable SBG exists, and the system
is isotropic.  On the other hand, if condition in
Eq.\ref{stable_orient} holds, i.e., $\chi<1$, $\xi_O$ is infinite and
SBG is stable.

The length scales $\xi_\perp^{NL}$ and $\xi_O$, derived in
Sec.\ref{anom_elasticity} and in this section, respectively, lead to
nontrivial, and in principle experimentally measurable crossover
behavior. At the shortest length scales, $q_\perp>>1/\xi_\perp^{NL}$,
nonlinear elasticity is unimportant and the dominant modes correspond
to $q_z\sim\lambda q_\perp^2$. At longer scales, for
$q_\perp<1/\xi_\perp^{NL}$ this quadratic behavior crosses over to
$q_z\xi_z^{NL}\sim(q_\perp\xi_\perp^{NL})^\zeta$, controlled by the
nontrivial anomalous elasticity fixed point studied in
Sec.\ref{anom_elasticity}. The anisotropy exponent $\zeta$ satisfies
$1<\zeta<2$ if Eq.\ref{stable_orient} is satisfied and orientational
order is stable, but is less than $1$ if this condition is violated
and $\xi_O$ is finite. In this latter case, the anomalous elasticity
scaling $q_z\xi_z^{NL}\sim(q_\perp\xi_\perp^{NL})^\zeta$ crosses over
to the isotropic scaling $q_z\sim q_\perp$ for lengths scales longer
than $\xi_O$. This orientational crossover length scale is,
reassuringly, precisely the orientational correlation length given in
Eq.\ref{xi_Ocalcs}, obtained from the condition of
$\overline{\langle|\delta{\bf n}({\bf r})|^2\rangle}=1$. It is
important to note, that even for $\zeta<1$, in the dominant regime of
wavevectors for $|{\bf q}|>1/\xi_O$, the condition $q_z<<q_\perp$ is
always satisfied. Hence, even in the absence of long ranged
orientational order beyond $\xi_O$, this justifies our use, throughout
the smectic regime ($|{\bf q}|>1/\zeta_O$), of the approximation
$q_z<<q_\perp$, which is indeed the usual approximation one makes in
treating smectic elasticity.

These dominant wavevector regimes in $q_\perp, q_z$ plane, various
important length scales, and the corresponding crossovers are
illustrated in Fig.11.
\begin{figure}[bth]
{\centering
\setlength{\unitlength}{1mm}
\begin{picture}(50,90)(0,0)
\put(-15,-64){\begin{picture}(50,30)(0,0)
\includegraphics{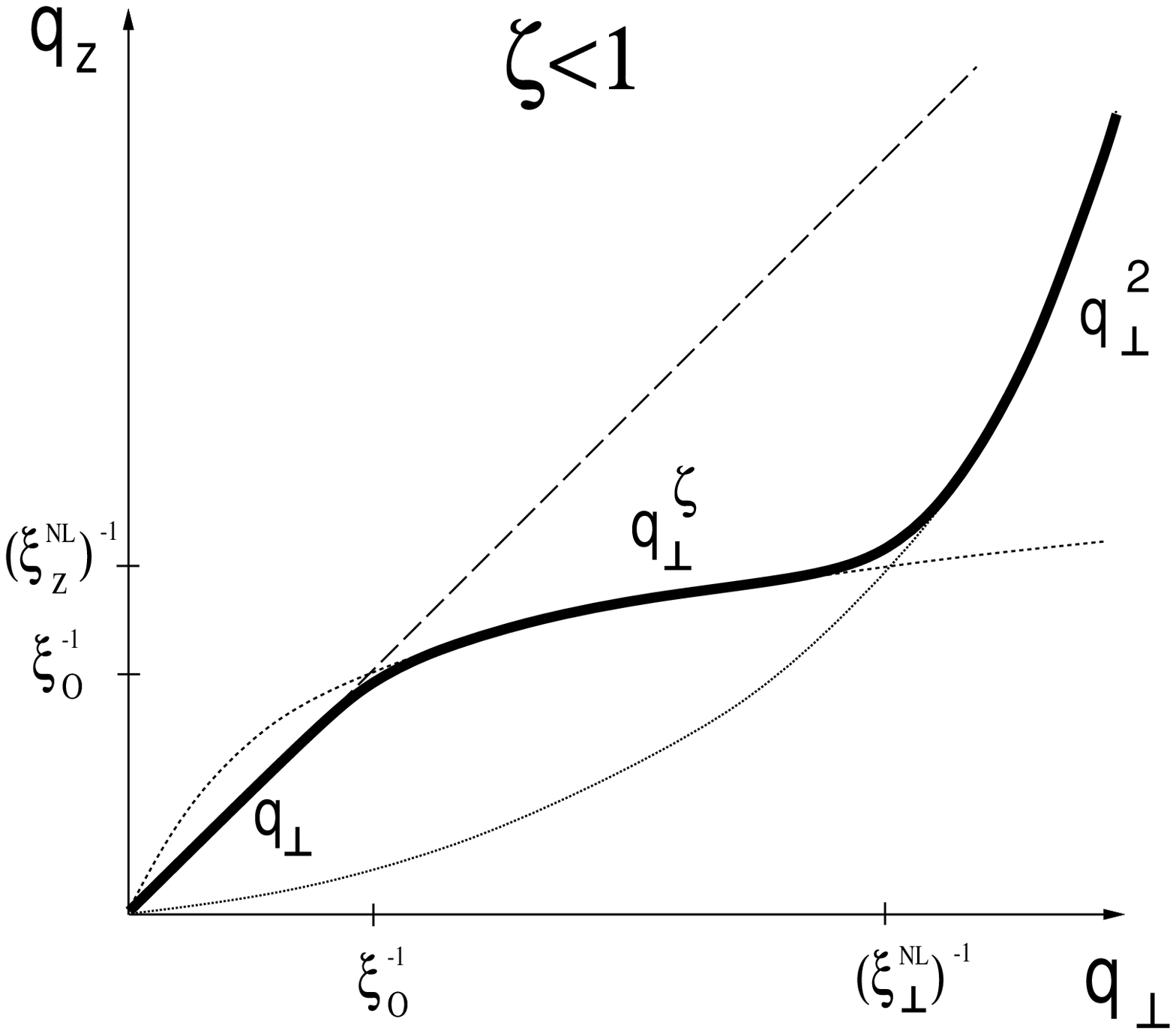}
\end{picture}}
\end{picture}}\\
Fig.11. {Crossover in the dominant wavevector regimes in $q_\perp, q_z$
plane, for $\zeta<1$.}
\label{crossover_qzqp}
\end{figure}

Combining the stability conditions Eq.\ref{stable_disl} and
Eq.\ref{stable_orient} with the rigorous bound
\begin{equation}
\eta_B + 5\eta_K > 4
\label{stable_orient2}
\end{equation}
that follows from requiring $\eta_{\Delta} > 0$ in the exact scaling
relation Eq.\ref{WI} in $d = 3$, we find that the three-dimensional
smectic Bragg glass phase can only be stable when $\eta_B$ and
$\eta_K$ lie in the shaded quadrilateral in Fig.1.

Our $\epsilon$-expansion results from Eqs.\ref{etaK3} and \ref{etaB3},
extrapolated to $d = 3$
\begin{mathletters}
\begin{eqnarray}
\eta_K&=&{2 \over 5}\;,\\
\eta_B&=&{12\over 5}\;,
\end{eqnarray}
\end{mathletters}
lie well outside this stable region.  Whether this extrapolation can
be trusted in $d = 3$ is, of course, an open question.  {\em We are
therefore left, despite our best efforts as displayed in this
manuscript, unable to decide whether or not the smectic Bragg glass
phase actually exists in $d = 3$.} This question can therefore only be
answered by experiments.

\subsection{Smectic correlation functions}
\label{anom_elasticity_correlations}

In addition to having important effects on dislocation unbinding, as
demonstrated in the previous subsection, anharmonic elasticity also
changes the smectic correlation functions from those derived for the
elastically harmonic model of Sec.\ref{tilt-only_model}. We study
these effects in this subsection.

We focus on the smectic layers displacement correlation function
defined by Eq.\ref{CTD}

\begin{equation}
C({\bf r}_\perp,z)=\overline
{\langle\left[u({\bf r_\perp},z)-u({\bf 0},0)\right]^2\rangle}\;
\label{CTDanom}
\end{equation}
To compute $C({\bf r}_\perp,z)$, we proceed as in
Sec.\ref{tilt-only_model}, except that instead of using the bare
values of elastic moduli $K$, $B$ and tilt-disorder variance
$\Delta_h$, we will use their wavevector-dependent counterparts,
$K({\bf k})$, $B({\bf k})$ and $\Delta_h({\bf k})$, given in
Eqs.\ref{KgBgDg}. We expect this procedure to take into account the
long length scale effects of anharmonic elasticity. We obtain:
\begin{equation}
C({\bf r}_\perp,z)
\approx\int{d^{d}k\over(2\pi)^{d}} 
{2(1-\cos({\bf k}\cdot{\bf r}))\Delta_h({\bf k})
k_\perp^2\over\big[K({\bf k}) k_\perp^4+B({\bf k}) k_z^2\big]^2}\;,
\label{Crz}
\end{equation}
where we have only kept the dominant disorder-induced
contribution. Performing an asymptotic analysis similar to that of the
previous subsection, we find that the harmonic elasticity result,
Eq.\ref{CDelta} is replaced by
\begin{equation}
C({\bf r}_\perp,z)\approx\lambda^2\left\{\begin{array}{lr}
\left({z\over\xi^{NL}_z}\right)^{\eta_B+\eta_K\over\zeta},&
z/\xi^{NL}_z>>(r_\perp/\xi^{NL}_{\perp})^\zeta\;,\\
\left({r_\perp\over\xi^{NL}_\perp}\right)^{\eta_B+\eta_K},&
z/\xi^{NL}_z<<(r_\perp/\xi^{NL}_{\perp})^\zeta\;,
\end{array}\right.
\label{CDelta_anom}
\end{equation}
valid for $d<5$ and on scales beyond the crossover scales
$\xi^{NL}_{z,\perp}$, when the smectic elasticity is controlled by the
nontrivial anomalous fixed point with nonvanishing $\eta_B, \eta_K$,
and the anisotropy exponent $\zeta\equiv 2-(\eta_B+\eta_K)/2<2$.

The above result for smectic layer undulations has obvious
implications for the smectic order parameter ($\psi({\bf r})$)
correlations. Within the Debye-Waller approximation, which amounts to
ignoring higher order cumulants, the smectic correlation function is
given by
\begin{mathletters}
\begin{eqnarray}
S({\bf r}_\perp,z)&\equiv&
\overline{\langle\psi^*({\bf r})\psi({\bf 0})\rangle}\;,\\
&\approx&e^{-{q_0^2\over2}C({\bf r})}\;.
\end{eqnarray}
\label{Sanom}
\end{mathletters}
It is short-ranged, confirming our earlier claims about the
destruction (in $d<5$) of the long-range smectic order by an
arbitrarily small amount of tilt disorder, even in the presence of
anharmonic elasticity.

Although the anomalous elasticity does not prevent the destruction of
the long-ranged smectic order, it does lead to some physically
appealing features in the above expression for $S({\bf r}_\perp,z)$,
and $C({\bf r})$, Eq.\ref{CDelta_anom}, which are absent when these
elastic anharmonic effects are not taken into account. In particular,
we observe that at the lower-critical dimension in which the
orientational order becomes short-ranged,
(i.e. $\eta_B+\eta_K\rightarrow 2$ and $\zeta\rightarrow 1$), the
smectic correlation function $S({\bf r})$ is isotropic and decays as a
Gaussian along both $z$ and $r_\perp$. In this case of short-ranged
orientational order this faster than exponential decay of smectic
correlations of course only persists up to the orientational
correlation length, beyond which we expect the order to decay simply
exponentially.

\section{Anomalous elasticity and translational periodic disorder in $3<d<7$}
\label{random_field_3-5}

The analysis of nonlinear elasticity of the preceding two sections
was confined to the tilt-only model in which the random {\em field}
disorder is neglected. In this section we study the effect of the
periodic field disorder in the presence of anharmonic elasticity and
show that, for all $d < 5$, and, in particular in $d = 3$, such
random-field disorder
\begin{equation}
H^{rf}_1=-{1\over T}\int \! d^d{\bf r}
|\psi_0|U({\bf r})\cos[q_0\big(u({\bf r})+z\big)]
\label{Hrf1}
\end{equation}
has no effect on the long-wavelength ``pure phonon'' tilt-only
disorder theory; that is, as long as dislocations are bound, the
anomalous elastic theory of the previous section is correct in
$d<5$. In the next section (Sec.\ref{failure}), we show that periodic
disorder does not induce dislocation unbinding.

We begin by showing that the periodic disorder itself induces
anomalous elasticity in all spatial dimensions $d<7$. Fortunately, the
induced corrections to $B$ and $K$, and to the $u$-fluctuations
themselves, are so weak (only logarithmic) in {\em all} spatial
dimensions $d < 7$ that, as soon as $d < 5$, they are completely
overwhelmed by the {\em power-law} divergent fluctuations and
renormalizations induced by the {\em tilt} disorder that we studied in
Sec.\ref{anom_elasticity}. These latter fluctuations and
renormalizations therefore dominate, and are unaffected by the
presence of the periodic random field disorder. Therefore, once again,
the static correlations that we derived in Sec.\ref{anom_elasticity}
from the $5-\epsilon$ expansion on the tilt-only disorder model are
correct, in the long-wavelength limit, even for the full model.

To derive these results, we consider the full disordered smectic
phonon model.  In addition to the two types of disorder (random tilt
and periodic random field disorders), described by the effective
Hamiltonian in Eq.\ref{Hr}, this model also includes the anharmonic
elasticity (considered in the previous section in the tilt-only phonon
model), but still ignores unbound dislocations. Generalizing
Eq.\ref{Hr} to anharmonic elasticity, as we did in
Sec.\ref{anom_elasticity} for the tilt-only model, we obtain the
effective Hamiltonian for such a phonon model of disordered smectic:
\begin{eqnarray}
H&=&\int d^d
r\bigg[\sum_{\alpha=1}^n\left({K\over2}(\nabla_\perp^2
u_\alpha)^2+{B\over2}\big(\partial_z u_\alpha-
{1\over2}(\bbox{\nabla}_\perp u_\alpha)^2\big)^2\right)\nonumber\\
&\hspace{-1.0cm}+&\hspace{-0.5cm}{1\over T}\hspace{-0.1cm}
\sum_{\alpha,\beta=1}^n
\left({\Delta_h\over4}|\bbox{\nabla_\perp}(u_\alpha-u_\beta)|^2 -
\Delta_V\cos[q_0(u_\alpha-u_\beta)]\right)\bigg]\;.\nonumber\\
\label{Hr_anharmonic_rf}
\end{eqnarray}

To study the long distance behavior of this model in general
dimensions $d$, for reasons that we now explain, we must first
generalize it further.

First, recall that in the Sec.\ref{stability_harmonic_elastic}
derivation of the model, the periodicity of the smectic phase required
the translational disorder (i.e., the random field $\Delta_V$) to be a
function that is invariant under discrete translations by smectic
lattice constant $a$ (i.e., periodic with period $a$). In writing down
Eq.\ref{Hr_anharmonic_rf}, we took this function to be the {\em
lowest} cosine harmonic. Our implicit justification for this was that
in $d=3$, in harmonic elastic theory, the eigenvalue $\lambda_n$ for
the $n$th harmonic (with wavevector $nq_0$)
\begin{equation}
H^{rf}_n=-{1\over T}\int \! d^d{\bf r}|\psi_0|U({\bf r})\cos[nq_0\big(u({\bf r})+z\big)]
\label{Hrfn}
\end{equation}
is given by a simple generalization of 
\begin{equation}
\lambda_1=4-\eta\;,
\end{equation}
(see Eq.\ref{gamma_tilde}) to be
\begin{equation}
\lambda_n=4-\eta n^2\;.\label{lambda_n}
\end{equation}
\top{-2.7cm}

\noindent
As long as $\eta$ is fixed under the RG flow, this shows that the {\em
higher} harmonics of the random pinning potential are less relevant,
i.e., they are irrelevant at the point at which the lowest $n=1$
harmonic becomes relevant. However, here we argue that in $d$
dimensions $\eta(\ell)$ flows according to
\begin{equation}
{d\eta(\ell)\over d\ell}=-\theta(\ell)\eta(\ell)\;,
\label{eta_flow}
\end{equation}
with the ``thermal eigenvalue'' $\theta>0$ for $d>d_c$ with $d_c>3$.
We can demonstrate this explicitly in three limiting cases described below.

The first is the truncated model Eq.\ref{Hr} analyzed in
Sec.\ref{stability_harmonic_elastic}, which ignores anharmonic
elasticity.  Using the exact expressions, Eqs.\ref{K_ell},\ref{B_ell}
for the flow of $K$ and $B$ (see discussion in
Sec.\ref{stability_harmonic_elastic}), and the constancy of $q_0$
under the RG, inside $\eta$, Eq.\ref{eta}, we obtain
\begin{equation}
\theta_{\rm harmonic}=d-3\;,\label{theta_harmonic}
\end{equation}
{\em exactly}, in this elastically {\em harmonic} model.

We can also compute $\theta$ near $d=5$ in a model defined by the
Hamiltonian in Eq.\ref{Hr_anharmonic} that includes anharmonic
elasticity but leaves out the random field disorder ($\Delta_V$),
which is, as we show below, subdominant for $d<5$. In this case,
combining Eqs.\ref{Banom_flow},\ref{Kanom_flow} with the flow of $q_0$
(governed by the eigenvalue $\phi=2-\omega$) and the fixed point value
for $g_2$, Eq.\ref{anomalous_fp}, inside Eq.\ref{eta}, we obtain
\begin{equation}
\theta_{d\rightarrow 5^-}=d-3-{\epsilon_5\over2}\;\label{theta_harmonic5}
\end{equation}
exact to order $\epsilon_5=5-d$.

Finally, as we will show in this section, (with details given in
Appendix B) for $5<d<7$,
\begin{equation}
\theta_{d\rightarrow7^-}(\ell)=d-3+O(1/\ell)\;,\label{theta_harmonic7}
\end{equation}

In these three regimes for $d>d_c\approx 3$, $\theta>0$, which leads
to $\eta(\ell\rightarrow\infty)\rightarrow0$. Based on
Eq.\ref{lambda_n}, the implication of this is that the spectrum of
random field harmonics is {\em degenerate}, i.e., an infinite class of
operators $H_n^{rf}$, Eq.\ref{Hrfn}, are all equally relevant and must
be treated simultaneously. Since these higher harmonics are generated
by the RG, we are forced to consider a more general model
\end{multicols}
\begin{equation}
H=\int d^d r\bigg[\sum_{\alpha=1}^n\left({K\over2}(\nabla_\perp^2
u_\alpha)^2+{B\over2}\big(\partial_z u_\alpha-
{1\over2}(\bbox{\nabla}_\perp u_\alpha)^2\big)^2\right)
+{1\over T}\sum_{\alpha,\beta=1}^n
\left({1\over4}\Delta_h\left(u_{\alpha} - u_{\beta}\right)
|\bbox{\nabla_\perp}(u_\alpha-u_\beta)|^2 -
\Delta_V\left(u_{\alpha} - u_{\beta}\right)\right)\bigg]\;,
\label{Hr_anharmonic_rf2}
\end{equation}
\bottom{-2.7cm}
\begin{multicols}{2}
\noindent
where $\Delta_h(u_{\alpha} - u_{\beta})$ and $\Delta_V(u_{\alpha} -
u_{\beta})$ are arbitrary {\em periodic} functions (with period $a$,
where $a$ is the smectic layer spacing), describing the
field-dependent variances of the tilt and translational
disorders. Rather than following the flow for an infinite number of
couplings (one for each Fourier component of the random potentials) we
apply functional RG (FRG) methods developed by D.S. Fisher in the
context of the random field Ising and XY models\cite{FisherFRG} to our
study of disordered smectics.

\subsection{Functional RG for $5<d<7$: irrelevance of random-tilt disorder}
We integrate out the short scale phonon modes within an infinitesimal
momentum shell near the uv cutoff and study how the elastic moduli
$B(\ell)$, $K(\ell)$ and the disorder variance {\em function}
$\Delta_V(u,\ell)$ evolve under this RG transformation.  As discussed
in detail in Sec.\ref{anom_elasticity}, the tilt disorder $\Delta_h$
(characterized by the dimensionless coupling $g_2$,
Eq.\ref{g2dimensionless}) is irrelevant by simple power-counting for
$d>5$. We will show at the end of Appendix B that this conclusion
persists, even in the presence of random-field
\top{-2.7cm}

\noindent
disorder, although the argument is subtler. Hence for
$5<d<7$ we can focus on the random-field-only model of a randomly
pinned smectic.  We relegate the details of the FRG procedure to
Appendix B. For $5<d<7$, the results are summarized by the FRG flow
equations
\begin{eqnarray}
{dK\over d\ell}&=&\left[d - 3 +
{\left(d^2 - 12d + 23\right)\tilde{\Delta}_V^{\prime\prime}(0,\ell)\over2 
\left(d^2 - 1\right)\lambda^2(\ell)}\right]K
\label{Kfrg_flow}\;,\\
{dB\over d\ell}&=&\left[d - 3 + {3\over2}
{\tilde{\Delta}_V^{\prime\prime}(0,\ell)\over\lambda^2(\ell)}\right]B
\label{Bfrg_flow}\;,
\end{eqnarray}
and
\end{multicols}
\vspace{1cm}
\begin{equation}
\partial_{\ell}\tilde{\Delta}_V(u,\ell)=\epsilon\tilde{\Delta}_V(u,\ell) +
{\eta(\ell)\over q_0^2}\tilde{\Delta}_V^{\prime\prime}(u,\ell)
-{3(d^2 - 6d + 11)\over 2(d^2-1)\lambda^2(\ell)}
\tilde{\Delta}_V^{\prime\prime}(0,\ell)\tilde{\Delta}_V(u,\ell)
+{1\over2}\left(\tilde{\Delta}_V^{\prime\prime}(u,\ell)\right)^2 
-\tilde{\Delta}_V^{\prime\prime}(u,\ell)
\tilde{\Delta}_V^{\prime\prime}(0,\ell)\;,
\label{DeltaVfrg_flow}
\end{equation}
\bottom{-2.7cm}
\begin{multicols}{2}
where
\begin{equation}
\tilde{\Delta}_V(u,\ell)\equiv {C_{d-1}\Lambda^{d-7}
\over4\sqrt{K^3(\ell)B(\ell)}}\,\Delta_V(u,\ell)\;,
\label{tildeDelta}
\end{equation}
is the appropriate measure of random field disorder,
$\epsilon\equiv7-d$,
\begin{mathletters}
\begin{eqnarray}
\lambda(\ell)&=&\sqrt{K(\ell)\over B(\ell)}\;,\label{lambda_l}\\
\eta(\ell) &=& {q_0^2\,T\over 4\pi\sqrt{K(\ell)  B(\ell)}}\;,\label{eta_l}
\end{eqnarray}
\end{mathletters}
and primes in Eq.\ref{DeltaVfrg_flow} indicate a partial derivative
with respect to $u$.

Note that we have written the flow equations for $K(\ell)$ and
$B(\ell)$ in general $d$, rather than specializing to $d =
7-\epsilon$. The reason for this is that for {\em all} $d > 5$ (i.e.,
wherever we can ignore {\em tilt} disorder), the recursion relations
Eqs.\ref{Kfrg_flow} and \ref{Bfrg_flow} become asymptotically {\em
exact} as $\tilde{\Delta}^{\prime\prime}(0,\ell)/\lambda ^2(\ell)
\rightarrow 0$, which, as we will show in a moment, it always does for
$5 < d < 7$.

To demonstrate the above assertion we first assume (and verify a
posteriori) that near $d=7$, the flow equations,
Eqs.\ref{Kfrg_flow}-\ref{DeltaVfrg_flow} lead to a perturbative fixed
point at which
$\tilde{\Delta}_V(u,\ell\rightarrow\infty)\equiv\tilde{\Delta}_V^\ast(u)$
is a universal function of order $\epsilon$ with a finite (negative)
second derivative $\tilde{\Delta}_{V\ast}^{\prime\prime}(0)\approx
-q_0^2 O(\epsilon)$. Then, combining
Eqs.\ref{Kfrg_flow},\ref{Bfrg_flow} into a recursion relation for
$\lambda(\ell)$, as defined by Eq.\ref{lambda_l}, we obtain
\begin{equation}
{d \lambda \over d\ell} = - 
{\tilde{\Delta}_V^{\prime\prime}(0,\ell)\over\lambda(\ell)} 
\left({d^2 + 6d-13\over 2(d^2-1)}\right)
\label{lambda_flow}
\end{equation}
which, assuming that
$\tilde{\Delta}_V^{\prime\prime}(0,\ell\rightarrow\infty)
\rightarrow\tilde{\Delta}_{V\ast}^{\prime\prime}(0)<0$, has the
solution
\begin{equation}
\lambda(\ell)=\lambda _0 \sqrt{1 + \ell/\ell_0}\;,
\label{lambda_sol}
\end{equation}
with
\begin{mathletters}
\begin{eqnarray}
\lambda_0&\equiv&\lambda(\ell = 0)\;,\label{lambda0}\\
\ell_0&=&{\lambda_0^2\over|\tilde{\Delta}_{V\ast}^{\prime\prime}(0)|}
{d^2-1\over d^2+6d-13}\;.\label{ell0}
\end{eqnarray}
\end{mathletters}

The main result of this analysis is that at large length scales,
$\ell\gg\ell_0$, $\lambda^2(\ell)$ {\em grows} as a {\em
universal} function of $\ell$
\begin{equation}
\lambda^2(\ell\rightarrow\infty)={d^2+6d-13\over d^2-1}
|\tilde{\Delta}_{V\ast}^{\prime\prime}(0)|\;\ell\;.
\label{lambda2flow}
\end{equation}
Hence, as asserted above, as $\lambda^2(\ell)$ grows, the graphical
corrections to the flow of $K(\ell)$ and $B(\ell)$ diminish,
Eqs.\ref{Kfrg_flow},\ref{Bfrg_flow} become asymptotically {\em exact},
and the third term in Eq.\ref{DeltaVfrg_flow} vanishes as
$\ell\rightarrow\infty$. All this is still subject to the assumption
that $\tilde{\Delta}_{V\ast}^{\prime\prime}(0)\neq0$.

We can begin demonstrating this latter assumption by combining
Eqs.\ref{Kfrg_flow},\ref{Bfrg_flow} into a recursion relation for
$\eta(\ell)$, given by Eq.\ref{eta_l}. Comparing the resulting
equation with the generic flow equation for $\eta(\ell)$,
Eq.\ref{eta_flow}, we find that $\theta(\ell)$ (for $\ell\gg\ell_0$)
is given by a {\em universal} function
\begin{equation}
\theta(\ell)=d-3-\left({d^2-3d+5\over
d^2+6d-13}\right){1\over\ell}\;.\label{theta_universal}
\end{equation}
We note that besides being universal, $\theta(\ell)$ is furthermore
independent of the fixed point value
$\tilde{\Delta}_{V\ast}^{\prime\prime}(0)$, as long as it is finite.
As a result of Eq.\ref{theta_universal} we conclude that for $d>3$,
$\eta(\ell)$ is strongly irrelevant.  Physically, this reflects the
strong irrelevance of thermal fluctuations for $d>3$, with the
properties of the pinned smectic determined by a competition between
elastic and random pinning energies (both infinitely larger than the
thermal energy $k_B T$). As a result, the second term on the right
hand side of Eq.\ref{DeltaVfrg_flow} flows to zero exponentially fast,
and can therefore be dropped. Likewise, the $1/\lambda(\ell)^2$ term
in Eq.\ref{DeltaVfrg_flow} also vanishes as $\ell\rightarrow\infty$
(albeit more slowly), and can also be neglected. This leads to a
closed recursion relation for $\tilde{\Delta}_V(u,\ell)$, which has
been extensively studied in the context of the random field XY
model\cite{FisherFRG} and other pinned periodic
media\cite{NarayanFisher,GL}.

The fixed point solution $\tilde{\Delta}_{\ast}(u)$ is readily found
by setting $\partial_{\ell}\tilde{\Delta}_V(u,\ell) = 0$, which leads
to an ordinary differential equation for $\tilde{\Delta}_{V\ast}(u)$:
\begin{equation}
\epsilon\tilde{\Delta}_{V\ast}(u)
+{1\over2}\left(\tilde{\Delta}_{V\ast}''(u)\right)^2
-\tilde{\Delta}_{V\ast}''(u)\tilde{\Delta}_{V\ast}''(0)=0\;.
\label{Delta_fixedpoint}
\end{equation}

The exact form of $\tilde{\Delta}_{V\ast}(u)$ is not important for our
conclusions (except for one exponent); all that matters is that a
fixed point solution $\tilde{\Delta}_{V\ast}(u)$ exists (which it does
for $5<d<7$). However, for completeness we summarize the fixed-point
results for $\tilde{\Delta}_{V\ast}(u)$. Focusing, for convenience, on
the random {\em field} correlator
$\tilde{\Delta}_{F\ast}(u)\equiv\tilde{\Delta}_{V\ast}''(u)$, which
satisfies
\begin{equation}
\epsilon\tilde{\Delta}_{F\ast}(u)
+(\tilde{\Delta}_{F\ast}'(u))^2
+\tilde{\Delta}_{F\ast}''(u)\big[\tilde{\Delta}_{F\ast}(u)-
\tilde{\Delta}_{F\ast}(0)\big]=0\;,\label{DeltaF_fixedpoint}
\end{equation}
we obtain
\begin{equation}
  \tilde{\Delta}_{F\ast}(u) =-\min_{n \in {\cal Z}} \,{\epsilon\over q_0^2}
  \left\{ {1 \over 6}(q_0 u-(2n+1)\pi)^2 - {\pi^2 \over 18}
  \right\}\;, \label{DeltaFfp}
\end{equation}
which has cusps at $u=n 2\pi/q_0=n a$, $n$ integer. It is easy to
show\cite{FisherFRG} that at $T=0$ ($\eta=0$) the cusp non-analyticity
in $\tilde{\Delta}_{F\ast}(u)$ develops at {\em finite} ``time''
$\ell$, while for finite $T$ ($\eta\neq0$) this singularity is cutoff
within a boundary layer of thickness $\sim{\eta(\ell)/\epsilon
q_0}$ at the value
$\tilde{\Delta}_{F\ast}''(0,\ell)\approx\epsilon^2\pi^2/9\eta(\ell)$.
We also see from the solution for
$\tilde{\Delta}_{F\ast}(u)=\tilde{\Delta}_{V\ast}''(u)$,
Eq.\ref{DeltaFfp}, that
$\tilde{\Delta}_{V\ast}''(0)=-\epsilon\pi^2/9q_0^2 < 0$, thereby,
together with Eq.\ref{lambda2flow}, demonstrating that
$\lambda^2(\ell\rightarrow\infty)\rightarrow\infty$, as assumed
earlier.

Having established the existence of the zero temperature fixed point
for $5<d<7$, characterized by the universal form of the disorder
correlator, Eq.\ref{DeltaFfp}, we now turn to the implications of
these results for the elastic properties of the randomly pinned
smectic.

Using Eqs.\ref{lambda_l}-\ref{ell0} inside the flow equations for
$K(\ell)$ and $B(\ell)$, Eqs.\ref{Kfrg_flow},\ref{Bfrg_flow}, and
solving for these elastic moduli, we obtain
\begin{mathletters}
\begin{eqnarray}
K(\ell)&=&e^{(d-3)\ell} K_p(\ell)
\label{Ksol}\;,\\
B(\ell)&=&e^{(d-3)\ell} B_p(\ell)
\label{Bsol}\;,
\end{eqnarray}
\end{mathletters}
where the physically measurable moduli $B_p(\ell)$ and $K_p(\ell)$ are
\begin{mathletters}
\begin{eqnarray}
 K_p(\ell)&=& K_0 \left(1 + {\ell\over  \ell_0}\right)^{{\gamma_K}}
\label{Kpsol}\;,\\
B_p(\ell)&=& B_0 \left(1 + {\ell \over \ell_0}\right)^{-\gamma_B}
\label{Bpsol}\;,
\end{eqnarray}
\label{KBp}
\end{mathletters}
with
\begin{equation}
\gamma_{K}(d) = {1 \over 2}\left({-d^2  + 12d - 23 \over d^2 + 6d - 13}\right)
\label{gamma_K}\;,
\end{equation}
and
\begin{equation}
\gamma_{B}(d) = {3 \over 2}\left({d^2 - 1 \over d^2 + 6d - 13}\right)
\label{gamma_B}\;.
\end{equation}

In the range of interest the exponents $\gamma_K(d)$ and $\gamma_B(d)$
are monotonically decreasing and increasing functions of $d$,
respectively, and do not change significantly for $5<d<7$:
$\gamma_K(5)={1/7}$, $\gamma_K(7)={1/13}$,
$\gamma_B(5)={6/7}$, $\gamma_B(7)={12/13}$. 

As advertised, we therefore find (for $5<d<7$) that the randomly
pinned smectic exhibits logarithmically weak anomalous elasticity,
with the bend modulus $K_p$ diverging and the compression modulus
$B_p$ vanishing logarithmically at large length scales
$e^\ell\rightarrow\infty$.

With these results in hand, we can calculate the corresponding
smectic time-persistent correlation function
\begin{equation}
C({\bf k})\equiv{\overline{\langle u({\bf k})\rangle\langle u({\bf
k}')\rangle}\over\delta^d({\bf k}+{\bf k}')}\;,
\end{equation}
which determines the long-range spatial smectic correlations.
Following the RG matching method, described in detail in
Sec.\ref{anom_elasticity}, we relate $C({\bf k})$ at the small
wavevectors of interest to us, to the same quantity computed in the
effective theory with rescaled $\ell$-dependent couplings and
evaluated at larger wavevectors $e^\ell k_{\perp}$,
$e^{\omega\ell}k_z$:
\begin{equation}
C({\bf k})= e^{(d - 1 + \omega)\ell} C(e^\ell k_{\perp},
e^{\omega\ell}k_z, B(\ell), K(\ell), \Delta_V(u,\ell))\label{Cmatch1}\;.
\end{equation}
Here let us consider the limit $k_z<<\lambda k_\perp^2$.  Now, we set
(for convenience) $\omega = 2$ and choose $e^{\ell^*} =
{\Lambda/k_{\perp}}$. The latter condition guarantees that no
infra-red divergences appear in the perturbative calculation of the
right hand side of Eq.\ref{Cmatch1} (since it is done at large
wavevector $e^{\ell^*}k_\perp=\Lambda$ at the Brillouin zone
boundary). At these short length scales ($\Lambda^{-1}$), phonon
displacements $u$ are genuinely small and the random potential
variance $\Delta_V(u_\alpha-u_\beta)$ (Eq.\ref{Hr_anharmonic_rf2}) can
be safely expanded in a convergent power series in
$u_\alpha-u_\beta$. Expanding the Hamiltonian
Eq.\ref{Hr_anharmonic_rf2} to the lowest (quadratic) nontrivial order
maps the full nonlinear FRG problem at these high wavevectors onto the
random tilt-only model Eq.\ref{Hr_anharmonic} with effective
$k$-dependent couplings $K(\ell^*)$, $B(\ell^*)$ and
$\Delta_h^{\mbox{eff}}(\ell^*)$ given by
\begin{equation}
\Delta_h^{\mbox{eff}}(\ell^*)
=-2\Delta_V''(0,\ell^*)/\Lambda^2\;,\label{Delta_heff}
\end{equation}
($\Delta_V''(0,\ell^*)<0$, as we show below).  Using this effective
theory and Eqs.\ref{Propdef},\ref{Gab} to compute the right hand side
of Eq.\ref{Cmatch1}, and taking $k_\perp$ sufficiently small, so that
$\ell^*$ is sufficiently large to guarantee that the fixed point
Eq.\ref{DeltaFfp} is reached, we readily obtain
\begin{mathletters}
\begin{eqnarray}
C({\bf k})&\approx&{-2e^{(d + 1)\ell^*}\Delta_V''(0,\ell^* ) \over
\left[K_p(\ell^*)e^{(d-3)\ell^*}\Lambda^4
+B_p(\ell^*)e^{(d - 3)\ell^*} e^{4\ell^*}k_z^2\right]^2}\;,\nonumber\\
&&\label{Cmatch2a}\\
&\approx&{-8\tilde{\Delta}_{V\ast}''(0)
\sqrt{B_p(\ell^*)K_p(\ell^*)^3}\,k_\perp^{7-d}
\over\left[K_p(\ell^*)k_\perp^4+B_p(\ell^*)k_z^2\right]^2C_{d-1}}
\;.\nonumber\\
&&\label{Cmatch2b}
\end{eqnarray}
\end{mathletters}
To obtain Eq.\ref{Cmatch2b} above, we used Eq.\ref{tildeDelta} to
relate $\Delta_V$ to $\tilde{\Delta}_V$ and Eq.\ref{KBp} to express
our results in terms of the physical elastic moduli $K_p$ and $B_p$.
Comparing our final expression for $C({\bf k})$ with the expression
for this quantity in Gaussian theory Eq.\ref{Gab} justifies our
calling $K_p(\ell^*)$ and $B_p(\ell^*)$ the ``physical'' $K$ and $B$
of the nonlinear theory of the randomly pinned smectic. Using
Eqs.\ref{KBp} for these, in the limit $\ell^*\gg\ell_0$, with
$\ell^*=\ln(\Lambda/k_\perp)$, and $k_z<<\lambda k_\perp^2$ we find
the {\em anomalous} moduli
\begin{mathletters}
\begin{eqnarray}
K({\bf k})&\propto&\left|\ln(k_{\perp}/\Lambda)\right|^{\gamma_K(d)}
\label{Kfrg_kperp}\;,\\
B({\bf k})&\propto&\left|\ln(k_{\perp}/\Lambda)\right|^{-\gamma_B(d)}
\label{Bfrg_kperp}\;.
\end{eqnarray}
\label{BKfrg_kperp}
\end{mathletters}
Similar analysis for an arbitrary direction in $k_\perp-k_z$ space
leads to Eqs.\ref{BKfrg}, advertised in the Introduction,
Sec.\ref{introduction}.

The Fourier transform of $C({\bf k})$ measures the spatial phonon correlations
\begin{eqnarray}
C({\bf r})&\equiv&\overline{\langle\left[u({\bf r_\perp},z)-
u({\bf 0_\perp},0)\right]^2\rangle}\;,\nonumber\\
&\approx&\int{d^{d}k\over(2\pi)^{d}}2(1-\cos({\bf k}\cdot 
{\bf r}))C({\bf k})\label{Cr_perp}\;.
\end{eqnarray}
Using Eq.\ref{Cmatch2b}, but ignoring the subdominant ($\ln(\ln r)$)
corrections arising from weak anomalous elasticity, we obtain for $z=0$
\begin{eqnarray}
C({\bf r}_\perp,0)&\approx& {-8\tilde{\Delta}_{V*}''(0)\over C_{d-1}}
\int_{k_\perp<r_\perp^{-1}}{d^d k\over(2\pi)^d}\,
{\sqrt{B({\bf k})K({\bf k})^3}\,k^{7-d}_{\perp}\over
\left[K({\bf k})k^4_{\perp}+B({\bf k})k^2_z\right]^2}\;,\nonumber\\
&\approx&{-4\tilde{\Delta}_{V*}''(0)\over C_{d-1}}
\int_{k_\perp<r_\perp^{-1}}{d^{d-1}k_\perp\over(2\pi)^{d-1}}
{1\over k_{\perp}^{d-1}}\;,\nonumber\\
&=&{4\epsilon\pi^2\over9 q_0^2}\ln(r_{\perp}/a)\;.\label{u2}
\end{eqnarray}
We therefore find that throughout the range of dimensions $5<d<7$, the
smectic displacements $u$ grow logarithmically (up to $\ln\ln$
corrections due to anomalous elasticity), with a universal amplitude
determined by Eq.\ref{DeltaFfp}. This super-universal logarithmic
roughness is analogous to that previously found for the random field
XY model\cite{VillainFernandez} and for the Abrikosov vortex lattice
in a dirty superconductor\cite{GL}.

The super-universal smectic Bragg glass phase found in this section is
quite exotic.  However, since it only exists for $5<d<7$, it
unfortunately has little experimental relevance for physical
3-dimensional disordered smectics.  Furthermore, because the {\em
periodic} disorder-driven layer fluctuations and the related anomalous
elasticity diverge only {\em logarithmically}, while the {\em tilt}
disorder, which becomes relevant for $d < 5$, leads to power-law
roughness, Eq.\ref{CDelta}, the former is completely swamped by the
latter in $d<5$.

\subsection{Functional RG for $3<d<5$: irrelevance of random-field disorder}

The above claim can readily be demonstrated by extending the
functional renormalization group analysis presented above and in
Appendix B to $d<5$.  The resulting set of flow equations for
$K(\ell)$, $B(\ell)$, and $\Delta_h(u,\ell)$, valid for $d<5$
(assuming the validity of the {\em elastic} model,
Eq.\ref{Hr_anharmonic_rf2}, i.e., that dislocations are bound) is
\begin{eqnarray}
{d K(\ell)\over d\ell}&=&\left[d - 3 + {1\over32}g(0,\ell)+
{\left(d^2 - 12d + 23\right)\tilde{\Delta}_V^{\prime\prime}(0,\ell)\over2 
\left(d^2 - 1\right)\lambda^2(\ell)}\right]K
\label{Kfrg_flow2}\;,\nonumber\\
&&\\
{d B(\ell)\over d\ell}&=&\left[d - 3 - {3\over16}g(0,\ell) + {3\over2}
{\tilde{\Delta}_V^{\prime\prime}(0,\ell)\over\lambda^2(\ell)}\right]B
\label{Bfrg_flow2}\;,\nonumber\\
\end{eqnarray}
\end{multicols}
\begin{equation}
\partial_{\ell}\Delta_h(u,\ell)=\left(d-1+{1\over64}
g(u,\ell)\right)\Delta_h(u,\ell) + \bigg({\eta(\ell)\over q_0^2}+
D(u)\bigg)\Delta_h''(u,\ell)+
c\left(\tilde{\Delta}_{V*}'''(u)\right)^2\sqrt{K(\ell)^3
B(\ell)}\Lambda^{5-d}/C_{d-1}\;,
\label{Delta_hfrg_flow2}
\end{equation}
\bottom{-2.7cm}
\begin{multicols}{2}
\noindent
where we have defined
\begin{equation}
g(u,\ell)\equiv\Delta_h(u,\ell)\left(B(\ell)\over
K^5(\ell)\right)^{1/2} C_{d-1}\Lambda^{d-5}
\;,\label{g2udimensionless_text}
\end{equation}
the ``position-dependent diffusivity''
\begin{equation}
D(u)\equiv c_2\left(\tilde{\Delta}_{V*}''(u)-\tilde{\Delta}_{V*}''(0)\right)\;,
\label{Diffusivity}
\end{equation}
and $c$ and $c_2$ are dimensionless constants of order unity.  Note
that these recursion relations reduce to those of the tilt-only model
considered in Sec.\ref{anom_elasticity}, if we take the tilt disorder
$\Delta_h(u,\ell)$ to be a constant independent of $u$, and drop the
random field disorder. In these relations, for convenience, we have
chosen the anisotropy exponent $\omega=2$, so that, as discussed
earlier, $q_0$ does not renormalize.

The recursion relation for the dimensionless random field disorder
$\tilde{\Delta}_V$ is given by Eq.\ref{DeltaVfrg_flow}. Note that
since we are now interested in $d<5$, $\tilde{\Delta}_V$ may grow
quite large (since its linear eigenvalue $\epsilon=7-d$ is no longer
small), so we cannot neglect the $O(\tilde{\Delta}_V^3)$
non-linearities in Eq.\ref{DeltaVfrg_flow}. Because we do not know
these, we cannot actually calculate the fixed point function
$\tilde{\Delta}_{V*}(u)$ that $\tilde{\Delta}_V(u,\ell)$ flows to upon
renormalization. However, as in $d>5$, we do not need to. Rather, we
only need to argue that it exists, is finite, and that its second
derivative at the origin, $\tilde{\Delta}_{V*}''(0)$ is finite and
negative.

To make this argument, we first note that if we drop the second and
third terms on the right hand side of Eq.\ref{DeltaVfrg_flow} and
ignore higher order nonlinear terms, the resulting equation certainly
has a fixed point solution; namely, our explicit solution
Eq.\ref{DeltaFfp}. Now let us consider the effects of the other terms.
The $\eta(\ell)$ term can only {\em reduce} $\tilde{\Delta}_V$, being
a ``diffusive'' type of term. So, while it may drive
$\tilde{\Delta}_V(u,\ell)$ to zero as $\ell\rightarrow\infty$, it
certainly {\em cannot} have the opposite effect of driving
$\tilde{\Delta}_V$ to infinity. Hence $\tilde{\Delta}_{V*}(u)$, and
therefore, $\tilde{\Delta}_{V*}''(0)$, remain finite in the presence
of this term.

If we assume that $\tilde{\Delta}_{V*}(u)$ {\em is} finite, we can
show a posteriori that the third term renormalizes exponentially (in
renormalization group ``time'' $\ell$) to zero, since, as we will show
in a moment, $\lambda(\ell)\rightarrow\infty$ exponentially in $\ell$.

This should be contrasted with the behavior of this term in $5<d<7$,
where it vanished much more slowly (like $1/\ell$) as
$\ell\rightarrow\infty$. The more rapid vanishing of this term in
$d<5$ is a consequence of the stronger renormalization of $B$ and $K$
due to the {\em tilt} disorder field ($\Delta_h$),\top{-2.7cm}

\noindent
which becomes relevant for $d<5$.

The crucial point here is that this much stronger renormalization of
$B$ and $K$ due to this relevance of the tilt field ($\Delta_h$) does
{\em not} destabilize the $\tilde{\Delta}_{V}$ fixed point. Since this
relevance of the tilt field is the only qualitative difference between
$d>5$ and $d<5$, showing that it does not destabilize the
$\tilde{\Delta}_{V}$ fixed point by itself shows that there is no more
reason to doubt the stability of this fixed point for $d<5$ than for
$5<d<7$. Indeed, if anything, the argument is {\em strengthened},
since the $1/\lambda(\ell)$ term vanishes even more rapidly for $d<5$.

Hence, the only possible problem can come from the unknown higher
(than second) order terms in $\tilde{\Delta}_{V}$, in
Eq.\ref{DeltaVfrg_flow}. Not knowing these terms, we cannot make a
very compelling argument that they do not destabilize the fixed
point. However, the same objection could have been raised for $d>5$
but well below $7$. Indeed, an analogous objection can be raised for
{\em any} fixed point that is found in an $\epsilon$-expansion once
one goes well below the upper-critical dimension. We simply assume (as
is done in all standard $\epsilon$-expansions) that the stability of
the FRG fixed point, rigorously derived for $\epsilon\rightarrow0$,
persists well below the upper-critical dimension, i.e., for $d<5$.

So it seems plausible to assume that a finite fixed point function
$\tilde{\Delta}_{V*}(u)$, with finite $\tilde{\Delta}_{V*}''(0)\leq0$,
exists for $d<5$. Let us now investigate the consequences of this
assumption. We will see in a moment that they are virtually
non-existent: the asymptotic RG flows of $B(\ell)$, $K(\ell)$, and
$\Delta_h(\ell)$, and long distance behavior of the static smectic
correlation functions that they imply, are {\em exactly} those we
found in Sec.\ref{anom_elasticity}, where we neglected the periodic
potential altogether.

To see this, let us take $\tilde{\Delta}_{V}(u,\ell)$ to have flown to
its fixed point, and combine the recursion relations for $B(\ell)$,
$K(\ell)$, and $\Delta_h(u,\ell)$ into recursion relations for the
dimensionless coupling function $g(u,\ell)$, defined in
Eq.\ref{g2udimensionless_text}, which evolves under the RG
transformation according to
\end{multicols}
\begin{equation}
\partial_{\ell}g(u,\ell)=\left(5-d-{5\over32}g(u,\ell)
+c_3{\tilde{\Delta}_{V*}''(0)\over
\lambda^2(\ell)}\right)g(u,\ell)+\left({\eta(\ell)\over q_0^2}
+D(u)\right)g''(u,\ell)+
c\ {\left(\tilde{\Delta}_{V*}'''(u)\right)^2\over\lambda^2(\ell)}\;,
\label{gfrg_flow_text}
\end{equation}
\bottom{-2.7cm}
\begin{multicols}{2}
\noindent
where we have defined $c_3\equiv
3/4+(d^2-12d+23)/(2(d^2-1))$. Similarly, from
Eqs.\ref{Kfrg_flow2},\ref{Bfrg_flow2}, we obtain the recursion
equation for $1/\lambda^2(\ell)\equiv B(\ell)/K(\ell)$
\begin{equation}
\frac{d\lambda^{-2}(\ell)}{d\ell}=-{7\over32}g(0,\ell)\lambda^{-2}(\ell)+
a_1\tilde{\Delta}_{V*}''(0)\lambda^{-4}(\ell)\;,
\label{lambda^2flow}
\end{equation}
where $a_1$ is a constant whose numerical values is unimportant for
our argument.

We will now prove that for $d<5$, at long scales (large $\ell$),
$g(u,\ell)$ flows to $g_2^*=32(5-d)/5$, the constant, $u$-independent
fixed point value of the dimensionless coupling $g_2$,
Eq.\ref{g2dimensionless}, that we found in our earlier, tilt-only
treatment of the disordered smectic, described in
Sec.\ref{anom_elasticity}.

To see this, we write
\begin{mathletters}
\begin{eqnarray}
g(u,\ell)&=&g_2^*+\delta g(u,\ell)\;,\\
&=&{32\over5}(5-d)+\delta g(u,\ell)\;,
\end{eqnarray}
\label{linearize}
\end{mathletters}
and expand the recursion relation Eq.\ref{gfrg_flow_text} to linear
order in $\delta g(u,\ell)$, obtaining:
\begin{equation}
\partial_{\ell}\delta
g(u,\ell)=\left[d-5+D(u)\partial_u^2\right]\delta g(u,\ell)
+{S(u)\over\lambda^2(\ell)}\;,
\label{delta_g_flow}
\end{equation}
where we have defined the ``source term''
\begin{equation}
S(u)\equiv c_3 g_2^*\tilde{\Delta}_{V*}''(0)+ 
c\left(\tilde{\Delta}_{V*}'''(u)\right)^2\;,
\label{S(u)}
\end{equation}
and have dropped the $\eta(\ell)$ term which decays away exponentially
to $0$ as $\ell\rightarrow\infty$.

To demonstrate the stability of the tilt-only $u$-independent fixed
point, we now only need to show that Eq.\ref{delta_g_flow} implies
that $\delta g(u,\ell)$ decays to $0$ as $\ell\rightarrow\infty$. To
do this we expand $\delta g(u,\ell)$ in eigenfunctions of the operator
$D(u)\partial_u^2$; that is, we write
\begin{equation}
\delta g(u,\ell)\equiv \sum_{n=0}^\infty g_n(\ell) \phi_n(u)\;,
\label{expand}
\end{equation}
where $\phi_n(u)$ are periodic functions of $u$ with period $a$
satisfying the eigenvalue equation
\begin{equation}
D(u){d^2\phi_n(u)\over d u^2}=\Gamma_n\phi_n(u)\;.
\label{eigenvalue_eqn}
\end{equation}

It is straightforward to see that all of the eigenvalues $\Gamma_n$
are $\leq 0$, provided that $D(u)\geq0$ for all $u$, as it is in
$7-\epsilon$ dimensions, where we can explicitly calculate it. We will
assume that $D(u)$ continues to be $\geq0$ for $d<5$, and, in fact,
all the way down to the physical dimension $d=3$.

Using the orthogonality of the $\phi_n(u)$ basis, with
\begin{equation}
\int_0^a{\phi_n(u)\phi_{n'}(u)\over D(u)} d u=I_n \delta_{n n'}\;,
\label{orthogonality}
\end{equation}
and
\begin{equation}
I_n\equiv\int_0^a{\phi_n^2(u)\over D(u)} d u\;,
\label{normalize}
\end{equation}
standard projection analysis leads to a set of decoupled recursion
relations for the $\ell$-dependent expansion coefficients $g_n(\ell)$
in Eq.\ref{expand}
\begin{equation}
{d g_n(\ell)\over d\ell}=\left(d-5+\Gamma_n\right) g_n(\ell)+
{S_n\over\lambda^2(\ell)}\;,
\end{equation}
where the source term is given by
\begin{equation}
S_n\equiv\int_0^a{S(u)\phi_n(u)\over I_n D(u)} d u\;.
\label{Sproject}
\end{equation}

Now, since all of the $\Gamma_n$ are $\leq0$, and we are considering
$d<5$, and since, as we show momentarily,
$\lambda(\ell)\rightarrow\infty$ as $\ell\rightarrow\infty$, we see
that all of the $g_n(\ell)$ flow to zero as $\ell\rightarrow\infty$,
provided only that all of the $S_n$, Eq.\ref{Sproject} are finite.  We
leave the demonstration of the latter fact\cite{D0comment} as a
straightforward homework exercise.  Hence as asserted, $g(u,\ell)$
flows to a stable fixed point $g^*(u)=g_2^*=32(5-d)/5$, identical to
that found in the tilt-only model of Sec.\ref{anom_elasticity}.

It is now easy to complete the proof, by demonstrating via
Eq.\ref{lambda^2flow} that near this fixed point $\lambda^{-2}(\ell)$
flows rapidly to zero:
\begin{mathletters}
\begin{eqnarray}
\lambda^{-2}(\ell)&=&\lambda^{-2}(0)e^{-{7g_2^*\over32}\ell}\;,\\
&=&\lambda^{-2}(0)e^{-{7(5-d)\over5}\ell}\;.
\end{eqnarray}
\end{mathletters}
Although this result holds only near $d=5$, it is straightforward to
generalize the above analysis to arbitrary $d<5$, by relating the
graphical corrections to $B$ and $K$ to the exact anomalous exponents
$\eta_B$ and $\eta_K$ of the tilt-only model. We find that
$g_2(\ell\rightarrow\infty)$ again flows to the same fixed point value
it would have in the absence of the periodic potential, while
$\lambda^{-2}(\ell)$ now flows to zero like
\begin{equation}
\lambda^{-2}(\ell)=\lambda^{-2}(0)e^{-(\eta_B+\eta_K)\ell}\;.
\end{equation}

So in all dimensions $d<5$, $\lambda^{-2}(\ell)$ flows exponentially
to zero, in contrast to $5<d<7$, where $\lambda^{-2}(\ell)$ vanishes
slowly, like $1/\ell$. This completes the a posteriori argument in our
earlier demonstration that $\tilde{\Delta}_{V}(u,\ell)$ flows to a
finite fixed point and that the tilt disorder $g(u)$ flows to a
stable, $u$-independent fixed point identical to that found for the
tilt-only model.

All that remains to complete our argument that the random periodic
potential has no effect on the static properties of disordered
smectics in $d<5$ is to consider the correlation functions
themselves. Using yet again the trajectory integral matching
formalism, we can show that
\begin{eqnarray}
C({\bf q})&\equiv&\overline{\frac{\langle u({\bf q})u({\bf q}')\rangle}
{\delta^d({\bf q}+{\bf q}')}}\;,\\
&=&\left(\Lambda\over q_\perp\right)^{d+1}
\Bigg[{T\over K(\ell^*)\Lambda^4+B(\ell^*)(\Lambda/q_\perp)^4q_z^2}+
\nonumber\\
&&\hspace{.6in}{\Delta_h(\ell^*)\Lambda^2-2\Delta_V''(0,\ell^*)
\over\left[K (\ell^*)\Lambda^4
+B(\ell^*)(\Lambda/q_\perp)^4q_z^2\right]^2}\Bigg]\;.\nonumber\\
\end{eqnarray}
with $\ell^*\equiv\ln(\Lambda/q_\perp)$. This only differs from our
expression for this correlation function in the tilt-only model,
Eq.\ref{Cz}, by the presence of the constant term
$\Delta_V''(0,\ell^*)$ in the numerator of the second term. Using
\begin{mathletters}
\begin{eqnarray}
\Delta_h(\ell)&=&g_2(\ell){K^{5/2}(\ell)\over
B^{1/2}(\ell)C_{d-1}\Lambda^{d-5}}\;,\\ &\rightarrow&
g_2^*{K^{5/2}(\ell)\over B^{1/2}(\ell)C_{d-1}\Lambda^{d-5}}\;,
\end{eqnarray}
\end{mathletters}
and
\begin{equation}
\Delta_V(u,\ell)\rightarrow_{\ell\rightarrow\infty}{4\Lambda^{7-d}\over
C_{d-1}}K^{3/2}(\ell)B^{1/2}(\ell)\tilde{\Delta}_{V*}(u)\;,
\end{equation}
it is easy to show that the ratio of this $\Delta_V''(0,\ell^*)$ term
to $\Delta_h(\ell_*)\Lambda^2$ is 
\begin{equation}
{\tilde{\Delta}_{V*}''(0)\over g_2^*}{B(\ell_*)\over K(\ell_*)} ={{\rm
const.}\over\lambda(\ell_*)^{2}}\;,
\end{equation}
which vanishes as $q_\perp\rightarrow0$ and $\ell_*\rightarrow\infty$,
since $\lambda^{-2}(\ell_*)$ does.

To summarize: {\em every} possible {\em static} effect of the periodic
potential $\Delta_V$ vanishes, as $\ell\rightarrow\infty$, like
$\lambda^{-2}(\ell)$, which vanishes exponentially as
$\ell\rightarrow\infty$. Hence, the periodic potential (the random
field disorder) has no effect on any {\em static} properties of the
disordered smectic in $d<5$. Our tilt-only model is sufficient to treat those.

All of the above arguments, of course, only apply if we ignore
dislocations. In the next section we include these in our analysis.

\section{Failure of periodic potential to induce dislocation unbinding}
\label{failure}

In Sections \ref{stability_harmonic_elastic}-\ref{tilt-only_model} and
\ref{random_field_3-5} we have demonstrated that in three-dimensional
(or more generally for $d<5$) randomly pinned smectics, the tilt
disorder ($\Delta_h$) asymptotically leads to significantly larger
layer disordering than the random field displacement disorder
($\Delta_V$). We thereby argued that static correlation functions for
a general disordered smectic can be accurately computed within a
tilt-disorder-only model, studied in Sections
\ref{tilt-only_model}-\ref{anom_elasticity_dislocations_correlations}.
However, the validity of these arguments and the conclusions that
follow from them critically rely on the stability of the {\em elastic}
model within which these assessments are made.  That is, it is in
principle possible that although the periodic disorder is less
important than the random tilt pinning in disordering smectic layers
in the {\em elastic} model, it could nevertheless drive dislocation
unbinding, thereby invalidating the elastic model and the conclusions
drawn from it.

Therefore, to make the arguments of the previous sections complete, it
is essential to assess whether the random-field disorder ($\Delta_V$)
{\em alone} can drive proliferation of dislocation loops, and this is
the subject of current section.

To this end we adapt D.S. Fisher's\cite{commentRealSpaceRG} real space
renormalization group analysis to disordered smectics and show that
the random periodic disorder {\em alone} does {\em not} induce a
proliferation of dislocation loops in three-dimensional smectics,
thereby concluding our justification for the tilt-only model of
randomly pinned smectics.

We first review Fisher's analysis as it applies to the
three-dimensional random-field XY model. Recall that within a
defect-free model the elastic energy is balanced against the pinning
energy, both much larger than the thermal energy $k_B T$. The
compromise in this competition leads to phase variations that grow
logarithmically with distance and typical ground-state energies that
vary as $L^\theta$, where $L$ is system size and $\theta$ is (minus)
of the thermal eigenvalue exponent, which, due to statistical
symmetry, is given exactly by
\begin{equation}
\theta_{xy}=d-2\;.
\end{equation}
We wish to know whether it energetically pays for the system to insert
a vortex loop. Confining our discussion to $d=3$, a vortex loop of
length $L$ will cost the system elastic energy $E_{\rm el}\sim J L\ln
L$, where the multiplicative logarithm is associated with the
long-range XY phase deformation around the vortex line and $J$ is the
spin-wave stiffness. For such a randomly placed vortex loop, which
relieves strains by allowing spin rotations of order $2\pi$, the
system can typically gain disorder energy $E_{\rm disorder}\sim
L^\theta\sim L$. Hence, this crude argument would suggest that,
because of the additional multiplicative logarithmic factor in $E_{\rm
el}$, it never pays to insert arbitrarily large vortex loops into the
random-field XY model. However, as argued by Fisher, a judicious
optimization of the vortex loop location and conformation can do
better energetically, and therefore it is essential to include in the
stability analysis the energy minimization over the vortex position
and conformation.

To this end we start out with a circular vortex loop and, as
illustrated in Fig.12, try to lower its energy by moving various
segments of length $\Lambda_w$ of it first by a distance $w$, then by
$2w$, etc.  We then calculate the renormalization of the line tension,
$\epsilon(\ell)$, by pinning at scale $2^\ell w$.
\begin{figure}[bth]
\begin{center}
\setlength{\unitlength}{1mm}
\begin{picture}(80,80)(0,0)
\put(-30,-55){\begin{picture}(30,30)(0,0)
\includegraphics{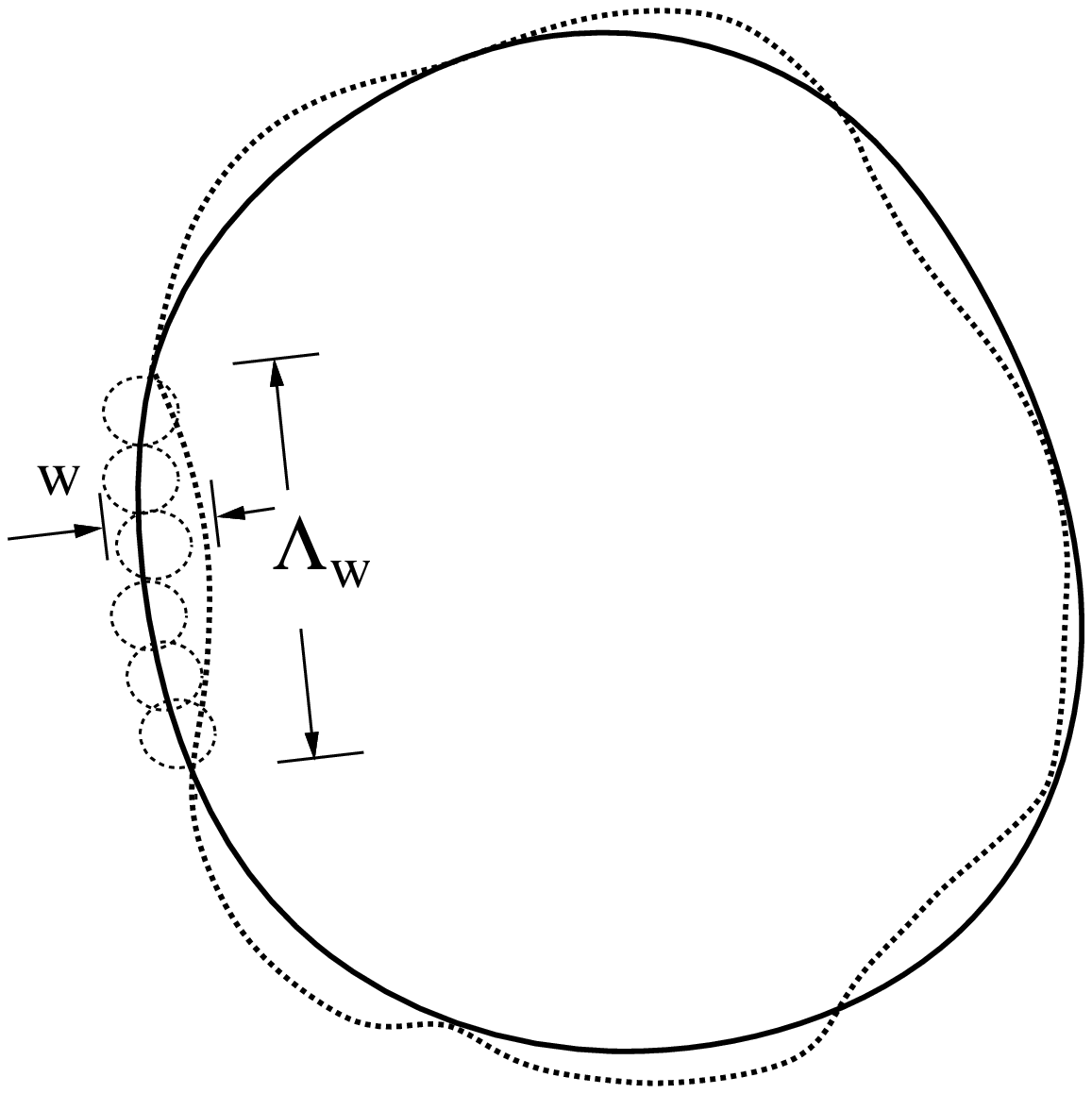}
\end{picture}}
\end{picture}
\end{center}
Fig.12. {Illustration of the optimization of a topological defect loop
on length scale $\Lambda_w$. The full and dashed large loops represent
the topological defect line before and after optimization on scale
$\Lambda_w$, respectively, and the small circles are the isotropic
correlated regions corresponding to displacements $w$.}
\label{RealSpaceRGrfxy}
\end{figure}
The idea is that each roughly spherical region of size $w$ is
statistically independent of all the others; each gains a pinning
energy of the order $w^{\theta}=w^{d-2}=w$ $({\rm in}\; d=3)$.  So the
total pinning energy gained from optimizing the conformation of the
vortex loop on scale $w$ is given by
\begin{mathletters}
\begin{eqnarray}
\delta E_{\rm pin}&\sim&-w \sqrt{N_{spheres}}\;,\\
&\sim&-w {\sqrt{\Lambda_w \over w}}\;,\\
&\sim&-\sqrt{\Lambda_w w}\;.
\label{EpinFisher}
\end{eqnarray}
\end{mathletters}
Balancing this energy against the stretching energy necessary to
deform the vortex loop
\begin{equation}
\delta E_{\rm stretch}\sim\epsilon {w^2\over \Lambda_w}
\label{EstretchFisher}
\end{equation}
gives
\begin{equation}
\Lambda^{3/2}_w\sim\epsilon w^{3/2}\;,
\end{equation}
which predicts the aspect ratio of the energetically most favorable
deformation of the vortex loop
\begin{equation}
\Lambda_w\sim\epsilon^{2/3}w\;.
\end{equation}
Using this in the expression for the pinning energy, we get the best
pinning energy gain of
\begin{equation}
\delta E_{\rm pin}\sim-\epsilon^{1/3}w\;.
\end{equation}
Associating this energetic reduction with the renormalization of the
vortex line tension $\epsilon(\ell)$ gives
\begin{eqnarray}
\delta \epsilon = 
{\delta E_{\rm el} + \delta E_{\rm pin} \over \Lambda_w} = J
\ln 2 - {\epsilon ^{1/3} w \over \epsilon ^{2/3} w}
\label{flowRG}
\end{eqnarray}
where the first term arises from the elastic energy increase due to
the factor of $2$ increase in length scale.  This result implies the
$RG$ recursion relation
\begin{equation}
{d\epsilon(\ell)\over d \ell} = J - {1\over\epsilon(\ell)^{1/3}}\;.
\label{flowRG2}
\end{equation}
Clearly, if the bare vortex line tension $\epsilon(0)\equiv\epsilon$
is sufficiently large, the line tension $\epsilon(\ell)$ at scale
$e^\ell$ grows like $\ell$ ($\approx\ln L$), and therefore it is
energetically unfavorable for vortex loops to unbind.  This therefore
argues for the stability of the topologically ordered ``Bragg'' glass
phase in a weakly disordered three-dimensional random-field XY model.

We now extend this analysis to smectics pinned by a random periodic
potential, with the main difference from the random-field XY model
being the anisotropy of the smectic state. We begin by calculating the
thermal eigenvalue exponents $-\theta$.  As was the case for the
random-field XY model, statistical symmetry guarantees that the
smectic elastic moduli $B$ and $K$ are not renormalized by
disorder.\cite{commentTheta} This therefore implies that the $\theta$
exponent can be computed {\em exactly} by estimating the elastic
energy due to layer displacement $u$ of order a lattice constant,
$O(a)$, inside an anisotropic volume $L_\perp\times L_\perp\times
(L_z\sim L_\perp^2)$.
\begin{mathletters}
\begin{eqnarray}
E&\sim&B \int dz d^2r_{\perp}\left(\partial_zu\right)^2\;,\\
&\propto& L^2_{\perp}  L^2_{\perp}{1\over L^4_{\perp}}\;,\\
&\sim&(L_{\perp})^0\;.
\end{eqnarray}
\label{theta1}
\end{mathletters}
In this estimate Eq.\ref{theta1} we have used the fact that, in the
{\em harmonic} smectic elastic theory, distances in the $z$-direction
scale like the square of those in the $\perp$ directions.  In three
dimensions, we therefore find that the ground state of a randomly
pinned, elastically {\em harmonic} smectic is described by
\begin{equation}
\theta_{\rm smectic} = 0\;.
\end{equation}

We now need to estimate the pinning energy gain from introducing a
dislocation loop and optimizing its configuration, analogously to the
analysis for a vortex loop above. The main difference here, however,
is that, because of the anisotropy of the smectic phase, there are
three dislocation line optimization cases to consider:
\begin{enumerate}
\item Vertical (along $z$) $w$ displacement of a horizontal
(perpendicular to $z$) segment (i.e., an edge dislocation).
\item Horizontal $w$ displacement of a horizontal segment.
\item Horizontal $w$ displacement of a vertical segment (i.e., a screw
dislocation).
\end{enumerate}
\subsection{Case 1: Vertical displacement of a horizontal dislocation ($w
\parallel z$)}

Because of the elastic anisotropy of the smectic state, in contrast to
the spherical correlated volumes of the random-field XY model, here
correlated regions are cigar-shaped (directed along $z$) characterized
by dimensions $L_{\perp}(w)\times L_{\perp}(w)\times w$, with
$L_{\perp}(w)\sim \sqrt{w}$, as shown in Fig.13.
\begin{figure}[bth]
\begin{center}
\setlength{\unitlength}{1mm}
\begin{picture}(50,50)(0,0)
\put(-50,-90){\begin{picture}(30,30)(0,0)
\includegraphics{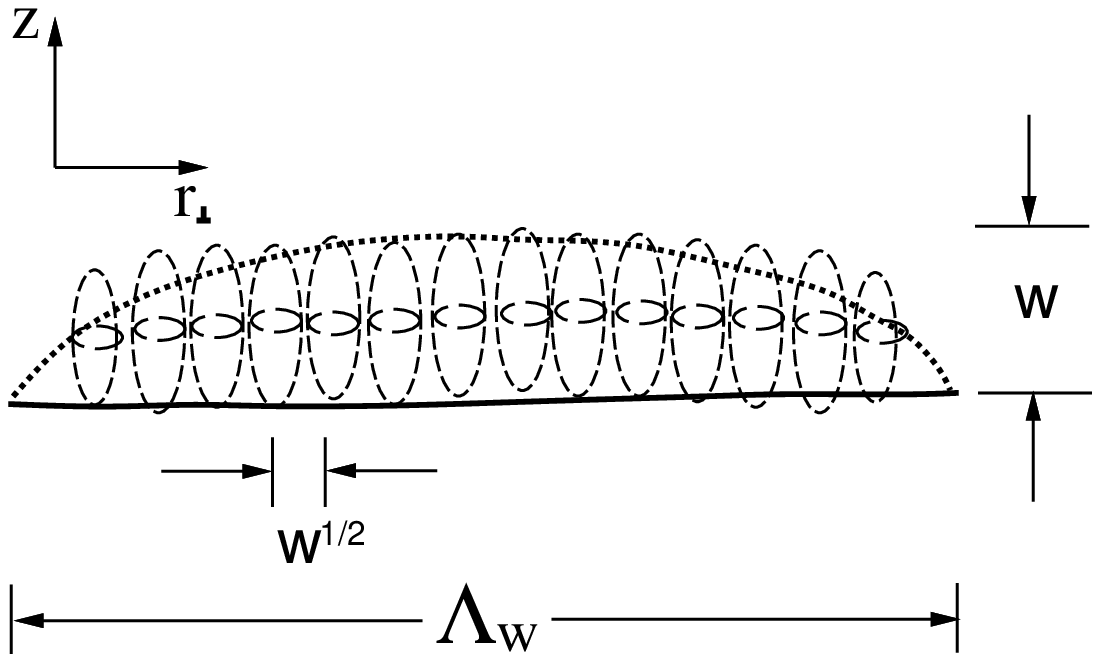}
\end{picture}}
\end{picture}
\end{center}
Fig.13. {Illustration of smectic anisotropic cigar-shaped correlation
regions inside the dislocation optimization length $\Lambda_w$ for a
displacement $w$ along $z$ of a dislocation running perpendicular to
$z$ (case 1).}
\label{case1RG}
\end{figure}
The corresponding pinning energy change due to optimization of a
dislocation segment of length $\Lambda_w$ is given by
\begin{mathletters}
\begin{eqnarray}
\delta E_{\rm pin}&\sim&- w^{\theta}\sqrt{\Lambda_w\over L_{\perp}(w)}\;,\\
&\sim&- w^0\sqrt{\Lambda_w \over \sqrt{w}}\;,\\
&\sim&-{\Lambda^{1/2}_w \over w^{1/4}}
\end{eqnarray}
\label{Epin1}
\end{mathletters}
Balancing this against the dislocation stretching energy, which, as
always, is
\begin{equation}
\delta E_{\rm stretch}\sim\epsilon{w^2 \over \Lambda_w}\;,
\label{Estretch1}
\end{equation}
we get
\begin{equation}
\Lambda_w = \epsilon^{2/3}w^{3/2}\;.
\label{Lambda_w1}
\end{equation}
We note that $\Lambda_w>>L_\perp(w)\sim\sqrt w$, consistent with the
assumption of many ``bubbles'' per segment $\Lambda_w$, on which this
calculation of $\delta E_{\rm pin}$ was based.  This leads to
\begin{mathletters}
\begin{eqnarray}
\delta E_{\rm pin}&\sim&{\Lambda^{1/2}_w \over w^{1/4}}\;,\\
&\sim&- \epsilon^{1/3}w^{1/2}\;,
\end{eqnarray}
\label{Epin_final1}
\end{mathletters}
and predicts a change of the line tension at scale $w$ given by
\begin{mathletters}
\begin{eqnarray}
\delta\epsilon&=&{\delta E_{\rm pin} \over \Lambda_w}\;,\\
&\sim&- {1 \over\epsilon^{1/3}w}\;.
\end{eqnarray}
\label{delta_epsilon1}
\end{mathletters}
Note that in the above we have ignored the elastic energy, because we
do not need it to prove the stability of smectic ``Bragg'' glass!
Even without the {\em stabilizing} contribution of the elastic energy,
Eq.\ref{delta_epsilon1} implies
\begin{equation}
{\delta\epsilon(\ell)\over d \ell} = 
- {1 \over \epsilon(\ell)^{1/3}}\, e^{-\ell}
\label{epsilon_l1}
\end{equation}
Clearly, if the bare $\epsilon (\ell = 0)=\epsilon$ is big enough,
$\epsilon(\ell \rightarrow \infty)$ will be non-zero. Hence {\em
periodic disorder}, by itself, cannot unbind horizontal loops by
displacing them vertically.

\subsection{Case 2: Horizontal displacement of a horizontal dislocation ($w
\perp z$)}

As illustrated in Fig.14, here, the volume of the correlated regions
is $w \times w \times w^2$, with the long axis along $z$.
\begin{figure}[bth]
\begin{center}
\setlength{\unitlength}{1mm}
\begin{picture}(50,50)(0,0)
\put(-50,-90){\begin{picture}(30,30)(0,0)
\includegraphics{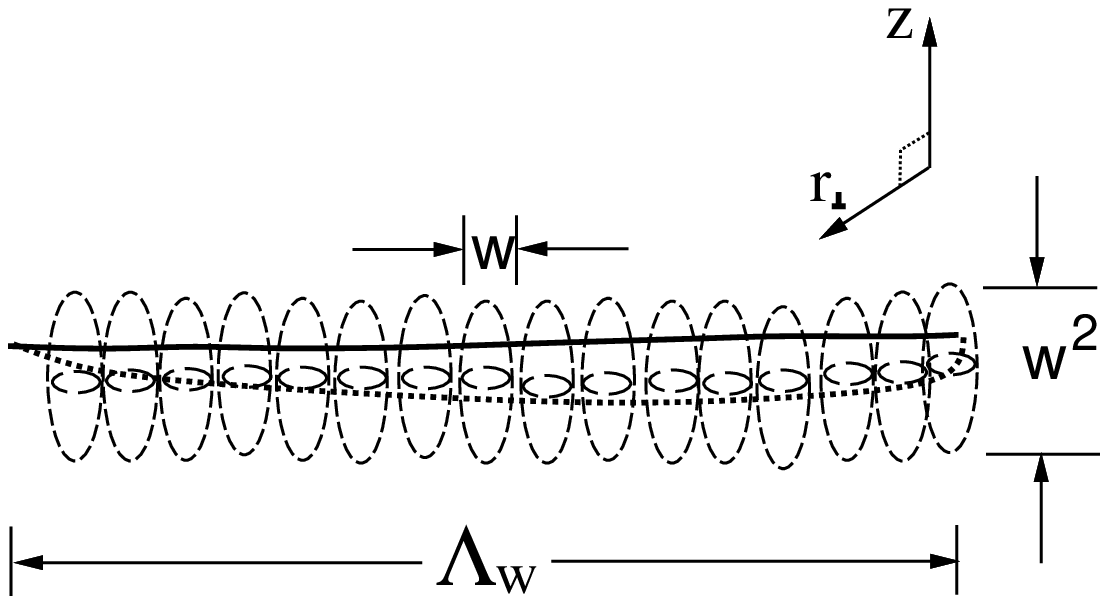}
\end{picture}}
\end{picture}
\end{center}
Fig.14. {Illustration of smectic anisotropic cigar-shaped correlation
regions inside the dislocation optimization length $\Lambda_w$ for
both the displacement $w$ and the dislocation running perpendicular to
$z$, i.e. along $r_\perp$ (e.g. along $x$ and $y$) (case 2).}
\label{case2RG}
\end{figure}
The width of the correlated volume {\em along} $\Lambda_w$ is now $w$,
so we have
\begin{equation}
\delta E_{\rm pin} \sim - \sqrt{\Lambda_w \over w}\;.
\label{Epin2}
\end{equation}
Equating this energy to the dislocation stretching energy $\epsilon
{w^2/\Lambda_w}$, we find
\begin{equation}
\Lambda_w\sim\epsilon^{2/3}w^{5/3}\;,
\label{Lambda_w2}
\end{equation}
which, upon inserting into $\delta E_{\rm pin}$, Eq.\ref{Epin2}, gives
\begin{equation}
\delta E_{\rm pin} \sim \epsilon^{1/3}w^{1/3}\;,
\label{Epin_final2}
\end{equation}
and implies
\begin{mathletters}
\begin{eqnarray}
\delta \epsilon&=& {\delta E_{\rm pin} \over \Lambda_w}\;,\\
&\sim&{1 \over \epsilon^{1/3}w^{4/3}}\;.
\end{eqnarray}
\label{delta_epsilon2}
\end{mathletters}
This leads to
\begin{equation}
{d\epsilon(\ell)\over d \ell}=-{1\over\epsilon(\ell)^{1/3}}\, e^{-4\ell/3}\;,
\label{epsilon_l2}
\end{equation}
which also demonstrates that there is {\em no} divergent downward
renormalization of $\epsilon$, implying that in the presence of only
periodic disorder, horizontal dislocation loops will {\em not} unbind
by displacing horizontally.

\subsection{Case 3: Horizontal displacement of a vertical dislocation ($w
\perp z$)} 

Finally we consider optimizing a z-directed screw dislocation, by
displacing it horizontally. This corresponds to a correlated region of
volume $w\times w\times w^2$, but, in contrast to Case 2, with the
axis along $\Lambda_w$ being the long dimension $w^2$, as illustrated
in Fig.15.
\begin{figure}[bth]
\begin{center}
\setlength{\unitlength}{1mm}
\begin{picture}(95,95)(0,0)
\put(-50,-90){\begin{picture}(30,30)(0,0)
\includegraphics{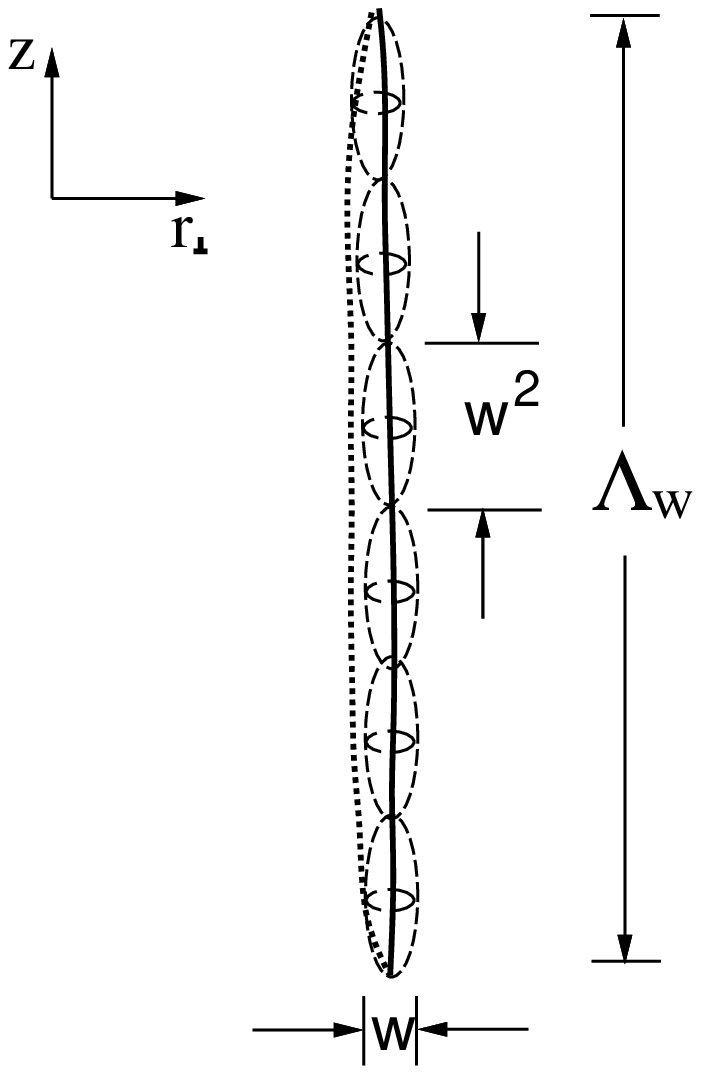}
\end{picture}}
\end{picture}
\end{center}
Fig.15. {Illustration of smectic anisotropic cigar-shaped correlation
regions inside the dislocation optimization length $\Lambda_w$ for
displacement $w$ along $r_\perp$ and dislocation running along $z$,
i.e. along $r_\perp$ (case 3).}
\label{case3RG}
\end{figure}
As a result, we have
\begin{equation}
\delta E_{\rm pin}\sim-\sqrt{{\Lambda_w \over w^2}}\;,
\label{Epin3}
\end{equation}
which when balanced against the dislocation stretching energy gives
\begin{equation}
\Lambda_w \sim \epsilon^{2/3} w^2\;.
\label{Lambda_w3}
\end{equation}
We note that for $\epsilon \gg 1$ this $\Lambda_w$ is much greater
than $w^2$, consistent with the assumption that goes into
Eq.\ref{Epin3}.  This gives
\begin{mathletters}
\begin{eqnarray}
\delta \epsilon&\sim&{\delta E_p \over \Lambda_w}\;,\\
&=& - {1 \over  \epsilon^{1/3}w^2}\;,
\end{eqnarray}
\label{delta_epsilon3}
\end{mathletters}
which implies
\begin{equation}
{d \epsilon(\ell)\over d \ell} = - {e^{-2\ell} \over  \epsilon(\ell)^{1/3}}
\label{epsilon_l3}
\end{equation}
whose solution for $\epsilon(\ell)$, again, is stable as $\ell
\rightarrow \infty$.

Since in all three cases we find that the effective dislocation line
tension is finite at long scales, we conclude that periodic
random-field disorder {\em alone} cannot unbind dislocation loops in
such a randomly pinned smectic.

\section{Experimental predictions and future research directions}
\label{experiments}

Our results imply many unequivocal dramatic predictions for
experiments. The first and most unequivocal of these is that
long-ranged, and even quasi-long-ranged, smectic translational order
is destroyed by the presence of even {\em arbitrarily} weak {\em
quenched} disorder. Furthermore, the decay of smectic translational
order induced by such disorder is far more rapid than in previously
studied ``Bragg glass'' systems\cite{VillainFernandez,GL,GH,DSFisher},
being exponential, rather than algebraic. The experimental signature
of this is equally unequivocal: broad X-ray scattering peaks with
finite peak height, as opposed to the infinitely sharp, divergent
peaks associated with quasi-long ranged order in thermal smectics.

This qualitative prediction is born out by all experiments on smectics
in aerogel to date: all show only broad, finite height X-ray
scattering peaks, which can be fit by assuming a finite X-ray
correlation length.\cite{review}

In strong contrast to {\em quenched} (frozen) disorder of this type,
it is straightforward to show analytically that weak short-ranged
``{\em annealed}'' disorder, in which the random environment can
rearrange to accomodate the smectic's preference for an ordered state,
has no {\em qualitative} effect on the long scale properties of liquid
crystal phases. Such weak annealed disorder only leads to {\em finite}
renormalizations of the effective parameters in our
model.\cite{RTunpublished} Consequently, the smectic phase, its
quasi-long-ranged smectic translational order, and the transition into
it are {\em stable to weak annealed} disorder, which might arise, for
example, due to a low concentration of microscopic inclusions.

Of course, strictly speaking, all physical random confining structure
are elastic, even aerogel, and therefore are able to deform to some
extent in response to the smectic, which resists distortions. It is
easy to see that this additional ``elastic compliance'' ingredient can
be described as an annealed component of the disorder, in addition to
the quenched part that is the main subject of our work.  In light of
the qualitative unimportance of weak annealed disorder, discussed
above, our theory for purely quenched random structures immediately
extends to deformable structures, which are quenched for large
deformations, i.e. described by a finite shear modulus.

A notable example of liquid crystals confined in random structure in
which these elastic annealed features appear are the recent
experiments on smectics in {\em
aerosil}.\cite{ZhouAerosil,GermanoAerosil} Since {\em typical} weak
hydrogen bonds that link the aerosil network together have
bond-breaking energies of order $k_BT$, a large fraction of the
aerosil network can rearrange itself in response to stresses imposed
on it by the smectic and thereby come to thermal equilibrium with
it. However, measurements\cite{GermanoAerosil} show that even these
tenuous, weakly-bonded structures are characterized by a small but a
{\em finite} shear modulus. Hence even aerosils are unable to perform
arbitrary rearrangements to equilibrate with the ordered
smectic. Therefore, while they have a large annealed component,
aerosils contain a small quenched component, which dominates the
qualitative physics of the confined smectic. We therefore expect a
destruction of the quasi-long-ranged translational order for smectics
confined in aerosils, with the X-ray correlation length obeying our
predictions for low density aerogels.

However, there are three important consequences of the large annealed
disorder component present in aerosils. Firstly, while {\em weak
annealed} disorder should only lead to a finite renormalization of
effective parameters, {\em strong} annealed disorder can destroy the
quasi-long-ranged smectic order by driving some of the effective
elastic moduli negative. Secondly, given the empirical fact that the
{\em bulk} NA transition appears to display non-universal
behavior,\cite{NAnonuniversal} we would expect this non-universality to also
manifest itself for smectics ``confined'' in low-density aerosils,
where the effective model parameters are functions of aerosil density
due to the presence of the annealed disorder component. Thirdly,
because of this large annealed disorder component in aerosils and a
correspondingly small quenched component, we expect aerosils to be
described by significantly smaller values of our effective quenched
disorder parameters $\Delta_V$ and $\Delta_h$ than for aerogels of the
same solid volume fraction. Consequently, experiments in aerosils are
more appropriate for exploring our weak quenched disorder theory and
to search for the delicate and ``fragile'' SBG and NEG phases.

These nontrivial qualitative predictions are clearly in agreement with
the experimental observations of smectics confined in
aerosils\cite{GermanoAerosil}: (i) for low aerosil densities these
experiments observe significantly longer (than in equivalent volume
fractions aerogels), finite smectic X-ray correlation lengths, (ii)
they measure heat capacities with aerosil density dependent (and
therefore nonuniversal) exponents, and (iii) in aerosils beyond a
critical density they find a behavior that is qualitatively similar to
smectics confined in aerogels.

It is important to note that in aerogels and aerosils, in contrast to
what one might have naively expected, the smectic correlation length
is {\em not} limited by some sort of ``finite size'' effect associated
with the ``pore size'' of the aerogel. That is, the smectic confined
in aerogel does {\em not} act as if it has simply been broken up into
many small volumes which do not interact with each other.

If this ``finite size'' scenario {\em were} the case, one would expect
to see the smectic translational correlation length $\xi^X$ grow with
decreasing temperature as $T$ approached the pure $NA$ transition
temperature $T_{NA}^{\rm pure}$ from above exactly as it did in the
pure (bulk) smectic. Indeed, one would expect it to track its
temperature dependent value $\xi_{\rm pure}^X(T)$ in the pure system
until $T$ got close enough to $T_{NA}^{\rm pure}$ that $\xi_{\rm
pure}^X(T)$= $\xi_{\rm aerogel}^X(T)=L_P$, the ``pore size'' of the
aerogel. At this point $\xi^X(T)$ would abruptly saturate, and remain
constant at $L_P$ for all lower temperatures.

This is emphatically {\em not} what is seen in experiments on
aerogels.\cite{review,Bellini,Clark} Rather the growth of $\xi^X$
continues right through $T_{NA}^{\rm pure}$, and only saturates at
some temperature well below $T_{NA}^{\rm pure}$. This is completely
consistent with our picture, in which $\xi^X$ is determined by a
competition between the smectic elasticity, which tries to keep the
system ordered, and the random forces exerted by the aerogel strands,
which disorder it. In this picture, the saturation of $\xi^X$ with
decreasing temperature occurs not {\em above} $T_{NA}^{\rm pure}$,
but, rather, well {\em below} $T_{NA}^{\rm pure}$; specifically, at
the temperature at which the smectic elastic compression modulus
$B(T)$ saturates.

Liquid crystals have also been confined in other random but
significantly more ordered structures. Some examples are
Anopore\cite{GermanoAnopore} and Vycor\cite{GermanoVycorIN}, which in
contrast to the tenuous aerogel strand structure present on all
scales, consist of a random distribution of empty pores, with fairly
regular wall structure and much narrower pore size distribution. Even
these systems can in principle be understood within the general model
presented here. However, they fall into the strong disorder regime of
our model, and we are therefore unable to utilize our weak disorder
approach to make quantitative predictions, aside from the observation
that the quasi-long-ranged smectic order should be destroyed in these
systems as well, consistent with experiments.

The qualitative agreement between our theory and experiment can be
made quantitative through our expression Eq.\ref{xiX}, which relates
$\xi^X$ to the smectic elastic constant $B(T)$. As discussed in the
Introduction, this expression implies that sufficiently far below
$T_{NA}^{\rm pure}$ (where our weak disorder theory applies), a simple
power-law relation between $\xi^X(T)$ and the ``bare'' layer
compressional modulus $B_{\rm bare}(T)$ of the pure system holds:
\begin{mathletters}
\begin{eqnarray}
\xi^X(T)&\propto&\left[B_{\rm bare}(T)\right]^{\zeta/2\chi}\;,\\
&\propto&\left[B_{\rm bare}(T)\right]^{1/\chi-1/2}\;,
\end{eqnarray}
\label{xiXTB}
\end{mathletters}

The bare $B_{\rm bare}(T)$ can be determined in a number of ways. One
way is simply to measure the layer compressional modulus $B_{\rm
pure}(T)$ of a pure (bulk, aerogel-free) sample of the same smectic
liquid crystal material, and pray that this is not changed when the
smectic is placed in aerogel.  However, since prayers are not always
answered, a direct, simultaneous in-situ determination of $B_{\rm
bare}(T)$ in the smectic inside the aerogel is clearly desirable.

This could be done by determining the squared magnitude of the smectic
order parameter $|\psi|^2$ from, e.g., the integrated intensity under
the (broadened) smectic X-ray peak. In mean-field theory, which should
hold for bare quantities sufficiently far below $T_{NA}$, $B_{\rm
bare}(T)\propto|\psi|^2$. Thus, we should, in this mean-field regime,
observe a universal power-law relation between the X-ray correlation
length $\xi^X$ and $|\psi|^2$: $\xi^X\propto |\psi|^{2/\chi-1}$. Note
that in the absence of anomalous elasticity, the second regime of
Eq.\ref{xiX} ($\xi^X=\xi_z^{NL}(a/\lambda)^2$) would apply, and we
would get $\xi^X\propto B\propto|\psi|^2$, the last proportionality
holding only in the mean-field regime. Thus, {\em any} departure of
the $\xi^X$-$B$ relation from linearity is evidence for anomalous
elasticity. Furthermore, observing such a power-law relation
determines $\chi$.

Preliminary analysis of $\xi^X$ versus $B$ data\cite{Science}, as
determined above, supports the relation Eq.\ref{xiXTB}, with
$\chi=0.8\pm0.1$. Note that $\chi<1$, favoring {\em stability} of the
smectic Bragg glass phase, although we have, as yet, no experimental
way of determining if the other necessary relation for the stability
of the SBG, namely $\eta_K<1$, is satisfied.

While our prediction for $\xi^X(T)$ is not as quantitative as we would
like, since we do not have quantitatively reliable predictions for
$\zeta$ and $\chi$ in $d=3$ (our $5-d=\epsilon$-expansion results not
being quantitatively trustworthy when $\epsilon=2$), it nonetheless
makes it possible, by experimentally determining $\zeta/\chi$ for {\em
one} smectic in {\em one} aerogel, to predict the temperature
dependence of $\xi^X$ for {\em any} smectic in {\em any} low density
aerogel. Furthermore, once $\zeta/\chi$ is known, $\zeta$ and $\chi$
separately are also known, since $\zeta=2-\chi$. Knowledge of these
exponents then makes other predictions possible.

In particular, it leads to our second prediction which concerns
whether or not the transition (which replaces the NA transition in
bulk smectics) from the nematic to the low-temperature phase (which
replaces the ordered smectic) is destroyed by disorder. Surprisingly,
this question can, in principle, have a different answer than the
question of whether or not smectic translational order is
destroyed. This is because our analysis shows that, even though
smectic translational order {\em is always} destroyed, a distinct
``smectic Bragg glass'' phase (SBG), in which dislocations remain
bound and long-ranged {\em orientational} (nematic) order persists,
maybe possible in the presence of sufficiently weak disorder. If it
is, then a thermodynamically sharp transition into it (with decreasing
temperature) replaces the NA transition of the pure system, as this
phase itself replaces the smectic-A phase. Consequently, an exotic and
quite unique feature of this transition is that, while thermodynamic
quantities (e.g., heat capacity) display singularities at the
transition to the putative smectic Bragg glass, the X-ray scattering
peak remains broad through the transition. Unfortunately, whether or
not this (SBG) phase exists, in principle, depends on whether or not
the imprecisely known anomalous exponents $\eta_B$ and $\eta_K$
satisfy the bounds Eq.\ref{bounds12}.

Fortunately, the above-described experimental determination of $\chi$
could answer this question, since one of the bounds
Eq.\ref{bound_orientations} is simply $\chi<1$. Thus, if the X-ray
measurements, described above, find $\chi<1$, a SBG phase can exist,
and a thermodynamically sharp remnant of the pure system's NA
transition could be seen in sufficiently low density aerogels. On the
other hand, if $\chi$ proves to be $>1$, no sharp transition should be
seen. This is a definite, falsifiable prediction that could be tested
experimentally: if $\chi>1$, no sharp transition should be seen. The
converse conclusion, that if $\chi<1$, a sharp transition should be
seen, need not hold, since it is possible that $\chi<1$, but the bound
$\eta_K<1$ is violated.

For the same reason, even if the current experimental evidence that
there is {\em no} transition even for arbitrarily low density aerogel
holds up, we can {\em not} conclude that $\chi>1$.

It is also important to note, that despite of the absence of the
long-ranged ordered smectic phase, it {\em is} possible, as always, to
have a thermodynamically sharp {\em first-order} transition {\em
within} the disordered (nematic) phase. At such transition, which
would be from a nematic with a short smectic correlation length to a
nematic with longer (but still finite) smectic correlation length,
thermodynamic quantities (like the heat capacity) could displaying
sharp non-analytic features (like latent heat). We believe that the
recent calorimetry experiments on 7O.4/aerogel samples by Haga and
Garland are examples of such a first order transition.\cite{HG} As a
consequence of our main result that only short range smectic order is
possible for smectics confined in quenched random environments, we
unambiguously predict a broad, finite X-ray scattering peak for the
system studied by Haga and Garland.

If the SBG phase {\em does} prove to be possible, then we can make far
more detailed tests of our predictions. In particular, the simple fact
that the SBG phase has long-ranged orientational order is, in itself,
a striking prediction that could be tested very easily by looking at
optical birefringence in smectics in aerogel. The existence of such
long-ranged order would make it possible even more quantitative tests
for our predictions of anomalous elasticity, by making it possible to
measure director fluctuations $\overline{\langle\delta n_i({\bf
q})\delta n_j(-{\bf q})\rangle}$, by light scattering. Since these are
related to smectic layer fluctuations by the usual smectic relation
$\delta{\bf n}({\bf r})=\bbox{\nabla}u({\bf r})$, we can immediately
write
\begin{mathletters}
\begin{eqnarray}
\overline{\langle\delta n_i({\bf q})\delta n_j(-{\bf q})\rangle}
&=&q_i q_j\overline{\langle|u({\bf q})|^2\rangle}\;,\\
&=&{\Delta_h({\bf q}) q_i q_j q_\perp^2\,V\over
\big[B({\bf q})q^2_z + K({\bf q})q^4_{\perp}\big]^2}\nonumber\\
&&+{k_B T q_i q_j\,V\over
B({\bf q})q^2_z + K({\bf q})q^4_{\perp}}\;,\label{nn_b}
\end{eqnarray}
\label{delta_n-delta_n}
\end{mathletters}
where $i$ and $j$ run over all indices but $z$, and $V$ is the volume
of the system.  The first, quenched, term in this expression,
Eq.\ref{nn_b}, dominates as ${\bf q}\rightarrow0$; thus fitting this
piece to the observed light scattering would enable a direct
experimental measurement of $K({\bf q})$, $B({\bf q})$, and
$\Delta_h({\bf q})$, and, hence, a direct test of the scaling
predictions Eq.\ref{KgBgDg} for these quantities, as well as a
determination of the anomalous exponents $\eta_K$, $\eta_B$, and
$\eta_\Delta$ . For instance, if we take $\bf q$ in the $x-z$ plane,
Eq.\ref{delta_n-delta_n} implies
\begin{equation}
\overline{\langle|\delta n_x({\bf q})|^2\rangle}
=q_x^{-4-\eta_\Delta+2\eta_K}f_n\left({q_z\xi_z^{NL}
\over (q_\perp\xi_\perp^{NL})^\zeta}\right)\;.\label{delta_n-delta_n2}
\end{equation}
Thus, a log-log plot of $\overline{\langle|\delta n_x({\bf
q})|^2\rangle}$ for $q_z=0$ has slope $-4-\eta_\Delta+2\eta_K$. The
simple observation that this slope is not $-4$ would confirm the
existence of anomalous elasticity, by showing that $\eta_\Delta$ and
$\eta_K$ were $\neq0$. The determination of the anisotropy exponent
$\zeta$ by collapsing the plots of $\overline{\langle|\delta n_x({\bf
q})|^2\rangle}$ versus $q_x$ for a variety of $q_z$'s would then fix
$\eta_B+\eta_K$. Our exact scaling relation between $\eta_\Delta$,
$\eta_B$ and $\eta_K$, Eq.\ref{WI} would then provide a third equation
to fix all $3$ exponents. That the results satisfied our bounds
Eq.\ref{bounds12} for the stability of the SBG phase would then
provide a non-trivial check on the consistency of our theory.

A more stringent test could be provided by measuring the second, $k_B
T$ term in Eq.\ref{delta_n-delta_n}. Although, as mentioned earlier,
this term is subdominant to the first, disorder term, it could
nonetheless be resolved by {\em time-dependent} light scattering. The
reason for this is that the first, disorder term of
Eq.\ref{delta_n-delta_n} represents the {\em static}, time-persistent
response of the smectic to the random tilt field of the aerogel, while
the $k_BT$ term represents thermal fluctuations about this
disorder-determined smectic layer configuration. That is, if we
consider the unequal time correlation function
$\overline{\langle\delta n_i({\bf q},t)\delta n_j(-{\bf
q},0)\rangle}$, we expect
\begin{eqnarray}
\overline{\langle\delta n_i({\bf q},t)\delta n_j(-{\bf q},0)\rangle}
&=&{\Delta_h({\bf q}) q_i q_j q_\perp^2\,V\over
\big[B({\bf q})q^2_z + K({\bf q})q^4_{\perp}\big]^2}\nonumber\\
&&+{k_B T q_i q_j\,f({\bf q},t) V\over
B({\bf q})q^2_z + K({\bf q})q^4_{\perp}}\;,\label{ntn0_b}
\end{eqnarray}
where $f({\bf q},t)$ is a function we do not know, but which
presumably will have some sort of slow, ``glassy'' decay, vanishing as
$t\rightarrow\infty$, and going to $1$, as $t\rightarrow0$, thereby
recovering the equal-time correlation function,
Eq.\ref{delta_n-delta_n}. Using these two limits immediately implies
\begin{eqnarray}
&\lim_{t\rightarrow\infty}&
\overline{\langle\delta n_i({\bf q},t)\delta n_j(-{\bf q},0)\rangle}
-\overline{\langle\delta n_i({\bf q},0)\delta n_j(-{\bf q},0)\rangle}
\nonumber\\
&=&{k_B T q_i q_j\,V \over
B({\bf q})q^2_z + K({\bf q})q^4_{\perp}}\;,\label{ntn0_thermal}
\end{eqnarray}
which allows a completely independent determination of $\eta_K$ and
$\eta_B$ from the approach based strictly on the equal-time
correlation function described earlier. Consistency of the two
approaches would demonstrate the validity of the exact scaling
relation between $\eta_K$, $\eta_B$, and $\eta_\Delta$.

We remind the avid experimentalist that the entire above discussion
{\em only} applies if the SBG phase is stable.  If the SBG phase does
{\em not} exist, then our experimental predictions are reduced to
Eq.\ref{xiX} for the X-ray correlation length, the anomalous
elasticity lengths given by Eqs.\ref{xi_zp_NL}, and the ``ghost'' of
non-analyticity in $T$ in these lengths, associated with the remnant
of $T_p$ below $T_{NA}$ (see Fig.6), defined by the temperature at
which the pure smectic order parameter correlation exponent $\eta$
passes $4$.

Presumably, there are experiments that could test our predictions for
the orientational correlation length $\xi_O$, but we will be unable to
make such predictions until we have developed a theory of the nematic
elastic glass behavior that holds on longer length scales. This is a
topic for future research.\cite{RTunpublished}

There are a number of other promising areas for future research. One
is the subject of smectic order inside {\em stretched} aerogels, which
we are currently studying.\cite{JSRT} In that system we have shown
that uniaxially stretching or compressing the aerogel matrix can
stabilize the smectic Bragg glass phase. The universality class of the
resultant SBG now depends upon whether the aerogel is stretched or
compressed: in the former case, it lies in the universality class of
the so-called ``vortex glass'' phase, which has recently been much
discussed in the context of random field XY-model\cite{GH,DSFisher}
and randomly pinned Abrikosov vortex lattices,\cite{GL} while
compression leads to a totally novel kind of ``$m=1$ Lifshitz SBG''. A
detailed discussion of this problem is in preparation.\cite{JSRT}

We expect a new glass phase similar in many respects to the ``m=1''
glass described above, but in a different universality class, to occur
when ``discotics'' (i.e. liquid crystals with {\em two solid}-like
directions and {\em one liquid}-like) are absorbed into {\em
unstressed} aerogel. Work on this interesting problem is also
currently underway.\cite{JSRT}

It is also interesting to consider what happens to a {\em cholesteric}
liquid crystal confined in aerogel.\cite{thankTom} Although the
symmetry of such a system is {\em identical} to that of a smectic
studied here, the addition of another long length scale (namely the
cholesteric pitch) may make it possible to access a novel type of
``random manifold'' regime of pinned elastic media. 

Probably the most interesting and challenging problem is the theory of
dynamics of these smectic Bragg glass systems, which would enable a
detailed understanding of recent dynamic light scattering
experiments.\cite{Bellini} If the smectic Bragg glass phase exists,
which we are quite certain it does in smectics in anisotropic aerogels
and discotics in isotropic aerogels,\cite{JSRT} it should have slow
glassy dynamics similar to those observed by Bellini et
al.\cite{Bellini}. Given its unusual {\em static} properties
(e.g. anomalous elasticity), its dynamics may be quite unusual, even
for glasses.

As noted in the Introduction, here we have studied smectics in aerogel
by perturbing around the low temperature translationally and
orientational ordered smectic phase. However, another complementary
approach to this problem, more convenient to understanding the effects
of disorder near and above the bulk NA transition, is based on de
Gennes model.\cite{RadzihovskyWard} This complementary high
temperature approach makes it possible to predict, e.g., the rounded
specific heat and X-ray scattering peaks observed in many experiments
on smectics in aerogel.\cite{review,Bellini,Clark,Wu,Kutnjak}

Finally, we note that the formalism we developed in
Sec.\ref{dislocations} for disordered smectics can be applied to a
variety of other disordered problems, including gauge glasses,
randomly pinned Abrikosov vortex lattices and other pinned periodic
media. These applications will be discussed in a future
publication.\cite{RTgauge_glass}

\section{Acknowledgments}
We thank Noel Clark and Tommaso Bellini for stimulating our interest
in this problem, Karl Saunders for a careful reading of the manuscript
and for correcting errors in equations, and acknowledge many long and
illuminating discussions with Carl Garland and Germano Iannacchione,
which led to a significant improvement of the manuscript. Leo
Radzihovsky was supported by the National Science Foundation CAREER
Award under \#DMR-9625111, by the MRSEC Program at the University of
Colorado at Boulder under \#DMR-9809555, and by the A.P. Sloan and
David and Lucile Packard Foundations. John Toner thanks the National
Science Foundation for financial support through Grant \#DMR-9634596,
and Prof. Lucy Pao for her kind hospitality while a portion of this
work was being completed, and for the pancakes.

\section{Appendix A: Details of the $d=3$ single harmonic 
Renormalization Group Analysis}
\label{single_cosRGapp}

Here we present the details of the derivation of the renormalization
group flow equations Eqs.\ref{gamma_ell}-\ref{K_ell} for the
three-dimensional disordered smectic model, in the harmonic elastic
approximation defined by the replicated Hamiltonian Eq.\ref{Hr}
\begin{equation}
H[u_\alpha]=H_0[u_\alpha]+H_\Delta[u_\alpha]\;,
\end{equation}
where the quadratic $H_0[u_\alpha]$ and the interaction
$H_\Delta[u_\alpha]$ parts of the Hamiltonian are respectively given by
\begin{mathletters}
\begin{eqnarray}
H_0&=&{1\over2}\int d^d q\sum_{\alpha,\beta}^n\big[\left(K 
q_\perp^4+ B q_z^2\right)\delta_{\alpha\beta}
-\Delta_h q_\perp^2\big]u_\alpha
u_\beta\;,\nonumber\\
&&\label{H0A}\\
H_\Delta&=&-\int d^d r{\sum_{\alpha\neq\beta}^n}'
\Delta_V\cos[q_0(u_\alpha({\bf r})-u_\beta({\bf r}))]\;.\label{HgA}
\end{eqnarray}
\end{mathletters}
In the above and throughout, in the interaction term $H_\Delta$, we
have for convenience excluded the $\alpha=\beta$ term inside the
summation. This exclusion is indicated by a prime, and is justified
since the $\alpha=\beta$ term is a constant. Also for simplicity,
throughout this appendix we have set $T=1$; the factors of $T$ can be
easily restored by the replacement $K\rightarrow K/T$, $B\rightarrow
B/T$ $\Delta_h\rightarrow \Delta_h/T^2$, and $\Delta_V\rightarrow
\Delta_V/T^2$. From $H_0$, the propagator $G_{\alpha\beta}({\bf q})$
can be easily obtained,
\begin{mathletters}
\begin{eqnarray}
G_{\alpha\beta}({\bf q})&=&G_T({\bf q})\;\delta_{\alpha\beta}
+G_{\Delta_h}({\bf q})\;,\label{GTGK}\\
&=&{\delta_{\alpha\beta}\over K q_\perp^4+B q_z^2}+{\Delta_h
q_\perp^2\over(K q_\perp^4+B q_z^2)^2}\;.\label{GabAppendix}
\end{eqnarray}
\end{mathletters}

In three dimensions, a direct perturbative calculation in $\Delta_V$ is
divergent in the thermodynamic limit even for an arbitrarily small
$\Delta_V$. Nevertheless physical observables can be computed utilizing
standard methods of the renormalization group\cite{Wilson}, in which
one avoids infrared divergences by integrating out degrees of freedom
a momentum ``shell'' at a time. In this procedure the goal is to
establish how the effective Hamiltonian functional changes after the
renormalization group transformation, which can be neatly summarized
by the renormalization group flow equations for the effective
parameters.

We focus on $\overline{Z^n}$ as the physical quantity to keep fixed
under the infinitesimal renormalization group
transformation. Following the standard RG procedure we write the
Fourier-transformed displacement field as $u_\alpha({\bf
q})=u^<_\alpha({\bf q})+u^>_\alpha({\bf q})$, where $u^<_\alpha({\bf
q})$ is the long wavelength set of modes, nonzero for $0<|{\bf
q}_\perp|<\Lambda e^{-\ell}$, with $q_z$ unrestricted, and
$u^>_\alpha({\bf q})$ are the short wavevector degrees of freedom that
have support only within a thin cylindrical momentum shell
\begin{mathletters}
\begin{eqnarray}
\Lambda e^{-\ell}<&|{\bf q}_\perp|&<\Lambda\;,\\
-\infty<&q_z&<\infty\;,
\end{eqnarray}
\label{shell2}
\end{mathletters}
We first integrate, perturbatively in $\Delta_V$, the high wavevector
part $u^>_\alpha$ out of $\overline{Z^n}$:
\begin{mathletters}
\begin{eqnarray}
\overline{Z^n}&=&\int[d
u^<_\alpha][d u^>_\alpha]\;e^{-H_0[u^<_\alpha+u^>_\alpha]-
H_\Delta[u^<_\alpha+u^>_\alpha]}\;,\\
&=&\int[d u^<_\alpha]\;e^{-H_0[u^<_\alpha]-\delta H[u^<_\alpha]}\;,
\end{eqnarray}
\end{mathletters}
where $\delta H[u^<_\alpha]$ is obtained by integrating out the short
wavevector degrees of freedom
\begin{mathletters}
\begin{eqnarray}
e^{-\delta H[u^<_\alpha]}&\equiv&\int[d u^>_\alpha]
\;e^{-H_0[u^>_\alpha]-H_\Delta[u^<_\alpha+u^>_\alpha]}\;,\\
&=&Z^>_0\left<e^{-H_\Delta[u^<_\alpha+u^>_\alpha]}\right>\;,\\
&=&Z^>_0\Big [1-\left <H_\Delta[u^<_\alpha+u^>_\alpha]\right>_>\\
&+&{1\over 2}\left<H_\Delta[u^<_\alpha+u^>_\alpha]^2\right >_>
+\ldots\Big]\;,\nonumber
\end{eqnarray}
\end{mathletters}
The perturbative correction to the non-constant part of the effective
Hamiltonian can therefore be expressed in terms of a cumulant
expansion
\begin{equation}
\hspace{-.3cm}
\delta H[u^<_\alpha]=\left <H_\Delta[u^<_\alpha+u^>_\alpha]\right >_>^c -
{1\over 2}\left<H_\Delta[u^<_\alpha+u^>_\alpha]^2\right >_>^c
+\ldots\;,
\end{equation}
where
\begin{equation}
\hspace{-.3cm}
\left<H_\Delta[u^<_\alpha+u^>_\alpha]^2\right >_>^c
=\left<H_\Delta[u^<_\alpha+u^>_\alpha]^2\right >_>-
\left<H_\Delta[u^<_\alpha+u^>_\alpha]\right >_>^2\;.
\end{equation}

To first order in the interaction $\Delta_V$ we obtain
\begin{eqnarray}
\big<H_\Delta\big>_>&=&-\Delta_V\int d^3
r\sum'_{\alpha\neq\beta}\big<\cos[q_0(u_\alpha-u_\beta)]\big>_>\;,\nonumber\\
&=&-\Delta_V\mbox{Re}\int d^3 r\sum'_{\alpha\neq\beta}
e^{i q_0(u^<_\alpha-u^<_\beta)}\left<e^{i
q_0(u^>_\alpha-u^>_\beta)}\right>_>\;,\nonumber\\
&=&-\Delta_V\int d^3 r\sum'_{\alpha\neq\beta}
\cos[q_0(u^<_\alpha-u^<_\beta)]\;e^{-q_0^2f_{\alpha\beta}}\;,\nonumber\\
&=&-\Delta_V\int d^3 r\sum'_{\alpha\neq\beta}
\cos[q_0(u^<_\alpha-u^<_\beta)]\;e^{-\eta\ell(1-\delta_{\alpha\beta})}
\;,\nonumber\\
\end{eqnarray}
where
\begin{equation}
f_{\alpha\beta}=\bigg(G^>_{\alpha\alpha}({\bf r}=0)-
G^>_{\alpha\beta}({\bf r}=0)\bigg)\;,
\end{equation}
in $G^>_{\alpha\alpha}(0)$ there is {\em no} implied sum over
$\alpha$ and
\begin{equation}
G_{\alpha\beta}^>({\bf r})=
\int_{\Lambda e^{-\ell}}^{\Lambda}{d^2
q_\perp\over(2\pi)^2}\int_{-\infty}^\infty{d q_z\over 2\pi}
G_{\alpha\beta}(q)e^{i{\bf q\cdot r}}\;.
\end{equation}
Using
\begin{equation}
G_{\alpha\alpha}({\bf q})-G_{\alpha\beta}({\bf q})=
{1\over K  q_\perp^4+B q_z^2}(1-\delta_{\alpha\beta})\;,
\label{G-G}
\end{equation}
which is obviously independent of the tilt-disorder variance
$\Delta_h$, and Fourier transforming the last equality in
Eq.\ref{G-G}, with $\eta$ as given in Eq.\ref{eta}, we obtain:
\begin{equation}
\eta={q_0^2\over 4\pi\sqrt{K  B}}\;.
\end{equation}
Hence, for an infinitesimal $\delta\ell$ the ``graphical'' correction
to $\Delta_V$ to first order in $\Delta_V$ is:
\begin{equation}
\delta\Delta_V^{(1)} = -\eta\Delta_V\;\delta\ell\;.
\end{equation}

We now proceed to compute the corrections to $\delta H[u^<_\alpha]$ to
second order in $\Delta_V$, which we anticipate will result in higher
order corrections to the random field disorder $\delta\Delta_V$, as
well as a correction to tilt disorder variance $\Delta_h$.  We compute
\begin{eqnarray}
\left<H_\Delta^2\right >_>^c&\equiv&
\left<H_\Delta[u^<_\alpha+u^>_\alpha]^2\right >_>^c\;,\nonumber\\
&=&\Delta_V^2\int_{\bf r r'}\sum'_{\alpha\neq\beta,\alpha'\neq\beta'}
T_{\alpha\beta}^{\alpha'\beta'}({\bf r}-{\bf r'})\;,
\label{H2}
\end{eqnarray}
where $\int_{\bf r}\equiv\int d^3 r$, the prime on the sum now
excludes $\alpha=\beta$ and/or $\alpha'=\beta'$ terms, and
\begin{eqnarray}
&&\hspace{-.5cm}
T_{\alpha\beta}^{\alpha'\beta'}({\bf r}-{\bf r'})\equiv\nonumber\\
&\equiv&\big<\cos[q_0(u_\alpha({\bf r})-u_\beta({\bf r}))]
\cos[q_0(u_{\alpha'}({\bf r'})-u_{\beta'}({\bf r'}))]\big>_>^c\;,\nonumber\\
&=&{1\over4}\sum_{q,q'}\left<e^{i q(u_\alpha({\bf r})-u_\beta({\bf
r}))+i q'(u_{\alpha'}({\bf r'})-u_{\beta'}({\bf r'}))}\right>_>^c\;,\nonumber\\
&=&{1\over4}\sum_{q,q'} e^{i q(u^<_\alpha({\bf r})-u^<_\beta({\bf
r}))+i q'(u^<_{\alpha'}({\bf r'})-u^<_{\beta'}({\bf r'}))}
S_{\alpha\beta}^{\alpha'\beta'}({\bf r}-{\bf r'})\;.\nonumber\\
\label{T}
\end{eqnarray}
In the above expression the $q$ and $q'$ variables independently range
over two values $\pm q_0$, and $S_{\alpha\beta}^{\alpha'\beta'}({\bf
\delta r})$ (with ${\bf\delta r}\equiv{\bf r}-{\bf r'}$) is the result
of the cumulant average over $u^>_\alpha$ fields given by
\begin{eqnarray}
&&\hspace{-.5cm}
S_{\alpha\beta}^{\alpha'\beta'}({\bf\delta r})\equiv\nonumber\\
&\equiv&
\left<e^{iq(u^>_\alpha({\bf r})-u^>_\beta({\bf r}))+ i q'(u^>_{\alpha'}({\bf
r'})-u^>_{\beta'}({\bf r'}))}\right>_>^c\;,\nonumber\\
&=&e^{-q_0^2 (f_{\alpha\beta}+f_{\alpha'\beta'})}\times\nonumber\\
&&\bigg[e^{-q q'\big(G^>_{\alpha\alpha'}({\bf\delta r})+
G^>_{\beta\beta'}({\bf\delta r})-
G^>_{\alpha\beta'}({\bf\delta r})-
G^>_{\beta\alpha'}({\bf\delta r})\big)}-1\bigg]\;,\nonumber\\
&=&e^{-q_0^2(f_{\alpha\beta}+f_{\alpha'\beta'})}
\bigg[e^{-q q'G^>_T({\bf\delta r})(\delta_{\alpha\alpha'}+\delta_{\beta\beta'}
-\delta_{\alpha\beta'}-\delta_{\beta\alpha'})}-1\bigg]\;,\nonumber\\
\end{eqnarray}
where to obtain the last equation above we used Eq.\ref{GTGK}.  The
real space propagator of the short wavelength modes $G^>_T({\bf\delta
r})$ is by definition a partial Fourier transform of the propagator in
momentum space, Eqs.\ref{GTGK},\ref{GabAppendix}, which for
infinitesimal renormalization group transformation is integrated only
over an infinitesimal shell of wavevectors $\Lambda\,
e^{-\delta\ell}<q_\perp<\Lambda$. This implies that the exponential in
the last equation can be expanded in powers of an infinitesimal
$G^>_T({\bf\delta r})$ proportional to $\delta\ell$
\end{multicols}
\begin{mathletters}
\begin{eqnarray}
S_{\alpha\beta}^{\alpha'\beta'}({\bf\delta r}) &=&e^{-q_0^2
(f_{\alpha\beta}+f_{\alpha'\beta'})} \bigg[-4q q'G^>_T({\bf\delta
r})\delta_{\alpha\alpha'}+ {q_0^4\over 2}\big(G^>_T({\bf\delta
r})\big)^2 (\delta_{\alpha\alpha'}+\delta_{\beta\beta'}
-\delta_{\alpha\beta'}-\delta_{\beta\alpha'})^2+\ldots\bigg]\;,\label{S1}\\ 
&=&e^{-q_0^2 (f_{\alpha\beta}+f_{\alpha'\beta'})}
\bigg[-4q q'G^>_T({\bf\delta r})\delta_{\alpha\alpha'}+
{2q_0^4}\big(G^>_T({\bf\delta r})\big)^2
\big(\delta_{\alpha\alpha'}+\delta_{\alpha\alpha'}\delta_{\beta\beta'}
-2\delta_{\alpha\beta'}\delta_{\beta\beta'}\big)+\ldots\bigg]\;,\label{S2}
\end{eqnarray}
\end{mathletters}
\begin{multicols}{2}
\noindent 
where to obtain the first terms in the respective square brackets of
Eqs.\ref{S1},\ref{S2}, above, we anticipated that
$S_{\alpha\beta}^{\alpha'\beta'}$ will be summed over $q$ and $q'$,
which allowed us to make simplifying changes in summation variables
$\alpha,\alpha',\beta,\beta'$. Inserting this last expression
Eq.\ref{S2} into $T_{\alpha\beta}^{\alpha'\beta'}(\delta{\bf r})$,
Eq.\ref{T}, and performing the sum over $q, q'$, we obtain
\end{multicols}
\begin{eqnarray}
&&\hspace{-0.3cm}T_{\alpha\beta}^{\alpha'\beta'}(\delta{\bf r})=
q_0^2\;e^{-q_0^2 G^>_T(0)
\big(2-\delta_{\alpha\beta}-\delta_{\alpha'\beta'}\big)}
\bigg[2G^>_T({\bf\delta r})\delta_{\alpha\alpha'}
A_{-\, \alpha\beta}^{\ \ \alpha'\beta'}({\bf r},{\bf r}')
+ q_0^2\big(G^>_T({\bf\delta r})\big)^2
\big(\delta_{\alpha\alpha'}+\delta_{\alpha\alpha'}\delta_{\beta\beta'}
-2\delta_{\alpha\beta'}\delta_{\beta\beta'}\big)
A_{+\, \alpha\beta}^{\ \ \alpha'\beta'}({\bf r},{\bf r}')
\bigg]\;.\nonumber\\
&&\label{T2}
\end{eqnarray}
where
\begin{equation}
A_{\pm\, \alpha\beta}^{\ \ \alpha'\beta'}({\bf r},{\bf r}')
=\Big(\cos[q_0(u^<_\alpha({\bf r})-u^<_\beta({\bf r})
-u^<_{\alpha'}({\bf r'})+u^<_{\beta'}({\bf r'}))]
\pm\cos[q_0(u^<_\alpha({\bf r})-u^<_\beta({\bf r})
+u^<_{\alpha'}({\bf r'})-u^<_{\beta'}({\bf r'}))]\Big)
\end{equation}
\begin{multicols}{2}
Naively, it appears that the $\delta_{\alpha\beta}$,
$\delta_{\alpha'\beta'}$ and
$-2\delta_{\alpha\beta'}\delta_{\beta\beta'}$ terms above contribute
to the renormalization of $\Delta_V$ to second order in $\Delta_V$.
However, upon summation over $\alpha,\beta,\alpha',\beta'$, these
terms, in fact, vanish, because in the sum over replica indices we
exclude the diagonal terms $\alpha=\beta$ and $\alpha'=\beta'$. If we
had kept these diagonal contributions in our definition of $H_\Delta$,
a (somewhat less obvious) cancellation would have taken place between
various terms with the final results, of course, unchanged.  Inserting
the resulting expression Eq.\ref{T2} into Eq.\ref{H2}, we obtain
\begin{equation}
\left<H_\Delta^2\right >_>^c
=I_1+I_2+I_3\;,
\label{H2i}
\end{equation}
where we have defined three contributions to
$\left<H_\Delta^2\right>_>^c$,
\end{multicols}
\begin{mathletters}
\begin{eqnarray}
I_1&=&\Delta_V^2\int_{\bf r, \delta r}\sum'_{\alpha\neq\beta,\alpha\neq\beta'}
q_0^2\;e^{-2q_0^2 G^>_T(0)}
\Big[2G^>_T({\bf\delta r})+q_0^2
\big(G^>_T({\bf\delta r})\big)^2\Big]
\cos[q_0(u^<_\alpha({\bf r})-u^<_\beta({\bf r})
-u^<_{\alpha}({\bf r'})+u^<_{\beta'}({\bf r'}))]
\label{I1}\\
I_2&=&\Delta_V^2\int_{\bf r, \delta r}\sum'_{\alpha\neq\beta,\alpha\neq\beta'}
q_0^2\;e^{-2q_0^2 G^>_T(0)}
\Big[-2G^>_T({\bf\delta r})+q_0^2
\big(G^>_T({\bf\delta r})\big)^2\Big]
\cos[q_0(u^<_\alpha({\bf r})-u^<_\beta({\bf r})+
u^<_{\alpha}({\bf r'})-u^<_{\beta'}({\bf r'}))]\;,
\label{I2}\\
I_3&=&\Delta_V^2\int_{\bf r, \delta r}\sum'_{\alpha\neq\beta}
q_0^4\;e^{-2q_0^2 G^>_T(0)}
\big(G^>_T({\bf\delta r})\big)^2\times\nonumber\\
&&\Big(\cos[q_0(u^<_\alpha({\bf r})-u^<_\beta({\bf r})
-u^<_{\alpha}({\bf r'})+u^<_{\beta}({\bf r'}))]
+\cos[q_0(u^<_\alpha({\bf r})-u^<_\beta({\bf r})
+u^<_{\alpha}({\bf r'})-u^<_{\beta}({\bf r'}))]\Big)\;,
\label{I3}
\end{eqnarray}
\label{I}
\end{mathletters}
\begin{multicols}{2}
We now carefully analyze each of the above terms. Although naively
$I_1$ and $I_2$ appear as 3-replica terms, in fact, as we will see
below, $I_1$ renormalizes both $\Delta_V$ and $\Delta_h$, and $I_2$ is
irrelevant.

$I_1$ naturally splits up into a $\beta=\beta'$ term and everything
else:
\begin{eqnarray}
I_1&=&\Delta_V^2\int_{\bf r, \delta r}
q_0^2\;e^{-2q_0^2 G^>_T(0)}\times\nonumber\\
&&\Big[2G^>_T({\bf\delta r})+q_0^2
\big(G^>_T({\bf\delta r})\big)^2\Big]
\Big(I_{1a}+I_{1b}\Big)\;,
\label{I1aI1b}
\end{eqnarray}
where
\begin{mathletters}
\begin{eqnarray}
I_{1a}&=&\sum'_{\alpha\neq\beta',\beta\neq\beta'}
\cos[q_0(u^<_\alpha({\bf r})-u^<_\beta({\bf r})
-u^<_{\alpha}({\bf r'})+u^<_{\beta'}({\bf r'}))]\;,\nonumber\\
&&\label{I1a}\\
I_{1b}&=&\sum'_{\alpha\neq\beta}
\cos[q_0(u^<_\alpha({\bf r})-u^<_\beta({\bf r})
-u^<_{\alpha}({\bf r'})+u^<_{\beta}({\bf r'}))]\;.\nonumber\\
\label{I1b}
\end{eqnarray}
\end{mathletters}
Since the function $G^>_T({\bf\delta r})$ multiplying the above
expressions is a short-ranged function of $\bf\delta r$ (with range on
the order $\Lambda^{-1}$, the inverse of the ultra-violet cutoff
$\Lambda$), we can safely expand above expressions in powers of $\bf
\delta r$.
\begin{eqnarray}
\hspace{-.3cm}I_{1a}&=&\hspace{-.3cm}
\sum'_{\alpha\neq\beta',\beta\neq\beta'}
\hspace{-.3cm}\cos[q_0(u^<_\alpha({\bf r})-u^<_\beta({\bf r})
-u^<_{\alpha}({\bf r}+{\bf\delta r})+
u^<_{\beta'}({\bf r}+{\bf\delta r}))]\;,\nonumber\\
&\approx&\hspace{-.3cm}
\sum'_{\alpha\neq\beta',\beta\neq\beta'}
\hspace{-.3cm}\cos[q_0\big(u^<_\beta({\bf r})-u^<_{\beta'}({\bf r})
+{\bf\delta r}\cdot
{\bbox{\nabla}}(u^<_{\alpha}({\bf r})-u^<_{\beta'}({\bf r}))\big)]\;,
\nonumber\\
&\approx&\sum'_{\beta\neq\beta'}
(n-2)\cos[q_0(u^<_\beta({\bf r})-u^<_{\beta'}
({\bf r}))]+\mbox{irrelevant terms}\;,\nonumber\\
\label{I1a2}
\end{eqnarray}
where the ``irrelevant terms'' are of the form of a product of
$\cos[q_0(u^<_\beta({\bf r})-u^<_{\beta'}({\bf r}))]$ and derivatives
of $u^<_{\alpha}({\bf r})-u^<_{\beta'}({\bf r})$, and are clearly less
important at long length scales than the term that we have
kept. Similarly for $I_{1b}$, we obtain,
\begin{eqnarray}
\hspace{-0.5cm}I_{1b}&=&\sum'_{\alpha\neq\beta}
\cos[q_0(u^<_\alpha({\bf r})-u^<_\beta({\bf r})
-u^<_{\alpha}({\bf r}+{\bf\delta r})+
u^<_{\beta}({\bf r}+{\bf\delta r}))]\;,\nonumber\\
&\approx&\sum'_{\alpha\neq\beta}
\cos[q_0({\bf\delta r}\cdot
\bbox{\nabla}(u^<_{\alpha}({\bf r})-u^<_{\beta}({\bf r})))]\;,\nonumber\\
&\approx&\sum'_{\alpha\neq\beta}
\bigg(1-{q_0^2\over2}|{\bf\delta r}\cdot
{\bbox{\nabla}}(u^<_{\alpha}({\bf r})-u^<_{\beta}({\bf r}))|^2\bigg)\nonumber\\
&&+\mbox{irrelevant terms}\;,
\label{I1b2}
\end{eqnarray}

Performing similar local expansion in powers of $\bf\delta r$ of the
term $I_2$, Eq.\ref{I2}, it is easy to see that $I_2$ contributes
terms of the form of a product $\cos[q_0(2u^<_\alpha({\bf
r})-u^<_{\beta}({\bf r})-u^<_{\beta'}({\bf r}))]$ and derivatives of
$u^<_{\alpha}({\bf r})-u^<_{\beta'}({\bf r})$, which are {\em all
irrelevant} at long scales, near the glass transition temperature, and
we therefore drop them.

Finally, the analysis of $I_3$ is also very similar. The {\em second}
cosine in Eq.\ref{I3} generates a higher harmonic and is therefore
irrelevant as in the analysis of $I_2$, while the {\em first} cosine
gives
\begin{eqnarray}
I_3&=&\Delta_V^2\int_{\bf r, \delta r}\sum'_{\alpha\neq\beta}
q_0^4\;e^{-2q_0^2 G^>_T(0)}
\big(G^>_T({\bf\delta r})\big)^2\times\nonumber\\
&&\cos[q_0(u^<_\alpha({\bf r})-u^<_\beta({\bf r})
-u^<_{\alpha}({\bf r}+{\bf\delta r})+
u^<_{\beta}({\bf r}+{\bf\delta r}))]\nonumber\\
&\approx&\Delta_V^2\int_{\bf r, \delta r}\sum'_{\alpha\neq\beta}
q_0^4\;e^{-2q_0^2 G^>_T(0)}
\big(G^>_T({\bf\delta r})\big)^2\times\nonumber\\
&&\bigg(1-{q_0^2\over2}|{\bf\delta r}\cdot
{\bbox{\nabla}}(u^<_{\alpha}({\bf r})-u^<_{\beta}({\bf r}))|^2\bigg)\nonumber\\
&+&\mbox{irrelevant terms}\;,\nonumber\\
\label{I3i}
\end{eqnarray}

Putting all this together into Eq.\ref{H2i} and dropping irrelevant
and constant terms (which renormalize the constant part of the free
energy), we obtain
\begin{eqnarray}
\left<H_\Delta^2\right >_>^c&\approx&
-\Delta_V^2 e^{-2q_0^2 G^>_T(0)}\times\nonumber\\
&\times&\int_{\bf r}\sum'_{\alpha\neq\beta}
\Big[A(2-n)\cos[q_0(u^<_\alpha({\bf r})-u^<_{\beta}({\bf r}))]\nonumber\\
&+&{1\over2}C_{ij}\partial_i\Big(u^<_{\alpha}
({\bf r})-u^<_{\beta}({\bf r})\Big)
\partial_j\Big(u^<_{\alpha}({\bf r})-u^<_{\beta}({\bf r})\Big)\bigg]\;,
\nonumber\\
\label{H2ii}
\end{eqnarray}
where we have defined
\begin{mathletters}
\begin{eqnarray}
A&\equiv&q_0^2\int_{\bf\delta r}\Big[2G^>_T({\bf\delta r})
+q_0^2\big(G^>_T({\bf\delta r})\big)^2\Big]\;,\label{A}\\
C_{ij}&\equiv&2q_0^4\int_{\bf\delta r}\Big[G^>_T({\bf\delta r})
+q_0^2\big(G^>_T({\bf\delta r})\big)^2\Big]\delta r_i \delta r_j\;.
\label{Cij}
\end{eqnarray}
\label{ACij}
\end{mathletters}

The first term in $A$ and $C_{ij}$ is just a Fourier transform of
$G^>_T({\bf r})$ evaluated at ${\bf q}=0$. Because the propagator of
the high wavevector modes, $G^>_T({\bf q})$ by definition only has
support near the lattice cutoff $q_\perp=\Lambda$, with $G^>_T({\bf
q}=0)=0$, the first terms in $A$ and $C_{ij}$ vanish.  With this
simplification $A$ and $C_{ij}$ can be evaluated via contour
integration.

For $A$ we easily find,
\begin{mathletters}
\begin{eqnarray}
A&=&q_0^4\int_{\bf\delta r}\big(G^>_T({\bf\delta r})\big)^2\;,\\
&=&q_0^4\int_{\Lambda e^{-\delta\ell}}^\Lambda
{d^2 q_\perp\over(2\pi)^2}\int_{-\infty}^\infty{d q_z\over 2\pi}
\big(G^>_T({\bf q})\big)^2\;,\\
&=&q_0^4\int_{\Lambda e^{-\delta\ell}}^\Lambda
{d^2 q_\perp\over(2\pi)^2}\int_{-\infty}^\infty{d q_z\over 2\pi}
{1\over(K  q_\perp^4+B q_z^2)^2}\;,\\
&=&{q_0^4\over8\pi\Lambda^4\sqrt{K^3B}}\,\delta\ell\;\\
&\equiv& A_1\delta\ell\;.
\end{eqnarray}
\end{mathletters}

The evaluation of $C_{ij}$ is a similar but a little more involved.
Because of the anisotropic scaling, $q_z^2\sim q_\perp^4<<q_\perp^2$,
enforced by the quadratic part of the elastic Hamiltonian,
Eq.\ref{Hr}, it is clear that we are only interested in $C_{ij}$ with
$i,j$ taking on values $x,y$, with
\begin{mathletters}
\begin{eqnarray}
C_{ij}^\perp&\equiv&2q_0^6 \int_{\bf\delta r}
\big(G^>_T({\bf\delta r})\big)^2\delta r_i^\perp \delta r_j^\perp\;,\\
&\equiv&C_\perp\delta^\perp_{ij}\;,
\end{eqnarray}
\label{Cperpij}
\end{mathletters}
where the second equation follows by rotational invariance in the
xy-plane, which enforces $C_{xx}=C_{yy}\equiv C_\perp$ and
$C_{xy}=C_{yx}=0$, with
\begin{mathletters}
\begin{eqnarray}
C_\perp&=&q_0^6\int d^2r dz\, r_\perp^2
\big(G^>_T(r_\perp,z)\big)^2\;,\\
&=&{c\,q_0^6\over4\pi\Lambda^6\sqrt{K^3B}}\delta\ell\;,\\
&\equiv& A_2\delta\ell\;.
\end{eqnarray}
\label{Cperp}
\end{mathletters}
In the above $c$ is a dimensionless constant of order $1$. Now
comparing the expression for $\left<H_\Delta^2\right >_>^c$,
Eq.\ref{H2ii}, with the ``bare'' Hamiltonian, Eq.\ref{Hr}, we find
\begin{mathletters}
\begin{eqnarray}
\delta\Delta_V^{(2)}&=&-{1\over2}(2-n)A_1\Delta_V^2\delta\ell\;,\\
_{n\rightarrow0}&=&-A_1\Delta_V^2\delta\ell\;,\nonumber\\
\delta\Delta_h&=&A_2\Delta_V^2\delta\ell\;.
\end{eqnarray}
\label{DeltaVDeltah}
\end{mathletters}
We note that the graphical correction to the random field disorder
$\Delta_V$ is {\em stabilizing} (negative), leading to a stable
Cardy-Ostlund--like glassy fixed line (Fig.6), {\em only} for
$n\rightarrow0$ ($n<2$), which can be physically understood from the
discussion given in Sec.\ref{stability_harmonic_elastic}.

After the above integration of the short wavelength modes
$u^>_\alpha$, the effective Hamiltonian is a functional of the long
wavelength modes $u^<_\alpha$, which have an ultra-violet cutoff
$\Lambda e^{-\ell}$ that is different from the cutoff $\Lambda$ of the
original theory. Therefore in order to identify the new effective
coupling constants $\Delta_V(\ell), \Delta_h(\ell),\ldots$ we need to
perform the second part of the renormalization group transformation,
that involves the rescaling of variables, which restores the cutoff in
the effective theory back to $\Lambda$. The rescaling that
accomplishes this is
\begin{eqnarray}
r_\perp&=&r_\perp'\;e^{\ell}\;,\nonumber\\
z&=&z'\;e^{\omega\ell}\;,\nonumber\\
u^<_\alpha({\bf r})&=&e^{\phi\ell}u_\alpha({\bf r'})\;,
\label{rescaleApp}
\end{eqnarray}
Because the random-field nonlinearity is a periodic function, it is
convenient (but not necessary) to take the arbitrary field dimension
$\phi=0$, thereby preserving the period $2\pi/q_0$ under the
renormalization group transformation.\cite{q0flow} Under this
transformation the resulting effective Hamiltonian functional can be
restored into its original form, Eq.\ref{Hr} with the effective
$\ell$-dependent couplings, that satisfy differential recursion
relations, Eqs.\ref{flows_random_field}.

\section{Appendix B: Details of the $3<d<7$ 
Functional renormalization group analysis, including anharmonic
elasticity}
\label{FRGapp}

In this appendix we present the details of the functional
renormalization group (FRG) calculation for the effective replicated
Hamiltonian, $H=H_0+H_{\rm int}$, given by Eq.\ref{Hr_anharmonic_rf2},
with the quadratic part $H_0$ 
\begin{equation}
H_0[u_\alpha]={1\over2}\int d^d r
\sum_{\alpha=1}^n\Big[{K}(\nabla_\perp^2
u_\alpha)^2+{B}(\partial_z u_\alpha)^2\Big]\;,
\label{H0App}
\end{equation}
and the anharmonic part of the effective Hamiltonian $H_{\rm
int}=H_{\rm anh}+H_\Delta$ given by
\begin{mathletters}
\begin{eqnarray}
H_{\rm anh}&=&{1\over2}\int d^d r\sum_{\alpha=1}^n
\Big[-B\partial_z u_\alpha(\bbox{\nabla}_\perp u_\alpha)^2
+{B\over4}(\bbox{\nabla}_\perp u_\alpha)^4\Big]\;,\nonumber\\
&&\label{HanhApp}\\
H_\Delta&=&\int d^d r\sum_{\alpha,\beta=1}^n
\Big[{1\over4}\Delta_h(u_\alpha-u_\beta)
|\bbox{\nabla_\perp}(u_\alpha-u_\beta)|^2-\nonumber\\
&&\Delta_V(u_\alpha-u_\beta)\Big]\;,\nonumber\\
\label{HDeltaApp}
\end{eqnarray}
\end{mathletters}
which incorporates both the elastic anharmonicities, previously
studied in Sec.\ref{anom_elasticity}, and the random field and tilt
disorders of a randomly pinned smectic liquid crystal. The result of
the analysis presented below is the set of the FRG flow equations
given in Eqs.\ref{Kfrg_flow}-\ref{DeltaVfrg_flow} of
Sec.\ref{random_field_3-5}.

Our RG analysis will closely follow the approach presented in the
previous RG sections, in particular paralleling the calculation and
notation for a single harmonic disorder in three dimensions, presented
in Appendix A. The difference here is that we are interested in the
effect of disorder in dimensions higher than $3$. As a result, as
discussed in detail in Sec.\ref{random_field_3-5}, in these higher
dimensions all harmonics of the random pinning potential are equally
relevant and need to be kept track of by studying the evolution of
arbitrary periodic functions $\Delta_V(u_\alpha-u_\beta)$ and
$\Delta_h(u_\alpha-u_\beta)$ under the RG transformation. Furthermore,
in contrast to Appendix A, here, in addition, we are including
anharmonic nonlinearities, which lead to the weak logarithmic
anomalous elasticity discussed in Sec.\ref{random_field_3-5} for
$5<d<7$, and to the power-law anomalous elasticity for $d<5$ discussed
in Sec.\ref{anom_elasticity}.

\subsection{{\em Assuming} tilt disorder is irrelevant for $d>5$}

We have argued in Secs.\ref{stability_harmonic_elastic} and
\ref{tilt-only_model}, that, although in a full model of a randomly
pinned smectic both random tilt and field disorders are present, for
$3<d<5$, the random tilt disorder always dominates over the random
field (periodic) disorder. However, as the power-counting suggests,
for $d>5$, the random tilt disorder becomes irrelevant and smectic
correlations are dominated by the random field disorder. In the second
subsection of this appendix we will explicitly demonstrate that this
conclusion remains valid even when potentially singular diagrammatic
corrections to $\Delta_h(u_\alpha-u_\beta)$ are taken into account.
Hence, for the remainder of this subsection, confining our discussion
to dimensions $d>5$, where random tilt disorder is irrelevant, we
leave it out of our analysis.

The logic of the renormalization group analysis here is the same as
that used in Appendix A: we separate the fields $u_{\alpha}$ into high
and low wavevector components $u^>_{\alpha}$ and $u^<_{\alpha}$,
respectively, and perturbatively integrate the $u^>_{\alpha}$ out of
the replicated partition function.  At leading order, the correction
to the quadratic Hamiltonian $H_0$ is formally given by a cumulant
expansion
\begin{equation}
\hspace{-.1cm}
\delta H[u^<_\alpha]=\left <H_{\rm int}[u^<_\alpha+u^>_\alpha]\right >_>^c -
{1\over 2}\left<H_{\rm int}[u^<_\alpha+u^>_\alpha]^2\right >_>^c
+\ldots\;,
\end{equation}
where as in Appendix A, we have set $T=1$, but in contrast to the
calculations there have not excluded diagonal terms in the replica
sum.

We first focus on the contributions to the random field disorder
$\Delta_V(u)$. We begin by noting that the renormalization of
$\Delta_V(u)$ gets {\em no} contributions from the elastic
anharmonicities, $\partial_z
u_{\alpha}\left(\bbox{\nabla}_{\perp}u_{\alpha}\right)^2$ and
$\left(\bbox{\nabla}_{\perp}u_{\alpha}\right)^4$, i.e. from the
vertices in $H_{\rm anh}$, Eq.\ref{HanhApp}. This is because the
graphs that look like they might renormalize
$\Delta_V(u_{\alpha}-u_{\beta})$ have $q$'s on the external legs and
hence renormalize {\em only} the tilt disorder (and other less
relevant operators), which, as we discussed above, is irrelevant for
$d>5$, which is the case we are considering here.

To first order in the disorder $\Delta_V(u_\alpha-u_\beta)$, keeping
only the relevant, non-constant terms, we obtain
\begin{eqnarray}
\delta H_\Delta^{(1)}&=&-\sum_{\alpha,\beta}\int_{\bf r}
\big<\Delta_V(u_\alpha-u_\beta)\big>_>\;,\nonumber\\
&\approx&-{1\over2}\sum_{\alpha,\beta}\int_{\bf r}
\Delta_V''(u_\alpha^<-u_\beta^<)\big<(u_\alpha^>-u_\beta^>)^2\big>_>\;,
\nonumber\\ 
&\approx&-d\ell{C_{d-1}\Lambda^{d-3}\over2\sqrt{B K}}
\sum_{\alpha,\beta}\int_{\bf r}\Delta_V''(u_\alpha^<-u_\beta^<) \;,
\end{eqnarray}
which when compared to the definition of $H_\Delta$ gives
\begin{equation}
\delta\Delta_V^{(1)}(u)\approx d\ell{C_{d-1}\Lambda^{d-3}\over2\sqrt{B K}}
\Delta_V''(u)\label{deltaDelta1}\;,
\end{equation}
with the primes indicating a partial derivative with respect to $u$.
To obtain the final result above, we used the Gaussian part of the
total effective Hamiltonian
\begin{eqnarray}
H_{\rm Gauss}&=&{1\over2}\int_{\bf r}
\Bigg\{\sum _{\alpha}\left[B\left(\partial _z u_{\alpha}\right)^2 
+ K\left(\nabla^2_{\perp}u_{\alpha}\right)^2\right]\nonumber\\
&&- \sum_{\alpha \beta}\Delta_V''(0)(u_{\alpha} - u_{\beta})^2\Bigg\}\;
\label{HGauss}
\end{eqnarray}
to calculate
\begin{equation}
\big<(u_\alpha^>-u_\beta^>)^2\big>_>
=2\int_{\bf q}^>\left[\big<|u_{\alpha}({\bf q})|^2\big>
-\big<u_{\alpha}({\bf q})u_{\beta}(-{\bf q})\big>\right]\;,
\label{uquqApp}
\end{equation}
using the corresponding propagator
\begin{equation}
\left< u_{\alpha}({\bf q})u_{\beta}(-{\bf q})\right> =
{\delta_{\alpha\beta}\over B q^2_z + Kq^4_{\perp}} 
-{2\Delta_V''(0)\over[B q^2_z+K q^4_{\perp}]^2}\;,
\label{uu_abApp}
\end{equation}
obtained from $H_{\rm Gauss}$ by manipulations virtually identical to
those described earlier for the tilt-only model,
Sec.\ref{anom_elasticity}.

The contribution to second order in $\Delta_V(u)$ is given by
\end{multicols}
\begin{eqnarray}
\delta H_\Delta^{(2)}&\approx&-{1\over4}
\sum_{\alpha_1,\beta_1,\alpha_2,\beta_2}\int_{\bf r_1,r_2}
\Delta_V''(u_{\alpha_1}^<({\bf r}_1)-u_{\beta_1}^<({\bf r}_1))
\Delta_V''(u_{\alpha_2}^<({\bf r}_2)-u_{\beta_2}^<({\bf r}_2))
I^{\alpha_2\beta_2}_{\alpha_1\beta_1}({\bf r}_1,{\bf r}_2)\;,
\nonumber\\
&&\label{deltaH2}
\end{eqnarray}
\begin{multicols}{2}
\noindent
where
\begin{mathletters}
\begin{eqnarray}
I^{\alpha_2\beta_2}_{\alpha_1\beta_1}
&=&\Big<\big(u_{\alpha_1}^>({\bf r}_1)-u_{\beta_1}^>({\bf r}_1)\big)^2
\big(u_{\alpha_2}^>({\bf r}_2)-u_{\beta_2}^>({\bf r}_2)\big)^2\Big>_>^c
\;,\nonumber\\
&&\label{Iab1}\\
&=&\Big<\big(u_{\alpha_1}^>({\bf r}_1)-u_{\beta_1}^>({\bf r}_1)\big)
\big(u_{\alpha_2}^>({\bf r}_2)-u_{\beta_2}^>({\bf r}_2)\big)\Big>_>^2
\;,\nonumber\\
&&\label{Iab2}\\
&=&\big(\delta_{\alpha_1\alpha_2}+\delta_{\beta_1\beta_2}
-\delta_{\alpha_1\beta_2}-\delta_{\beta_1\alpha_2}\big)^2
\Big[G_T^>({\bf r}_1-{\bf r}_2)\Big]^2\;,\nonumber\\
&&\label{Iab3}\\
&=&4\big(\delta_{\alpha_1\alpha_2}
+\delta_{\alpha_1\alpha_2}\delta_{\beta_1\beta_2}
-2\delta_{\alpha_1\beta_2}\delta_{\beta_1\beta_2}\big)
\Big(G_T^>(\delta{\bf r})\Big)^2\;,\nonumber\\
&&\label{Iab4}
\end{eqnarray}
\label{Iab}
\end{mathletters}
$\delta{\bf r}\equiv{\bf r}_1-{\bf r}_2$, we used the Wick
decomposition theorem for Gaussian random variables (a, b, c, d)
\begin{equation}
\langle a b c d\rangle =\langle ab\rangle \langle cd\rangle 
+ \langle ac\rangle\langle bd\rangle + \langle ad\rangle\langle bc\rangle\;,
\label{Wick}
\end{equation}
to go from Eq.\ref{Iab1} to Eq.\ref{Iab2}, and used the symmetry of
the summand in $\delta H_\Delta^{(2)}$, Eq.\ref{deltaH2} under
$\alpha,\beta$ interchange to get Eq.\ref{Iab4}.

Substituting the last expression for
$I^{\alpha_2\beta_2}_{\alpha_1\beta_1}$, Eq.\ref{Iab4}, into $\delta
H_\Delta^{(2)}$ above and using the short-range property of
$G_T^>(\delta{\bf r})$ to expand ${\bf r}_1$ as ${\bf r}_1={\bf
r}_2+\delta{\bf r}$, keeping only the most relevant terms, we obtain
\end{multicols}
\begin{eqnarray}
\hspace{-0.3cm}
\delta H_\Delta^{(2)}&\approx&-\delta\ell\,G_2\int_{\bf r}\sum_{\alpha,\beta}
\Big[\Delta_V''(u_{\alpha}^<({\bf r})-u_{\beta}^<({\bf r}))^2
-2\Delta_V''(u_{\alpha}^<({\bf r})-u_{\beta}^<({\bf r}))\Delta_V''(0)
+\sum_{\gamma}\Delta_V''(u_{\alpha}^<({\bf r})-u_{\beta}^<({\bf r}))
\Delta_V''(u_{\alpha}^<({\bf r})-u_{\gamma}^<({\bf r}))\Big]
\nonumber\\
&&\label{deltaH2b}
\end{eqnarray}
\begin{multicols}{2}
\noindent
where the constant $G_2$ is defined by
\begin{mathletters}
\begin{eqnarray}
G_2\,\delta\ell&=&\int_{\bf\delta r}\big(G^>_T({\bf\delta r})\big)^2\;,\\
&=&\int_{\Lambda e^{-\delta\ell}}^\Lambda
{d^{d-1} q_\perp\over(2\pi)^{d-1}}\int_{-\infty}^\infty{d q_z\over 2\pi}
\big(G^>_T({\bf q})\big)^2\;,\\
&=&\int_{\Lambda e^{-\delta\ell}}^\Lambda
{d^{d-1} q_\perp\over(2\pi)^{d-1}}\int_{-\infty}^\infty{d q_z\over 2\pi}
{1\over(K q_\perp^4+B q_z^2)^2}\;,\\
&=&{C_{d-1}\Lambda^{d-7}\over8\sqrt{K^3B}}\,\delta\ell\;.
\end{eqnarray}
\label{G2}
\end{mathletters}
It is easy to see by power-counting that the three-replica (last) term
in $\delta H_\Delta^{(2)}$, Eq.\ref{deltaH2b} is irrelevant relative
to the two-replica terms and can therefore be neglected. Comparing the
resulting expression for $\delta H_\Delta^{(2)}$ with $H_\Delta$,
Eq.\ref{HDeltaApp} we find
\begin{equation}
\delta\Delta_V^{(2)}(u)\approx\delta\ell{C_{d-1}
\Lambda^{d-7}\over8\sqrt{K^3B}}
\Big(\Delta_V''(u)^2-2\Delta_V''(u)\Delta_V''(0)\Big)
\label{deltaDelta2}\;.
\end{equation}
Combining the first and second order contributions to $\Delta_V(u)$,
Eqs.\ref{deltaDelta1}, \ref{deltaDelta2}, with the length and field
rescalings, Eq.\ref{rescaleApp}, necessary to bring the ultra-violet
cutoff back to $\Lambda$ we obtains the recursion relation for
$\Delta_V(u)$ given by
\end{multicols}
\begin{equation}
\partial_{\ell}\Delta_V(u,\ell)=(d+1){\Delta}(u,\ell) +
{\eta(\ell)\over q_0^2}{\Delta}_V^{\prime\prime}(u,\ell)
+\left[{1\over2}\left({\Delta}_V^{\prime\prime}(u,\ell)\right)^2 
-{\Delta}_V^{\prime\prime}(u,\ell)
{\Delta}_V^{\prime\prime}(0,\ell)\right]
{C_{d-1}\Lambda^{d-7}\over4\sqrt{K^3(\ell)B(\ell)}}\;,
\label{DeltaV_flowApp}
\end{equation}
\begin{multicols}{2}
\noindent
where, for simplicity, in the above we have set the field rescaling
$\phi=0$ and used $\omega=2$.

We now turn to the renormalization of the elastic moduli $B$ and
$K$. As discussed in Sec.\ref{stability_harmonic_elastic}, statistical
symmetry forbids renormalization of these by disorder alone (The same
argument applies to the XY spin stiffness in the random field XY
model). However, anharmonic elasticity, special to smectic liquid
crystals, conspires with the random field $\Delta_V(u)$ to renormalize
$B$ and $K$.

Calculations very similar to those for the tilt-only model treated in
Sec.\ref{anom_elasticity} show that the graphical corrections to $B$
and $K$ found in that section can be taken over for the random field
disorder, with the identification
\begin{equation}
\Delta_h^{\mbox{eff}}=-2\Delta_V''(0)/\Lambda^2\;.\label{Delta_heffApp}
\end{equation}
In addition to a detailed perturbative RG calculation, this can also
be easily seen by comparing
\begin{equation}
\Delta_V(u_\alpha-u_\beta)\approx {\rm
const.}+{1\over2}\Delta_V''(0)(u_\alpha-u_\beta)^2\;,
\end{equation}
to the disorder part of the the tilt-only Hamiltonian.

Using this mapping we immediately find
\begin{mathletters}
\begin{eqnarray}
{dK\over d\ell}&=&\left[d - 3 +
{d^2 - 12d + 23\over8(d^2 - 1)}\left({B\over K^5}\right)^{1/2}
\hspace{-.5cm}\Delta_V^{\prime\prime}(0)C_{d-1}\Lambda^{d-7}\right]
\hspace{-0.1cm}K\nonumber\\
&&\label{Kfrg_flowApp}\\
{dB\over d\ell}&=&\left[d - 3 + {3\over8}
\left({B\over K^5}\right)^{1/2}
\hspace{-.5cm}\Delta_V^{\prime\prime}(0)C_{d-1}\Lambda^{d-7}\right]
\hspace{-0.1cm}B\;.\nonumber\\
&&\label{Bfrg_flowApp}
\end{eqnarray}
\end{mathletters}
Now identifying the dimensionless measure of the random field disorder 
\begin{equation}
\tilde{\Delta}_V(u,\ell)\equiv {C_{d-1}\Lambda^{d-7}
\over4\sqrt{K^3(\ell)B(\ell)}}\,\Delta_V(u,\ell)\;,
\label{tildeDeltaApp}
\end{equation}
and using it inside Eqs.\ref{DeltaV_flowApp}, \ref{Kfrg_flowApp} and
\ref{Bfrg_flowApp}, we obtain the FRG flow Eqs.\ref{DeltaVfrg_flow},
\ref{Kfrg_flow} and \ref{Bfrg_flow} given in the
Sec.\ref{random_field_3-5}.

\subsection{Proving that tilt disorder is irrelevant for $d>5$}

In the previous subsection we have ignored the tilt disorder for
$d>5$, arguing for its irrelevance based on power-counting in these
higher dimensions. However, the (supernaturally) alert reader might be
concerned (as we were at first) that the singular behavior of
$\tilde{\Delta}_V^*(u)$, Eq.\ref{DeltaFfp} (in particular, the
divergence of its fourth derivative with respect to $u$,
$\tilde{\Delta}_V^{(iv)}(u,\ell)$ as $\ell\rightarrow\infty$) could
potentially invalidate the simple power-counting argument that tilt
disorder is irrelevant for $d>5$. Here, we demonstrate that this does
not happen, by deriving a {\em functional} renormalization group
recursion relation for $\Delta_h(u,\ell)$ to {\em linear} order in
$\Delta_h(u,\ell)$ itself at the (FRG) fixed point
$\tilde{\Delta}_V^*(u)$, Eq.\ref{DeltaFfp}, found in
Sec.\ref{random_field_3-5}. We obtain
\end{multicols}
\begin{equation}
\partial_{\ell}\Delta_h(u,\ell)=\left(d-1\right)\Delta_h(u,\ell) +
\bigg({\eta(\ell)\over q_0^2}+ D(u)\bigg)\Delta_h''(u,\ell)+
c\left(\tilde{\Delta}_{V*}'''(u)\right)^2\sqrt{K^3(\ell)
B(\ell)}\Lambda^{5-d}/C_{d-1}\;,
\label{Deltahfrg_flowApp}
\end{equation}
\bottom{-2.7cm}
\begin{multicols}{2}
\noindent
where we have defined the ``position-dependent diffusivity''
\begin{equation}
D(u)\equiv c_2\left(\tilde{\Delta}_{V*}''(u)-\tilde{\Delta}_{V*}''(0)\right)\;,
\label{Diffusivity_app}
\end{equation}
and $c$ and $c_2$ are dimensionless constants of order unity.

As in the tilt-only model, the important quantity proves not to be
$\Delta_h(u,\ell)$, but rather the dimensionless combination
\begin{equation}
g(u,\ell)\equiv\Delta_h(u,\ell)\left(B(\ell)\over
K^5(\ell)\right)^{1/2} C_{d-1}\Lambda^{d-5}
\;,\label{g2udimensionless_app}
\end{equation}
whose recursion relation readily follows from those for $K(\ell)$,
$B(\ell)$, and $\Delta_h(u,\ell)$,
Eqs.\ref{Kfrg_flowApp},\ref{Bfrg_flowApp}, and \ref{Deltahfrg_flowApp},
respectively:
\end{multicols}
\begin{equation}
\partial_{\ell}g(u,\ell)=\left(5-d+c_3{\tilde{\Delta}_{V*}''(0)\over
\lambda^2(\ell)}\right)g(u,\ell)+\left({\eta(\ell)\over q_0^2}
+D(u)\right)g''(u,\ell)+ 
c\ {\left(\tilde{\Delta}_{V*}'''(u)\right)^2\over\lambda^2(\ell)}\;,
\label{gfrg_flow}
\end{equation}
\begin{multicols}{2}
\noindent
where $c_3\equiv 3/4+(d^2-12d+23)/(2(d^2-1))$. 

The solution to this inhomogeneous linear partial differential
equation can, as usual, be written as a sum of the solution to the
{\em homogeneous} equation obtained by dropping the last term, plus
any solution of the inhomogeneous equation.

The solution to the homogeneous equation can be written as an
expansion in a complete basis
\begin{equation}
g_{hom}(u,\ell)=\sum_{n=0}^\infty g_n(\ell)\phi_n(u)\;,
\label{g_hom}
\end{equation}
where the set $\phi_n(u)$ are the eigenfunctions of the operator
$D(u)\partial_u^2$ defined and discussed in
Sec.\ref{random_field_3-5}.

The expansion coefficients $g_n(\ell)$ can be readily
found from Eq.\ref{gfrg_flow} to be
\begin{equation}
g_n(\ell)=g_n(0)e^{-\lambda_n\ell}\exp\bigg(-\int_0^\ell
d\ell'{c_3|\tilde{\Delta}_{V*}''(0)|\over\lambda^2(\ell')}\bigg)\;,
\label{gn_ell}
\end{equation}
with
\begin{equation}
\lambda_n\equiv d-5+|\Gamma_n|\;,
\end{equation}
which is positive for all $n$ for $d>5$, and we have dropped the
$\eta(\ell)$ term, which vanishes exponentially fast as
$\ell\rightarrow\infty$. Since the second exponential in
Eq.\ref{gn_ell} is a monotonically decreasing function of $\ell$,
every $g_n(\ell)$, and, hence, the entire homogeneous solution
Eq.\ref{g_hom} for $g(u,\ell)$, vanishes as $\ell\rightarrow\infty$.

Thus, the only possible problem with neglecting tilt disorder (for
$d>5$) must come from the {\em inhomogeneous} solution for
$g(u,\ell)$. We will now show that this vanishes as well, thereby
proving that tilt disorder is irrelevant for $d>5$.

We do this by finding an asymptotic (large $\ell$) solution
$g_I(u,\ell)$ to the full, inhomogeneous equation of the form
\begin{equation}
g_I(u,\ell)=\sum_{n=1}^\infty {f_n(u)\over\ell^n}\;,
\label{gI_ell}
\end{equation}
which clearly vanishes as $\ell\rightarrow\infty$. 

Inserting Eq.\ref{gI_ell} into Eq.\ref{gfrg_flow}, and keeping in mind
that $1/\lambda^2(\ell)$ is itself $\propto 1/\ell$, as
$\ell\rightarrow\infty$ (see Eq.\ref{lambda_sol}), we find an equation
for $f_1(u)$ by equating coefficients of the leading order $1/\ell$
terms on both sides of Eq.\ref{gfrg_flow}. The resulting equation
reads
\begin{equation}
D(u) f''_1(u)-(d-5)f_1(u)=
-{c A(d)|\tilde{\Delta}_{V*}'''(0)|^2\over|\tilde{\Delta}_{V*}''(0)|}\;,
\label{f1}
\end{equation}
where we have defined $A(d)\equiv (d^2-1)/(d^2+6d-13)$ and used
Eq.\ref{lambda_sol} for $\lambda(\ell)$. Once $f_1(u)$ is known, the
remaining $f_n$'s for $n>1$ can be determined by equating coefficients
of $1/\ell^n$ on both sides of Eq.\ref{gfrg_flow} after inserting the
ansatz Eq.\ref{gI_ell}. This gives
\begin{equation}
D(u)f''_n(u)-(d-5)f_n(u) = (c_3 A(d)+1-n)f_{n-1}(u)\;.
\label{fn}
\end{equation}

Now, to complete the proof that $g_I(u,\ell)$ vanishes as
$\ell\rightarrow\infty$, we need merely show that all $f_n$'s obtained
from Eq.\ref{f1} and Eq.\ref{fn} are finite for all $n$.  If $f_1(u)$
is finite, then clearly $f_2(u)$ is also finite, since no derivatives
of $f_1(u)$ (which could potentially diverge if $f_1(u)$ is cusped)
appear in Eq.\ref{fn}. This finiteness argument then obviously
recursively carries through via Eq.\ref{fn} to {\em all} of other
$f_n(u)$'s.

Although $f_1(u)$ can be found explicitly applying standard
integration methods to Eq.\ref{f1}, our only goal here is to
demonstrate that $f_1(u)$ is finite.  To do this we simply expand
$f_1(u)$ in eigenfunctions $\phi_n(u)$ of the operator
$D(u)\partial_u^2$:
\begin{equation}
f_1(u)=\sum_{n=0}^\infty a_n \phi_n(u)\;,
\label{f_1expand}
\end{equation}
Inserting this expansion into Eq.\ref{f1}, multiplying both sides by
$\phi_m(u)$, integrating from $u=0$ to $u=a$, and using the eigenvalue
equation, Eq.\ref{eigenvalue_eqn} and the orthogonality relation
Eq.\ref{orthogonality}, we obtain
\begin{equation}
a_m={c A(d)\over\Big(d-5+|\Gamma_m|\Big)|\tilde{\Delta}_{V*}''(0)|}\ S_m\;,
\label{a_m}
\end{equation}
where the source
\begin{equation}
S_m\equiv\int_0^a{|\tilde{\Delta}_{V*}'''(u)|^2\phi_m(u)\over I_m D(u)} d u
\label{S_m}
\end{equation}
was shown to be finite for all $m$ (including $m=0$) in
Sec.\ref{random_field_3-5}. 

Hence, $f_1(u)$ will be finite for all $u$, {\em provided} that the sum
\begin{equation}
\sum_{n=0}^\infty {S_n\phi_n(u)\over d-5+|\Gamma_n|}=
{|\tilde{\Delta}_{V*}''(0)|\over c A(d)}f_1(u)
\label{sum_finite}
\end{equation}
converges for all $u$. 

To prove that it does, let us follow the obvious convention of ordering
the eigenmodes $n$ such that $|\Gamma_n|$ is a monotonically
increasing function of $n$. Then, for large $n$, we can use the WKB
solution for the eigenfunctions $\phi_n(u)$
\begin{equation}
\phi_n(u)=D^{1/4}(u)\sin\Big[n q_0\overline{D}^{1/2}
\int_0^u{d u'\over D^{1/2}(u')}\Big] 
\label{WKB1}
\end{equation}
and eigenvalues $\Gamma_n$
\begin{equation}
\Gamma_n=-\overline{D} n^2 q_0^2\;,
\label{WKB2}
\end{equation}
where we have defined the suitably averaged diffusion constant
\begin{equation}
\overline{D}\equiv\Big[{1\over a}\int_0^a{d u'\over
D^{1/2}(u')}\Big]^{-2}\;,
\label{WKB3}
\end{equation}
to prove the convergence of the sum in
Eq.\ref{sum_finite}. Eq.\ref{WKB1} shows that $\phi_n(u)$ is bounded
from above for all $u$ (by $[\mbox{max}_u D(u)]^{1/4}$ ), which in
turn implies that the $S_n$'s are also bounded above (since
$\tilde{\Delta}_{V*}'''(u)$ is finite for all $u$, and so is
$\phi_n(u)$) as $n\rightarrow\infty$. Then using Eq.\ref{WKB2} for the
$\Gamma_n$, we see that the large $n$ behavior of the summand in
Eq.\ref{sum_finite} is bounded above by $\mbox{const.}/n^2$. Hence,
the sum converges, and so $f_1(u)$ is in fact finite for all $u$;
consequently all $f_n(u)$'s are also finite. This therefore
demonstrates that $g(u,\ell)$ vanishes as $\ell\rightarrow\infty$,
proving our assertion that tilt disorder is irrelevant for $d>5$.

\section{Appendix C: Derivation of smectic ``coulomb gas''
dislocation theory, without Euler-Lagrange equation}
\label{CoulombGasApp}
In this appendix we derive the Coulomb gas description of smectic
dislocation loops characterized by an effective Hamiltonian,
Eq.\ref{Hd}. However, here, in contrast to the derivation of the main
text, we accomplish this {\em without} minimizing the Hamiltonian with
respect to smooth smectic deformations. 

Our starting point is the elastic Hamiltonian for a tilt-only model of
a randomly pinned smectic, given in Eq.\ref{Hu_V=0}
\begin{equation}
H[u]=\int \! d^d{\bf r} \bigg[{B\over 2}(\partial_z {u})^2 +
{K\over 2}(\nabla^2_\perp {u})^2 + {\bf h}({\bf
r})\cdot\bbox{\nabla_\perp} u\bigg]\;.
\label{HelApp}
\end{equation}
We include dislocations in above description by allowing the layer
displacement $u({\bf r})$ to be a multi-valued function, such that,
\begin{equation}
\bbox{\nabla}\times\bbox{\nabla} u={\bf m}\;.\label{dislocationApp}
\end{equation}
We then decompose $\bbox{\nabla} u$ into a sum of a singular, {\em
purely transverse} part $\bf v_d$ satisfying
\begin{equation}
\bbox{\nabla}\times{\bf v}_d={\bf m}\;,\label{vApp}
\end{equation}
and a nonsingular, purely longitudinal part
$\bbox{\nabla} u_p$, according to
\begin{equation}
\bbox{\nabla} u={\bf v}_d+\bbox{\nabla} u_p\;.\label{decomposeApp}
\end{equation}
The difference between this approach and our earlier derivation in
Sec.\ref{dislocations} is in our arbitrary choice of the separation of
$\bbox{\nabla} u$ into phonon and dislocation parts. In
Sec.\ref{dislocations}, we chose ${\bf v}_d$ to minimize $H$ given
$\bf m$; here, we simply choose it to be purely transverse. We will
now demonstrate that this difference between these two {\em arbitrary}
choices has no effect on the final answer, as it should not.

Substituting this decomposition into $H[u]$, Eq.\ref{HelApp}, we
obtain
\begin{equation}
H_{\rm tot}[u_p,{\bf v}_d]=H_{\rm el}[u_p]+H_{\rm v}[{\bf v}_d]
+H_{\rm int}[u_p,{\bf v}_d]\;,
\end{equation}
where (dropping the subscript $d$ on $\bf v$),
\begin{mathletters}
\begin{eqnarray}
H_{\rm v}&=&\int\! d^d{\bf r}\bigg[{B\over 2}v_z^2 
+ {K\over2}(\bbox{\nabla}_\perp\cdot{\bf v}_\perp)^2 
+ {\bf h}({\bf r})\cdot{\bf v}_\perp\bigg]\;,\label{HvAppa}\\
&=&\int\! d^d{\bf r}\bigg[{B\over 2}v_z^2 
+ {K\over2}(\partial_z v_z)^2 
+ {\bf h}({\bf r})\cdot{\bf v}_\perp\bigg]\;,\label{HvAppb}
\end{eqnarray}
\label{HvApp}
\end{mathletters}
and
\begin{mathletters}
\begin{eqnarray}
H_{\rm int}&=&\int\! d^d{\bf r}\bigg[B v_z\partial_z u
+ K(\bbox{\nabla}_\perp\cdot{\bf v}_\perp)\nabla_\perp^2 u\bigg]\;,
\label{HintAppa}\\
&=&\int\! d^d{\bf r}\bigg[B v_z\partial_z u
- K \partial_z v_z\nabla_\perp^2 u\bigg]\;.
\label{HintAppb}
\end{eqnarray}
\label{HintApp}
\end{mathletters}
In going from Eq.\ref{HvAppa} to Eq.\ref{HvAppb} and from
Eq.\ref{HintAppa} to Eq.\ref{HintAppb}, we used the purely transverse
property of ${\bf v}$, $\bbox{\nabla}\cdot{\bf
v}=0$, or equivalently,
\begin{equation}
\bbox{\nabla}_\perp\cdot{\bf v}_\perp+\partial_z v_z=0\;,
\label{transverse_vApp}
\end{equation}
to eliminate ${\bf v}_\perp$ in favor of $v_z$.

We now integrate over the single-valued phonon degree of freedom
$u_p$. After a simple Gaussian integration, some algebra, and Fourier
transformation, we obtain an effective Hamiltonian
\begin{equation}
H_d=\int_{\bf q}\left[{K q^4\over2\Gamma_q}|v_z({\bf q})|^2
+ {\bf b}({\bf q})\cdot{\bf h}(-{\bf q})\right]\;,
\label{HdApp}
\end{equation}
where, $\Gamma_q= q_z^2 + \lambda^2 q_\perp^4$, and ${\bf b}({\bf q})$
is given by
\begin{equation}
b_i({\bf q})=\left[P^T_{\perp ij}({\bf q})+
{q^2\over\Gamma_q}P^L_{\perp ij}({\bf q})\right]\, v_{\perp j}({\bf q})\;,
\label{bApp}
\end{equation}
with,
\begin{mathletters}
\begin{eqnarray}
P^T_{\perp ij}({\bf q})&=&\delta_{i j}^\perp 
-{q_i^\perp q_j^\perp\over q_\perp^2}\;,\label{PperpApp}\\
P^L_{\perp ij}({\bf q})&=&{q_i^\perp q_j^\perp\over q_\perp^2}
\;,\label{PlongApp}
\end{eqnarray}
\label{PperplongApp}
\end{mathletters}
the transverse and longitudinal projection operators in the space
($\perp$) perpendicular to $\hat{\bf z}$. 

Now Fourier transforming Eq.\ref{vApp}, and solving it for $\bf v({\bf
q})$, keeping in mind that $\bf v$ is purely transverse, we find
\begin{equation}
v_i({\bf q})=i\epsilon_{ijk} q_j m_k/q^2\;.\label{vsolutionApp}
\end{equation}
Using this solution to eliminate $\bf v$ inside $H_d$, Eq.\ref{HdApp},
above, in favor of the dislocation density $\bf m$, we obtain the
final expression for the dislocation loop Hamiltonian, Eq.\ref{Hd},
quoted in the main text.

\section{Appendix D: Analysis of fluctuations in the {\em dual}
model of disordered smectics in type I limit}
\label{fluctuations_typeI}
In this appendix we analyze the type I regime of the dual disordered
smectic liquid crystal model derived and studied in
Sec.\ref{dislocations}. In this regime, we will compute {\em exactly}
the effective free energy and as a by-product obtain from it the
fluctuation corrections to the reduced dual transition temperature,
given in Eq.\ref{tR}.

Our starting point is the replicated dual ``action'' $S_r$ given in
Eq.\ref{Sr}, together with the quenched ``gauge field'' $\bf a$
variance, given by Eq.\ref{aa2}. We build on the ideas developed in
Refs.\onlinecite{HLM}, generalizing them here to disordered
systems. As discussed in these references and in
Sec.\ref{dislocations}, in the type I regime the order parameter
$\psi$ is significantly stiffer than the gauge field $\bf A$ and can
therefore be accurately treated in a mean-field approximation. This
amounts to taking $\psi({\bf r})$ to be constant in space. The
constant $\psi$ is then calculated by minimizing the resultant free
energy, which can now be computed {\em exactly}, without further
approximations. As is clear from the discussion in
Sec.\ref{dislocations}, to do so, we must calculate
\begin{equation}
\overline{Z^n}={1\over N_a}\int' [d {\bf a}] \prod_{\alpha=1}^n
\left[d{\bf A}_\alpha\right] 
e^{-S_{mf}\left[\psi,{\bf A}_\alpha,{\bf a}\right]}\;,
\label{ZrApp}
\end{equation}
with
\end{multicols}
\begin{eqnarray}
S_{mf}[\psi,{\bf A}_\alpha,{\bf a}]&=&S_r[\psi,{\bf A}_\alpha,{\bf a}]
+\sum_{\bf q}{\Gamma^2_q q^2 \over2\Delta_h
q^2_{\perp}q^2_z} P^{\perp} _{ij} a_i ({\bf q})a_j (-{\bf q})\;,\\
&=&nV \left(t\left|\psi \right|^2 + u \left|\psi \right|^4\right)
+{1\over 2}\int_{\bf r}\Big[\sum_\alpha|{\bf A}_{\alpha}|^2
-n|{\bf a}|^2 + i2\sum_{\alpha}{\bf A}_{\alpha}\cdot{\bf a}\Big]
|\psi|^2\nonumber\\
&&+{1 \over 2}\sum_{{\bf q},\alpha}
\Big[G^{-1}_A({\bf q}) P^{\perp}_{ij}({\bf q}) 
A_{\alpha i}({\bf q}) A_{\alpha j}(-{\bf q}) 
+ G^{-1}_a({\bf q}) P^{\perp}_{ij}({\bf q}) a_i({\bf q}) a_j(-{\bf q})\Big]\;,
\label{Smf}
\end{eqnarray}
\begin{multicols}{2}
The prime on the integral in Eq.\ref{ZrApp} indicates that it is
constrained to be taken only over the {\em transverse} parts of ${\bf
A}_\alpha$ and $\bf a$, $V$ is the volume of the system, $N_a$ is the
normalization factor coming from the probability distribution of ${\bf
a}$, and will be chosen such that $\overline{Z^n}\rightarrow 1$ as
$n\rightarrow0$, and
\begin{mathletters}
\begin{eqnarray}
G^{-1}_A&=&{\Gamma_q\over Kq_{\perp}^2}
\;,\label{G-1_A}\\
G^{-1}_a&=&{\Gamma^2_q q^2 \over\Delta_h q^2_{\perp}q^2_z}
\;,\label{G-1_a}
\end{eqnarray}
\end{mathletters}
can be read off from Eqs.\ref{Sr} and \ref{aa2}, taking the appropriate
long wavelength limit in the latter.

The great virtue of the effective ``action'' $S_{mf}$ is that it is
{\it quadratic} in ${\bf A}_{\alpha}$ and ${\bf a}$ and therefore
allows exact evaluation of the functional integrals over them in
Eq.\ref{ZrApp}. Taking these Gaussian integrals, we obtain
\end{multicols}
\begin{equation}
\ln\overline{Z^n}=-V\Big[n\left(t|\psi|^2 + u|\psi|^4\right)
+{n(d-2)\over2}\int_{\bf q}\ln\left(G^{-1}_A({\bf q})+|\psi|^2 \right)
+{(d-2)\over2}\int_{\bf q}\ln\left(G^{-1}_a({\bf q}) 
- n|\psi|^2 \right)\Big]
-\ln N_a\;,\label{lnZn}
\end{equation}
\begin{multicols}{2}
\noindent
where the factors of $d-2$ correspond to the fact that there are $d-2$
transverse ``gauge-field'' components of ${\bf A}_\alpha$ and $\bf a$,
due to the transversality constraint on the functional integral
discussed above.

We observe that, as expected, the normalization factor $N_a$ for the
probability distribution of $\bf a$, given by
\begin{equation}
\ln N_a = - V{(d-2)\over2}\int_{\bf q}\ln\big(G^{-1}_A({\bf q}) 
\big)\;,
\label{lnNa}
\end{equation}
guarantees that $\ln\overline{Z^n}\rightarrow 0$ as $n\rightarrow0$,

Incorporating $N_a$ into Eq.\ref{lnZn} gives
\end{multicols}
\begin{equation}
\ln\overline{Z^n}=-V n\Big[\left(t|\psi|^2 + u|\psi|^4\right)
+{(d-2)\over2}\int_{\bf q}\ln\left(G^{-1}_A({\bf q}) +|\psi|^2 \right)
+{(d-2)\over2n}\int_{\bf q}\ln\left(1 - n G_a({\bf q})|\psi|^2 \right)\Big]\;.
\label{lnZnNa}
\end{equation}
Expanding this in $n$, we find:
\begin{equation}
\ln\overline{Z^n}=-V n\Big[t|\psi|^2 + u|\psi|^4
+{d-2\over2}\int_{\bf q}\ln\left(G^{-1}_A({\bf q}) +|\psi|^2 \right)
-{d-2\over2}|\psi|^2\int_{\bf q}G_a({\bf q})\Big]
+ O\left(n^2\right)\;.
\label{lnZnNa2}
\end{equation}
\begin{multicols}{2}
Now using the standard replica expression for the average
free energy, we finally obtain
\begin{mathletters}
\begin{eqnarray}
\overline{F}_{MF}&=&\stackrel{\rm lim}{n\rightarrow 0} -k_BT
{\overline{Z^n} - 1\over n}\;, \\ 
&=&\stackrel{\rm lim}{n\rightarrow 0} -k_BT
{\ln\overline{Z^n}\over n}\;,\\ 
&=& k_B T V\Big[t_R|\psi|^2+u|\psi|^4
+{d-2\over2}\int_{\bf q}\ln\big(G^{-1}_A+|\psi|^2\big)\Big]\;,
\nonumber\\
\label{Fmf_final}
\end{eqnarray}
\end{mathletters}
where the renormalized reduced temperature $t_R$ is given by
\begin{equation}
t_R=t-{d-2\over2}\int_{\bf q} G^{-1}_a({\bf q})\;,
\end{equation}
which, using Eq.\ref{G-1_a}, is precisely the result used in
Eq.\ref{tR} of Sec.\ref{dislocations} to assess the stability of the
disordered smectic to dislocation unbinding.  Thus, the result of our
more rigorous and systematic perturbation theory is recovered by this
fluctuation corrected mean field theory. The annealed term in
Eq.\ref{Fmf_final} is analogous to the term that leads to a
fluctuation-driven first-order transition in type I
superconductors.\cite{HLM} Here, however, its effects are innocuous:
expanding the integrand in $|\psi|^2$ leads only to a finite
renormalization of $t$, plus a non-analytic $|\psi|^{d + 1}$ piece,
which only leads to a finite renormalization of $u$ in $d = 3$.

\end{multicols}

\begin{references}
\bibitem{sr} We expect the disorder to be short-range correlated in
space. One might expect that the continuity of the random aerogel
strands should lead to correlated disorder, with a
wavevector-dependent variance that possibly diverges as
$q\rightarrow0$. However, at the NA transition the order parameter is
the Fourier transform of the density at a large wavevector $q$ far
from $q=0$. We therefore expect that the disorder that couples to such
an order parameter to be short-range correlated in space.
\bibitem{commentGaussian} We have taken the distribution of disorder
to be Gaussian which relies on the assumption that at long length
scales universality will ensure that small deviations of the true
distribution from a Gaussian are not important, as long as the
distribution is short-ranged, which we expect on physical grounds.
\bibitem{q0flow} The choice of the field-rescaling exponent $\phi$ is
of course arbitrary. If instead of the convenient $\phi=0$ choice we
left it arbitrary, then to return the Hamiltonian into the form it had
before the RG transformation, we would have to also allow the smectic
wavevector $q_0$ to flow as $q_0(\ell)=q_0 e^{\phi\ell}$. In the end
the flow equations for the {\em dimensionless} coupling constants
would involve $q_0(\ell)$ and would be independent of the choice of
$\phi$.
\bibitem{comment_vd} This choice of having the nonsingular part of
$v_d$ minimize the elastic Hamiltonian is not necessary, but is
convenient because for such choice the dislocation degrees of freedom
$v_d$ decouple from the phonons $u$; see Appendix C.
\bibitem{D0comment} All $S_n$ are finite despite the vanishing of
$D(u)$ as $u\rightarrow 0^+$ and $u\rightarrow a^-$.
\bibitem{commentTheta} In Sec.\ref{random_field_3-5}, for a
periodic-disorder-only model (strictly only valid for $d>5$), we
explicitly demonstrated that this conclusion remained valid up to
logarithmic corrections even in the presence of nonlinear elasticity.
\end{references}
\end{document}